\documentclass[pdflatex,sn-mathphys-num]{sn-jnl}

\usepackage{graphicx}
\usepackage{amsmath,amssymb,amsfonts}
\usepackage{amsthm}
\usepackage{booktabs}
\usepackage{siunitx}
\usepackage{subcaption}
\usepackage{algorithm}
\usepackage{algpseudocode}
\usepackage{placeins}

\makeatletter
\@ifundefined{orcidlogo}
  {\newcommand{\orcidlogo}{}}
  {\renewcommand{\orcidlogo}{}}
\makeatother

\begin{document}

\title[Numerical simulations of random-matrix state reduction]
{Random-matrix reduction in projective quantum mechanics: Numerical simulations}

\author*[1]{\fnm{Alexey A.} \sur{Kryukov}}\email{kryukov@uwm.edu}

\affil*[1]{\orgdiv{Department of Mathematics \& Natural Sciences},
\orgname{University of Wisconsin--Milwaukee},
\orgaddress{\city{Milwaukee}, \state{WI}, \country{USA}}}

\abstract{
We present numerical simulations supporting the random-matrix state-reduction
framework developed in the companion theoretical paper. The simulations test
the main derived features of the model: isotropic diffusion generated by
Gaussian Unitary Ensemble Hamiltonians in projective state space, the
restriction of this diffusion to Brownian motion on the classical submanifold,
Born-rule frequencies for detector-defined outcome classes, and stroboscopic
Newtonian motion for macroscopic systems under repeated environmental
monitoring. We also compare GUE and GOE random Hamiltonians and show that GOE
fails to produce the required isotropic complex projective diffusion. Further
simulations examine finite-resolution detector records in the double-slit
experiment, Zeno stability of recorded equivalence classes, effective
irreversibility from high-dimensional state-space dynamics and loss of
{\bf (RM)} path information, and tensor-product particle-device dynamics in
the device limit. The results show that microscopic state reduction, stable
measurement records, effective irreversibility, and macroscopic classicality
can be described as different coarse-grained manifestations of the same
stochastic unitary mechanism.

}


\maketitle

\section{Introduction}
\label{sec:introduction}

The companion theoretical paper \cite{KryukovRMMain} develops a geometric
framework in which classical configuration space and classical phase space
arise as localized submanifolds of projective quantum state space. The
random-matrix conjecture \({\bf (RM)}\) asserts that measurement and
environmental interactions are described, after coarse-graining, by random
Hermitian Hamiltonians drawn from the Gaussian Unitary Ensemble. The resulting
stochastic unitary motion is homogeneous and isotropic in projective Hilbert
space with respect to the Fubini--Study metric.

The present paper gives numerical tests of several consequences of this
framework. We first test the finite-dimensional implication that GUE
Hamiltonians generate homogeneous and isotropic infinitesimal tangent steps in
complex projective space. We compare this behavior with GOE Hamiltonians, whose
real orthogonal invariance is insufficient to produce the required complex
projective isotropy.

We then examine the restriction of the same GUE-induced state-space diffusion
to localized classical submanifolds. In one dimension, projection onto the
tangent space of \(M_1^\sigma\) gives Gaussian increments in the classical
position coordinate and Brownian scaling under repeated steps. We also test the
\((\tau,s)\)-coordinates used to describe localization and reduction. These
coordinates define detector-relevant equivalence classes, and the simulations
show that the projected increments are Gaussian in the induced
Fubini--Study metric.

The subsequent simulations illustrate the microscopic measurement regime.
An unbiased reduction-coordinate walk reproduces Born-rule frequencies, and
finite-resolution screen records reproduce both coherent and which-slit
double-slit patterns. We also examine the formation of the interference pattern
as an accumulation of detector records and the apparent Zeno stability of a
recorded equivalence class under continuing \({\bf (RM)}\) motion.

The macroscopic regime is tested through conditioned stroboscopic records.
When the tangential diffusion between returns is small compared with the
detector resolution, and when the state returns frequently to the localized
sector, the recorded positions remain concentrated around Newtonian
trajectories. This gives a numerical version of macroscopic classicality as a
conditioned stochastic process on state space.

We also test the effective irreversibility of the {\bf (RM)} dynamics. For
each realized Hamiltonian history the evolution remains unitary and reversible,
but once the realized path is not retained, recurrence to a prescribed ray is
strongly suppressed by the high-dimensional geometry of projective state space.
We also compare exact unitary inversion with antiunitary time reversal for GUE
histories and examine the additional irreversibility produced when a
measurement record retains only a finite-resolution detector-defined
equivalence class.

Finally, we test the tensor-product structure needed for particle-device
systems and multi-particle measurements. The simulations show that the
projected GUE increments in the particle and device coordinates are
orthogonal and Gaussian in the induced Fubini--Study metric. In the device
limit, the particle state can move toward its detector-defined outcome class
while the device state remains in the same macroscopic equivalence class with
overwhelming probability. Components corresponding to displaced device classes
then have vanishingly small weight, providing the numerical counterpart of a
stable measurement record.

\section{Finite-dimensional projective set-up}
\label{sec:finite-dimensional-setup}

Let
\[
\mathcal H=\mathbb C^N,
\qquad
\mathbb{CP}^{N-1}=\mathbb P(\mathcal H)
\]
be the corresponding finite-dimensional projective Hilbert space. A normalized
state vector is denoted by \(\psi\), with \(\|\psi\|=1\). A short random
Hamiltonian step is
\[
\psi \longmapsto e^{-iH\Delta t}\psi .
\]
For the infinitesimal test considered here, the scale \(\Delta t\) is
irrelevant, and the first-order displacement is represented by
\[
\delta\psi=-iH\psi .
\]
The vertical phase direction is removed by projecting orthogonally to \(\psi\):
\[
\delta\psi_\perp
=
\delta\psi
-
\psi\langle \psi,\delta\psi\rangle .
\]
The Fubini--Study norm of the infinitesimal projective displacement is, to
first order,
\[
\|\delta\psi_\perp\|.
\]
Thus the distribution of \(\delta\psi_\perp\) is the local distribution of
projective tangent steps generated by the random Hamiltonian ensemble.

For the first finite-dimensional isotropy and homogeneity tests below, we take
\[
N=20,
\qquad
\dim_{\mathbb R}T_\psi\mathbb{CP}^{N-1}=2(N-1)=38.
\]
Later tensor-product simulations use dimensions specified separately in the corresponding sections. All results are based on independent samples of random Hamiltonians.

\section{Isotropy test at a fixed projective point}
\label{sec:isotropy-test}

\subsection{Set-up}

We first test isotropy at the projective point represented by
\[
\psi=e_0=(1,0,\ldots,0).
\]
For each random Hamiltonian \(H\), we compute
\[
\delta\psi_\perp
=
-iH\psi
-
\psi\langle\psi,-iH\psi\rangle .
\]
The nonzero transverse components are written in real and imaginary parts,
producing a vector in \(\mathbb R^{38}\). Isotropy of the infinitesimal
tangent-step distribution means that the covariance matrix of these real
tangent vectors is a scalar multiple of the identity.

For GUE, the off-diagonal entries \(H_{j0}\), \(j>0\), are circular complex
Gaussian variables. Hence the real and imaginary tangent coordinates should
have equal variance. For GOE, the off-diagonal entries \(H_{j0}\) are real.
Then \(-iH\psi\) has only imaginary transverse components at \(e_0\), and half
of the real tangent directions have zero variance. Therefore GOE cannot give an
isotropic tangent-step distribution in complex projective Hilbert space.

\subsection{Numerical results}

Using \(50{,}000\) independent samples, the covariance eigenvalues are
summarized in Table~\ref{tab:isotropy-summary}. The GUE covariance has no zero
eigenvalues and is nearly flat. The GOE covariance has \(19\) zero eigenvalues,
corresponding to half of the real tangent directions.

\begin{table}[h]
\centering
\caption{Covariance eigenvalues of infinitesimal projective tangent steps for
GUE and GOE Hamiltonians at \(\psi=e_0\).}
\label{tab:isotropy-summary}
\begin{tabular}{lccccc}
\toprule
Ensemble & Tangent dim. & Zero eigenvalues & Min eig. & Max eig. &
Mean \(\|\delta\psi_\perp\|^2\) \\
\midrule
GUE & 38 & 0  & 0.472 & 0.526 & 19.015 \\
GOE & 38 & 19 & 0.000 & 1.029 & 18.982 \\
\bottomrule
\end{tabular}
\end{table}

\begin{figure}[h]
\centering
\includegraphics[width=0.78\textwidth]{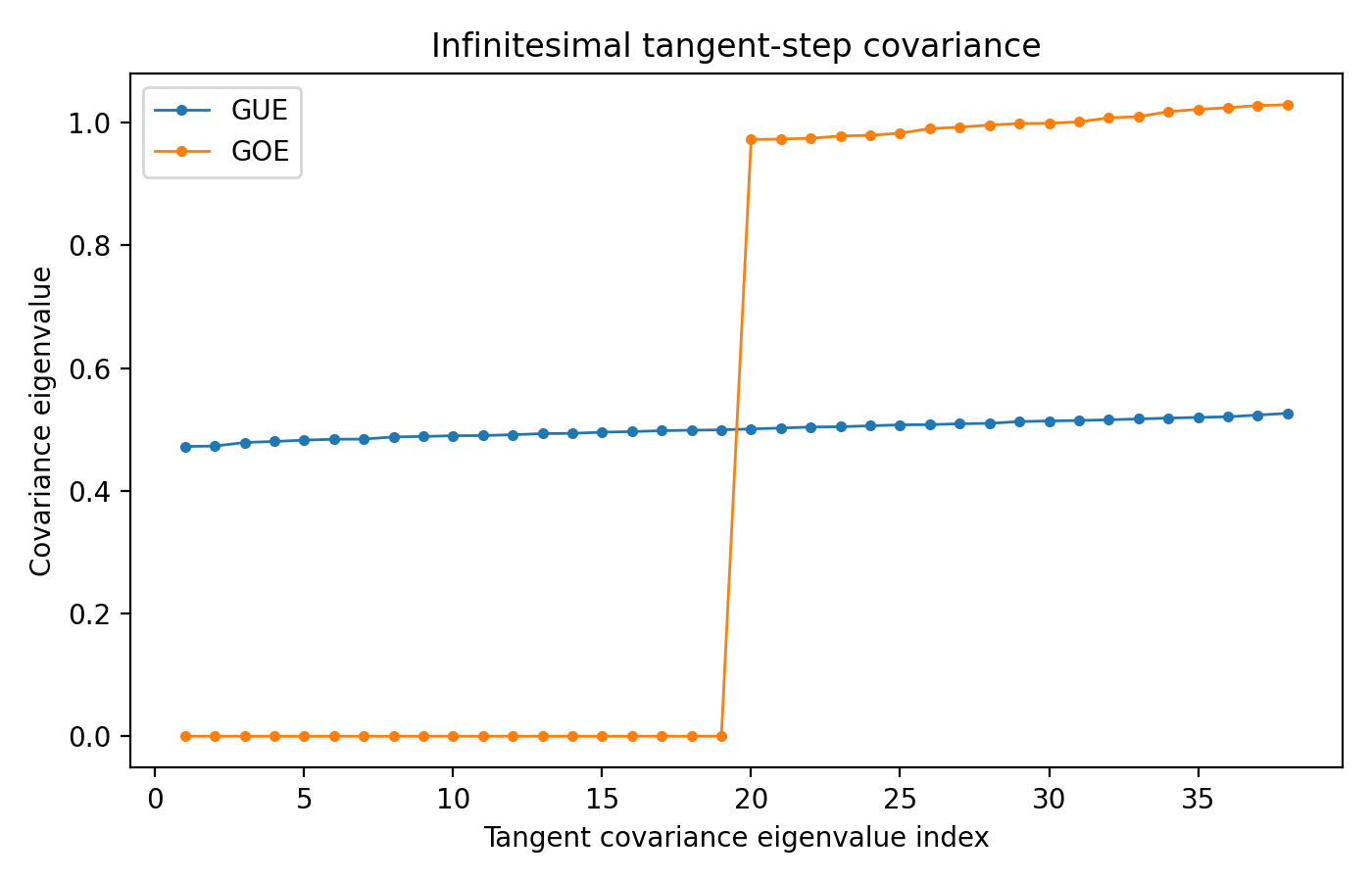}
\caption{Eigenvalues of the covariance matrix of infinitesimal tangent
displacements. GUE fills all real tangent directions approximately equally,
whereas GOE leaves half of the real tangent directions with zero variance.}
\label{fig:isotropy-eigenvalues}
\end{figure}

\begin{figure}[h]
\centering
\includegraphics[width=0.78\textwidth]{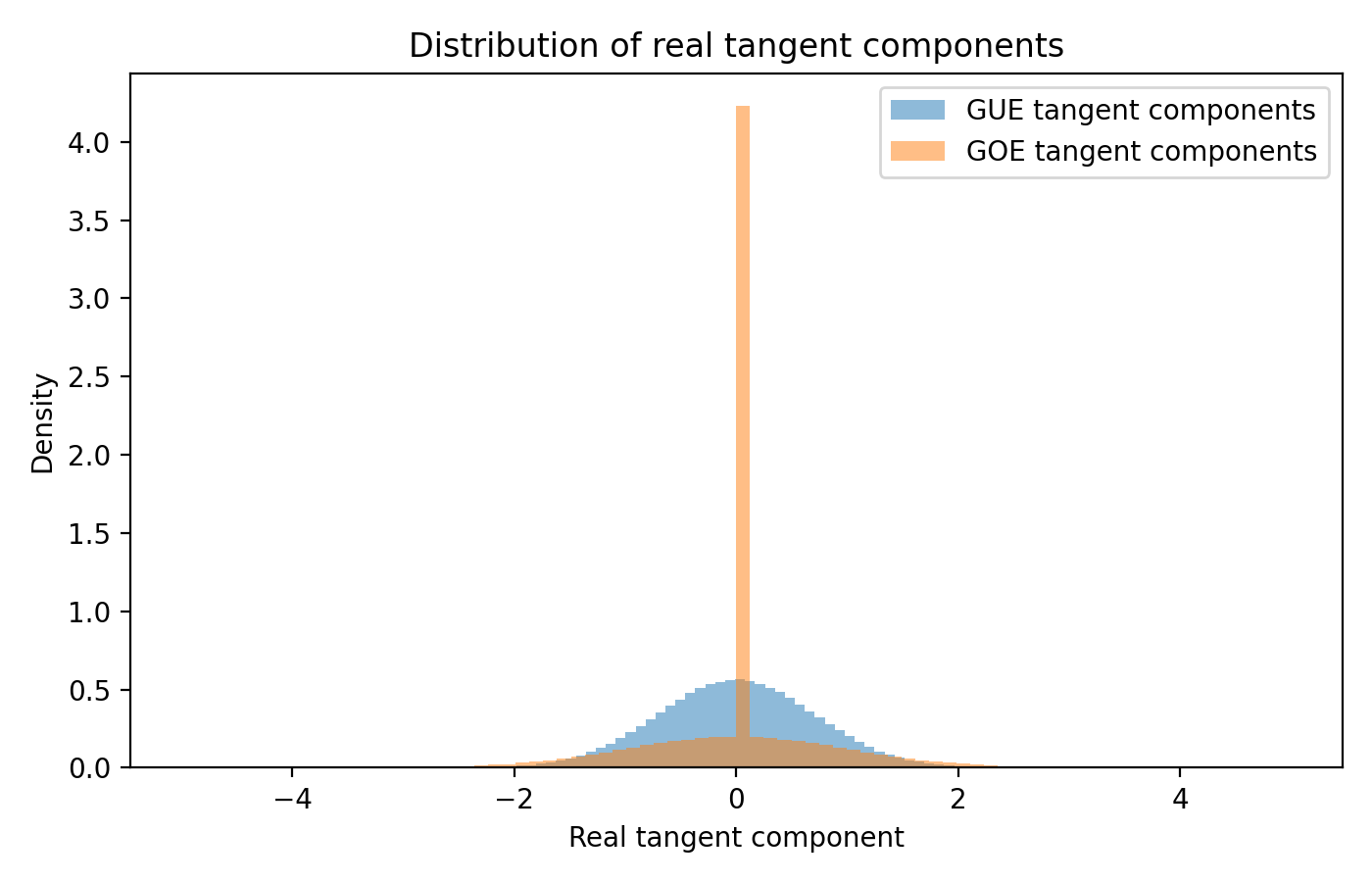}
\caption{Pooled real tangent components for GUE and GOE infinitesimal steps.
The spike at zero for GOE reflects the absence of fluctuations in half of the
real tangent directions at the chosen projective point.}
\label{fig:component-histogram}
\end{figure}

Figure~\ref{fig:isotropy-eigenvalues} shows the covariance eigenvalues. The GUE
eigenvalues are approximately constant, while the GOE eigenvalues split into
zero and nonzero groups. Figure~\ref{fig:component-histogram} shows the pooled
real tangent components for the two ensembles.

The total mean value of \(\|\delta\psi_\perp\|^2\) is nearly the same for GUE
and GOE, but the directional structure is different. GUE distributes this
variance across all real tangent directions, while GOE concentrates it in only
half of them. This confirms that GUE, not GOE, gives the required complex
projective isotropy.

\section{Homogeneity test across projective states}
\label{sec:homogeneity-test}

\subsection{Set-up}

Homogeneity means that the distribution of projective infinitesimal steps is
independent of the initial point of projective state space. To test this, we
compare the distribution of the projective step size
\[
\|\delta\psi_\perp\|^2
\]
at three different normalized states in \(\mathbb C^{20}\):
\[
\psi_1=e_0,
\qquad
\psi_2=\frac{1}{\sqrt N}(1,1,\ldots,1),
\qquad
\psi_3=\psi_{\mathrm{rand}},
\]
where \(\psi_{\mathrm{rand}}\) is a randomly chosen complex unit vector. For
each state and each ensemble, \(20{,}000\) independent Hamiltonians were
sampled. The resulting step-size distributions were compared using two-sample
Kolmogorov--Smirnov tests, using the distribution at \(e_0\) as the reference.

\subsection{Numerical results}

The results are shown in Table~\ref{tab:homogeneity-ks}. For GUE, the
step-size distributions at the three projective points are statistically
indistinguishable at this sample size. For GOE, the distribution changes
significantly when the initial state is genuinely complex.

\begin{table}[h]
\centering
\caption{Kolmogorov--Smirnov comparison of step-size distributions at different
initial states. The reference state is \(e_0\).}
\label{tab:homogeneity-ks}
\begin{tabular}{llcc}
\toprule
Ensemble & Comparison & KS statistic & \(p\)-value \\
\midrule
GUE & \(e_0\) vs. real superposition & 0.0098 & 0.290 \\
GUE & \(e_0\) vs. complex random state & 0.0066 & 0.774 \\
GOE & \(e_0\) vs. real superposition & 0.0139 & 0.0416 \\
GOE & \(e_0\) vs. complex random state & 0.1386 & \(6.46\times10^{-168}\) \\
\bottomrule
\end{tabular}
\end{table}

\begin{figure}[h]
\centering
\includegraphics[width=0.82\textwidth]{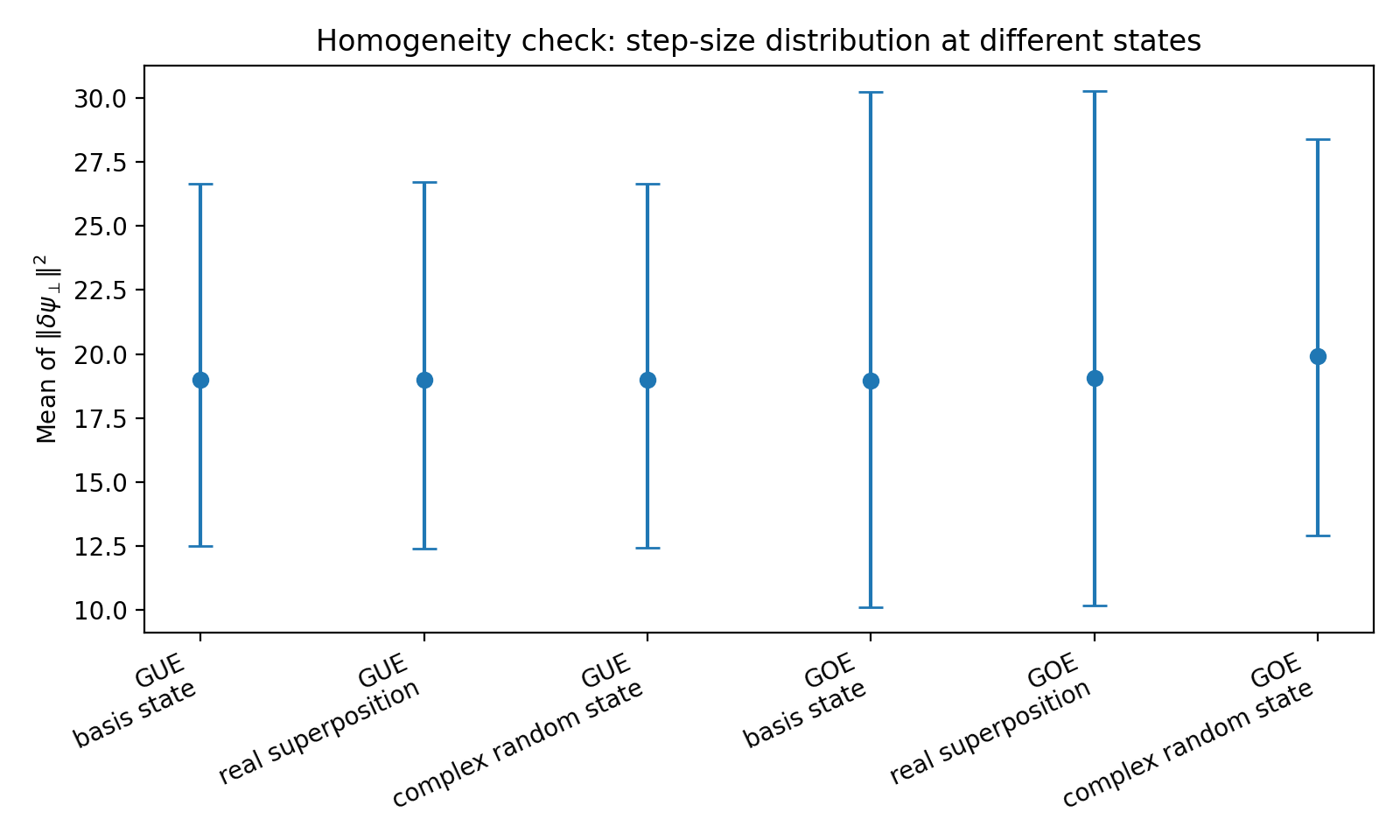}
\caption{Homogeneity check for the distribution of
\(\|\delta\psi_\perp\|^2\). For GUE, the distribution is independent of the
tested initial projective state. For GOE, the distribution changes for a
genuinely complex state, reflecting the preferred real structure of the
ensemble.}
\label{fig:homogeneity-step-norms}
\end{figure}

Figure~\ref{fig:homogeneity-step-norms} shows the mean step-size values with
five-percent and ninety-five-percent quantile ranges. The GUE values are stable
across the tested projective states. GOE shows a marked change for a complex
initial state.

\section{Interpretation of the GUE/GOE tests}
\label{sec:gue-goe-interpretation}

The two tests confirm the basic finite-dimensional behavior required by the
random-matrix conjecture \({\bf (RM)}\). First, GUE Hamiltonians generate an
isotropic infinitesimal tangent-step distribution at a fixed projective point.
Second, the step-size distribution is homogeneous across projective state
space. These are the finite-dimensional numerical signatures of homogeneous
and isotropic diffusion in \(\mathbb{CP}^{N-1}\) with the Fubini--Study
geometry.

The GOE comparison clarifies why real orthogonal invariance is insufficient.
GOE Hamiltonians may have a comparable total mean step size, but the direction
distribution is not the complex projective one. At \(e_0\), half of the real
tangent directions have zero variance. At a genuinely complex initial state,
the step-size distribution itself changes. Thus GOE selects a preferred real
structure and fails to produce the homogeneous and isotropic complex
projective walk needed in the \({\bf (RM)}\) framework.

\subsection{Summary for the subsequent simulations}

The GUE/GOE comparison verifies the basic numerical consequence of the
\({\bf (RM)}\) assumption used in the remaining sections. In finite-dimensional
projective Hilbert space, GUE random Hamiltonians generate infinitesimal
stochastic unitary steps whose tangent covariance is isotropic and whose
step-size distribution is independent of the initial projective state. GOE
Hamiltonians do not satisfy these properties in the complex projective setting.
The following simulations therefore use GUE-induced motion to test its
restriction to Brownian motion on the classical submanifold, its projected
\((\tau,s)\)-description, Born-rule frequencies for detector-defined outcome
classes, and stroboscopic Newtonian motion.

\section{Brownian motion on the classical submanifold}
\label{sec:brownian-restriction}

\subsection{Purpose of the simulation}

The companion theoretical paper argues that the same \({\bf (RM)}\)-induced
state-space diffusion has two complementary manifestations. In the full
projective state space it gives homogeneous and isotropic stochastic unitary
motion, while its tangential restriction to the localized classical submanifold
\(M_3^\sigma\) gives ordinary Brownian motion. In the present section we test
this statement numerically in the one-dimensional case.

We work with the classical position submanifold \(M_1^\sigma\), represented by
localized Gaussian states
\[
g_{a,\sigma}(x)
=
\left(\frac{1}{2\pi\sigma^2}\right)^{1/4}
\exp\left[-\frac{(x-a)^2}{4\sigma^2}\right].
\]
The coordinate \(a\) labels the classical position. The tangent direction to
\(M_1^\sigma\) at \(g_{a,\sigma}\) is represented by
\[
\frac{\partial g_{a,\sigma}}{\partial a}
=
\frac{x-a}{2\sigma^2}\,g_{a,\sigma}.
\]
The goal is to show that the tangential component of the \({\bf (RM)}\)-induced
state-space increment gives a Gaussian random increment \(\Delta a\), with a
distribution independent of \(a\). Repeated increments should then converge to
Brownian motion on \(M_1^\sigma\simeq \mathbb R\).

\subsection{Numerical set-up}

The Hilbert space \(L_2(\mathbb R)\) was approximated on a uniform grid
\[
x\in[-8,8],
\qquad
N=512,
\qquad
\sigma=0.75.
\]
The Gaussian \(g_{a,\sigma}\) and its derivative
\(\partial_a g_{a,\sigma}\) were discretized and normalized using the discrete
approximation to the \(L_2\)-inner product.

For a GUE-induced infinitesimal state-space displacement, we write
\[
\delta\psi=-iH\psi ,
\]
with the vertical phase direction removed:
\[
\delta\psi_\perp
=
\delta\psi
-
\psi\langle\psi,\delta\psi\rangle .
\]
The induced classical increment \(\Delta a\) is obtained by projecting
\(\delta\psi_\perp\) onto the tangent vector \(\partial_a g_{a,\sigma}\):
\[
\Delta a
=
\frac{
\operatorname{Re}
\left\langle
\partial_a g_{a,\sigma},
\delta\psi_\perp
\right\rangle
}{
\left\|
\partial_a g_{a,\sigma}
\right\|^2
}.
\]
This is the infinitesimal coordinate displacement on \(M_1^\sigma\) induced by
the state-space random step.

The simulation used \(50{,}000\) independent samples at each of the three
centers
\[
a=-2,\qquad a=0,\qquad a=2.
\]
These three choices test translation invariance along \(M_1^\sigma\).

\subsection{Gaussianity and translation invariance of the increments}

The projected increments \(\Delta a\) are Gaussian to high numerical accuracy.
Table~\ref{tab:sim2-increment-summary} gives the sample means and standard
deviations. The means are close to zero and the standard deviations are the
same, within sampling error, at the three tested centers.

\begin{table}[h]
\centering
\caption{Projected tangential increments on \(M_1^\sigma\). The distribution is
Gaussian with approximately the same variance at different centers \(a\).}
\label{tab:sim2-increment-summary}
\begin{tabular}{lcccc}
\toprule
Center \(a\) & Samples & Mean \(\Delta a\) & Std. dev. & Variance \\
\midrule
\(-2\) & \(50{,}000\) & \(0.0052\)  & \(1.0584\) & \(1.1202\) \\
\(0\)  & \(50{,}000\) & \(-0.0048\) & \(1.0651\) & \(1.1345\) \\
\(2\)  & \(50{,}000\) & \(-0.0067\) & \(1.0625\) & \(1.1288\) \\
\bottomrule
\end{tabular}
\end{table}

The normality and translation-invariance checks are summarized in
Table~\ref{tab:sim2-ks}. For each value of \(a\), the standardized increments
were compared with the standard normal distribution. In addition, the increment
distributions at \(a=-2\) and \(a=2\) were compared with the distribution at
\(a=0\). The Kolmogorov--Smirnov tests show no statistically significant
departure from normality or translation invariance.

\begin{table}[h]
\centering
\caption{Kolmogorov--Smirnov tests for Gaussianity and translation invariance
of the projected increments.}
\label{tab:sim2-ks}
\begin{tabular}{llcc}
\toprule
Test & Comparison & KS statistic & \(p\)-value \\
\midrule
Normality & \(a=-2\) vs. \(N(0,1)\) & \(0.0030\) & \(0.767\) \\
Normality & \(a=0\) vs. \(N(0,1)\)  & \(0.0022\) & \(0.973\) \\
Normality & \(a=2\) vs. \(N(0,1)\)  & \(0.0024\) & \(0.939\) \\
Translation invariance & \(a=0\) vs. \(a=-2\) & \(0.0063\) & \(0.266\) \\
Translation invariance & \(a=0\) vs. \(a=2\)  & \(0.0034\) & \(0.937\) \\
\bottomrule
\end{tabular}
\end{table}

\begin{figure}[h]
\centering
\includegraphics[width=0.78\textwidth]{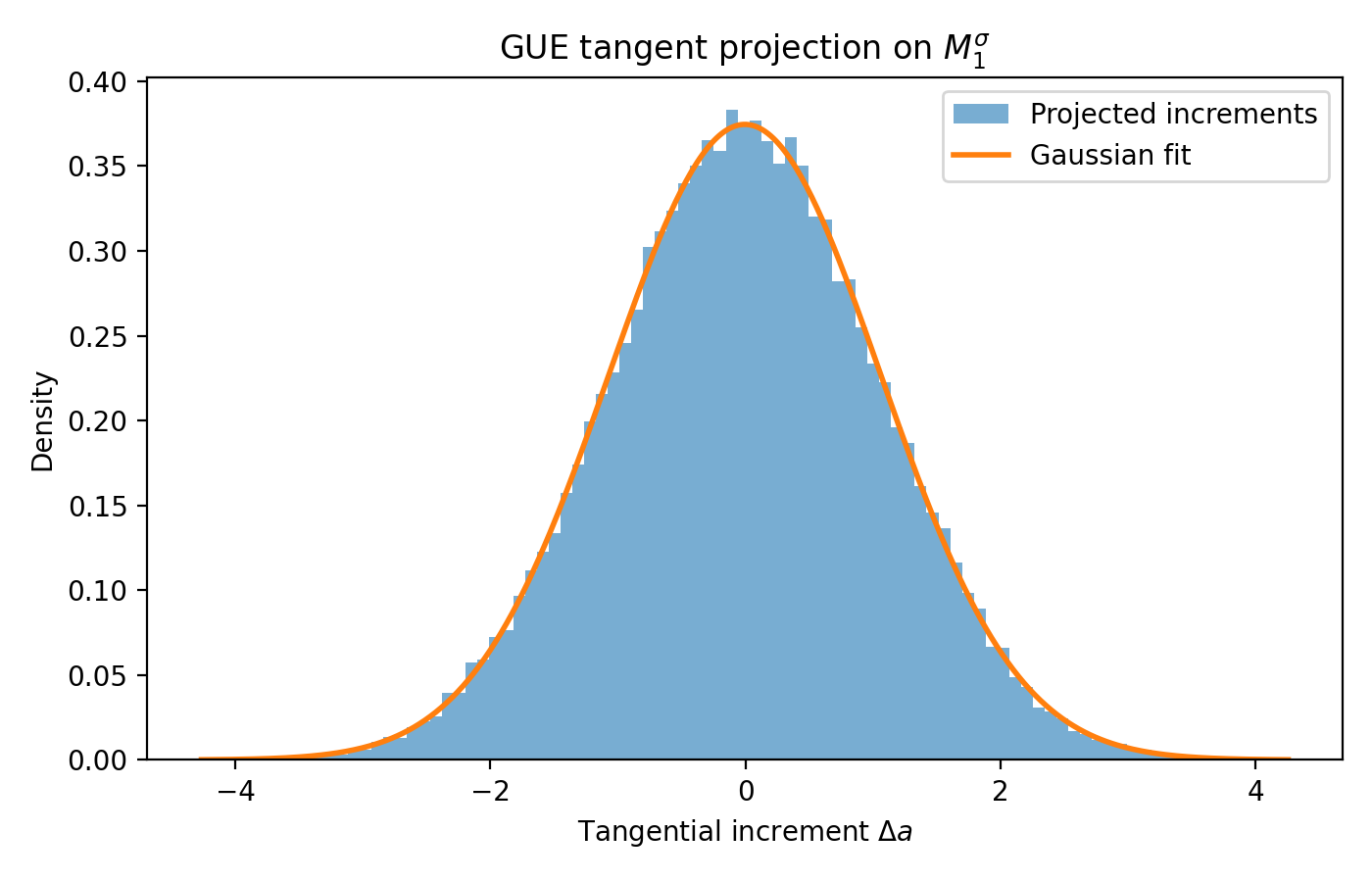}
\caption{Histogram of tangential increments \(\Delta a\) obtained by projecting
GUE-induced state-space increments onto the tangent direction of
\(M_1^\sigma\) at \(a=0\). The distribution agrees with a Gaussian fit.}
\label{fig:sim2-increment-histogram}
\end{figure}

Figure~\ref{fig:sim2-increment-histogram} shows the histogram of projected
increments at \(a=0\), together with a Gaussian fit.

\subsection{Brownian scaling under repeated projected steps}

To test the Brownian limit, we used the measured one-step standard deviation at
\(a=0\),
\[
\sigma_{\Delta a}=1.0651,
\]
and simulated \(30{,}000\) random walks with
\[
n=400
\]
independent projected increments. Brownian scaling predicts that the endpoint
distribution should be Gaussian with variance
\[
n\sigma_{\Delta a}^2.
\]
The predicted endpoint standard deviation is therefore
\[
\sqrt{n}\,\sigma_{\Delta a}=21.3023.
\]
The simulated endpoint standard deviation was
\[
21.2708,
\]
in excellent agreement with the Brownian prediction.

\begin{table}[h]
\centering
\caption{Endpoint distribution after \(400\) projected GUE increments.}
\label{tab:sim2-endpoint}
\begin{tabular}{lcc}
\toprule
Quantity & Simulation & Brownian prediction \\
\midrule
Number of paths & \(30{,}000\) & -- \\
Number of steps & \(400\) & -- \\
Endpoint mean & \(0.0358\) & \(0\) \\
Endpoint standard deviation & \(21.2708\) & \(21.3023\) \\
Endpoint variance & \(452.45\) & \(453.79\) \\
\bottomrule
\end{tabular}
\end{table}

\begin{figure}[h]
\centering
\includegraphics[width=0.78\textwidth]{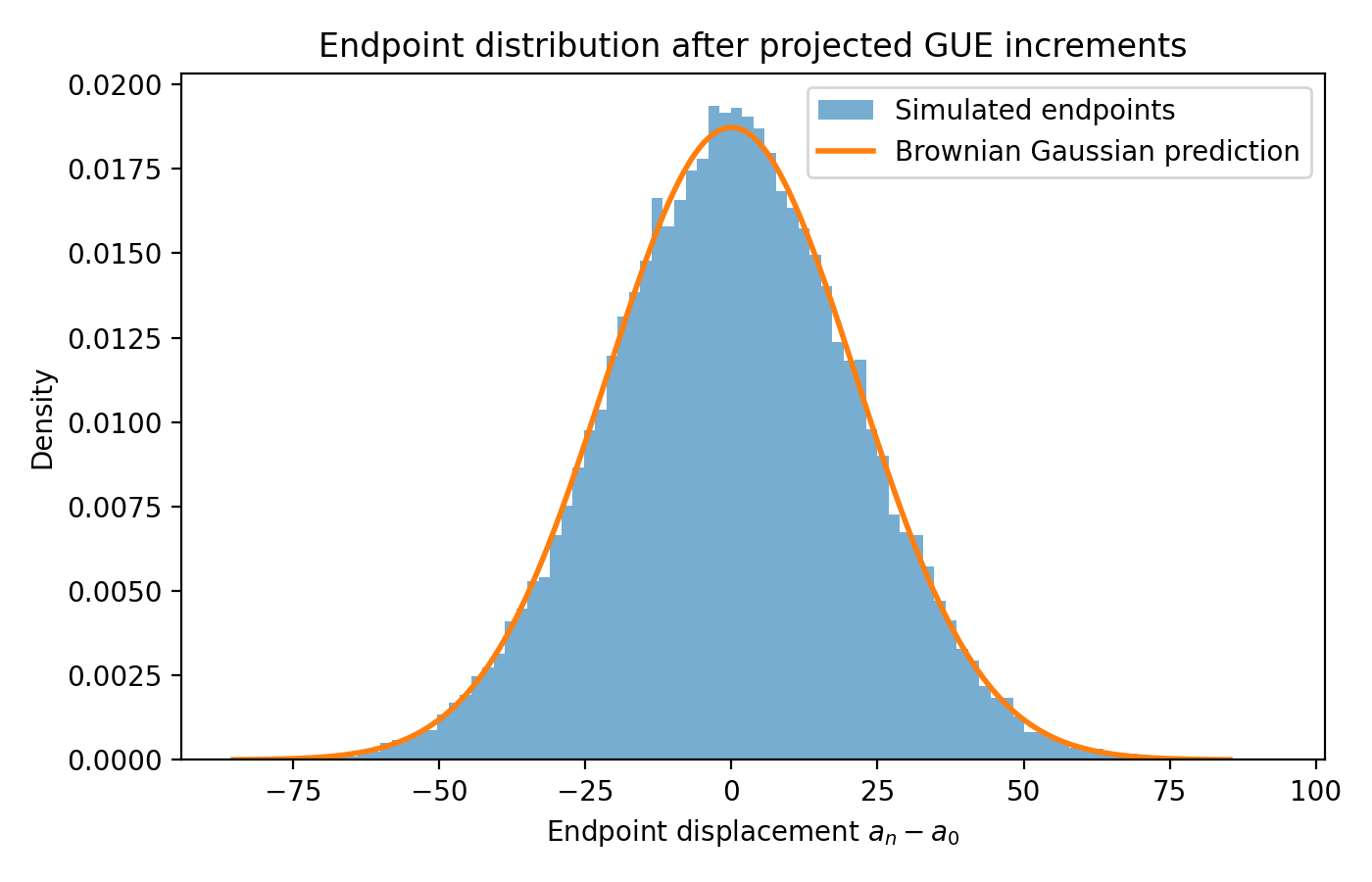}
\caption{Endpoint distribution after \(400\) projected GUE-induced tangent
increments on \(M_1^\sigma\). The histogram agrees with the Gaussian
distribution predicted by Brownian scaling.}
\label{fig:sim2-endpoint-distribution}
\end{figure}

Figure~\ref{fig:sim2-endpoint-distribution} compares the simulated endpoint
distribution with the predicted Brownian Gaussian.

\subsection{Interpretation}

The simulation confirms the expected restriction of \({\bf (RM)}\)-induced
state-space diffusion to the classical submanifold. The tangential projection
of the GUE-induced infinitesimal displacement onto \(M_1^\sigma\) gives a
Gaussian random increment in the classical coordinate \(a\). The distribution
is independent of \(a\), confirming translation invariance along the classical
submanifold. Repeated projected increments give a Gaussian endpoint
distribution with variance proportional to the number of steps.

Thus, in the one-dimensional numerical model,
\[
{\bf (RM)}\text{-induced diffusion in projective state space}
\quad\longrightarrow\quad
\text{Brownian motion on }M_1^\sigma
\]
when the state-space increments are restricted to their tangential components
along the classical submanifold. This verifies numerically the mechanism used in
the companion paper to identify the normal distribution of classical
measurement errors as the restriction of the same state-space diffusion that
produces Born-rule probabilities in the full projective space.

\section{Calibrated \({\bf (RM)}\) motion in \((\tau,s)\)-coordinates}
\label{sec:tau-s-calibration}

\subsection{Purpose of the simulation}

The preceding simulations verified two basic ingredients of the framework:
GUE Hamiltonians generate homogeneous and isotropic infinitesimal motion in
projective state space, and the tangential restriction of this motion to
\(M_1^\sigma\) gives Brownian motion in the classical position coordinate. We
now use these facts to introduce the coordinate system used to describe
localized detector records.

For a localized one-dimensional state, define
\[
\tau=\langle x\rangle,
\qquad
\delta_x=
\left(
\langle x^2\rangle-\langle x\rangle^2
\right)^{1/2},
\qquad
s=\ln(\delta_x/\sigma),
\]
where \(\sigma\) is the detector resolution scale. The coordinate \(\tau\)
labels the recorded position, while \(s\) measures localization relative to the
detector resolution. The localized sector is
\[
s\leq 0.
\]
A detector-defined outcome class is therefore represented, at finite
resolution, by a condition of the form
\[
s\leq 0,
\qquad
\tau\in I_j,
\]
where \(I_j\) is the position interval corresponding to the \(j\)-th outcome.

The purpose of the present simulation is to check the local behavior of the
GUE-induced walk in the \((\tau,s)\)-coordinates and to verify the induced
Fubini--Study metric used in the reduction-coordinate description.

\subsection{Gaussian surface and tangent coordinates}

We use the two-dimensional Gaussian surface
\[
\psi_{\tau,s}(x)=g_{\tau,\delta}(x),
\qquad
\delta=\sigma e^s,
\]
where
\[
g_{\tau,\delta}(x)
=
\left(\frac{1}{2\pi\delta^2}\right)^{1/4}
\exp\left[-\frac{(x-\tau)^2}{4\delta^2}\right].
\]
The tangent directions are
\[
\partial_\tau g_{\tau,\delta}
=
\frac{x-\tau}{2\delta^2}\,g_{\tau,\delta},
\]
and
\[
\partial_s g_{\tau,\delta}
=
\left(
-\frac{1}{2}
+
\frac{(x-\tau)^2}{2\delta^2}
\right)
g_{\tau,\delta}.
\]
The first direction changes the mean position, while the second changes the
width, or localization scale, of the state.

The Hilbert space \(L^2(\mathbb R)\) was approximated by functions on a
uniform grid
\[
x\in[-10,10],
\qquad
N=256,
\qquad
\sigma=0.75.
\]
Here \(N\) is the number of grid points, and hence the dimension of the
finite-dimensional numerical approximation. The corresponding grid spacing is approximately
\[
\Delta x=\frac{20}{255}\approx0.078,
\]
so the Gaussian width \(\sigma=0.75\) is resolved by about ten grid intervals.
The interval \([-10,10]\) is sufficiently large compared with the Gaussian
centers and widths used in the simulation, so boundary effects from the finite
grid are negligible.

At each sampled point \((\tau,s)\), the two tangent vectors were orthogonally
projected to the projective tangent space, and their Gram matrix
\[
G_{ij}
=
\operatorname{Re}\langle T_i,T_j\rangle
\]
was computed, where \(T_1=\partial_\tau g_{\tau,\delta}\) and
\(T_2=\partial_s g_{\tau,\delta}\), after removal of the vertical phase
direction.

For \(s=0\), the numerical metric coefficients were
\[
G_{\tau\tau}=0.4444,
\qquad
G_{ss}=0.5000,
\qquad
G_{\tau s}\approx 0.
\]
Thus the \(\tau\)- and \(s\)-directions are orthogonal, as required by the
geometric decomposition used in the theoretical paper.

\subsection{Projected GUE increments}
\label{projectedGUE}

At each point \((\tau,s)\), infinitesimal GUE-induced projective increments
were sampled and projected onto the two tangent directions. Equivalently, for a
state \(\psi_{\tau,s}\), we sample the projective displacement
\[
\delta\psi_\perp
=
-iH\psi_{\tau,s}
-
\psi_{\tau,s}
\langle
\psi_{\tau,s},
-iH\psi_{\tau,s}
\rangle,
\qquad
H\in \mathrm{GUE},
\]
and write its projection onto the \((\tau,s)\)-surface as
\[
\delta\psi_\perp
\simeq
\Delta\tau\,\partial_\tau g_{\tau,\delta}
+
\Delta s\,\partial_s g_{\tau,\delta}.
\]
The coefficients \(\Delta\tau\) and \(\Delta s\) are the induced increments in
the recorded-position and localization coordinates.

For numerical stability, we report the normalized increments
\[
\sqrt{G_{\tau\tau}}\,\Delta\tau,
\qquad
\sqrt{G_{ss}}\,\Delta s.
\]
The simulation used \(20{,}000\) independent samples at each of the four points
\[
(\tau,s)=(-2,0),\quad (0,0),\quad (2,0),\quad (0,0.5).
\]

\subsection{Results in the \((\tau,s)\)-coordinates}

The projected increments are Gaussian and essentially uncorrelated in the
orthogonal \((\tau,s)\)-directions. Table~\ref{tab:tau-s-summary} gives the
main covariance data.

\begin{table}[h]
\centering
\caption{Projected GUE increments in the \((\tau,s)\)-coordinates. The last
column gives the correlation between the normalized \(\Delta\tau\) and
\(\Delta s\) increments.}
\label{tab:tau-s-summary}
\begin{tabular}{lcccc}
\toprule
Point & \(G_{\tau\tau}\) & \(G_{ss}\) & \(G_{\tau s}\) &
Corr\((\Delta\tau,\Delta s)\) \\
\midrule
\((\tau,s)=(-2,0)\) & 0.4444 & 0.5000 & \(2.0\times10^{-17}\) & 0.0059 \\
\((\tau,s)=(0,0)\)  & 0.4444 & 0.5000 & \(4.7\times10^{-18}\) & \(-0.0033\) \\
\((\tau,s)=(2,0)\)  & 0.4444 & 0.5000 & \(-3.9\times10^{-17}\) & 0.0060 \\
\((\tau,s)=(0,0.5)\) & 0.1635 & 0.5000 & \(-3.7\times10^{-18}\) & \(-0.0054\) \\
\bottomrule
\end{tabular}
\end{table}

The normality of the projected increments was tested by comparing the
standardized \(\tau\)- and \(s\)-increments with the standard normal
distribution. The Kolmogorov--Smirnov \(p\)-values are shown in
Table~\ref{tab:tau-s-normality}. All values are large, confirming that the
increment distributions are well described by Gaussians.

\begin{table}[h]
\centering
\caption{Kolmogorov--Smirnov normality tests for the normalized projected
increments.}
\label{tab:tau-s-normality}
\begin{tabular}{llcc}
\toprule
Point & Coordinate & KS statistic & \(p\)-value \\
\midrule
\((\tau,s)=(-2,0)\) & \(\tau\) & 0.0038 & 0.932 \\
\((\tau,s)=(-2,0)\) & \(s\)    & 0.0051 & 0.671 \\
\((\tau,s)=(0,0)\)  & \(\tau\) & 0.0044 & 0.837 \\
\((\tau,s)=(0,0)\)  & \(s\)    & 0.0035 & 0.969 \\
\((\tau,s)=(2,0)\)  & \(\tau\) & 0.0035 & 0.968 \\
\((\tau,s)=(2,0)\)  & \(s\)    & 0.0042 & 0.875 \\
\((\tau,s)=(0,0.5)\) & \(\tau\) & 0.0031 & 0.989 \\
\((\tau,s)=(0,0.5)\) & \(s\)    & 0.0034 & 0.973 \\
\bottomrule
\end{tabular}
\end{table}

\begin{figure}[h]
\centering
\includegraphics[width=0.68\textwidth]{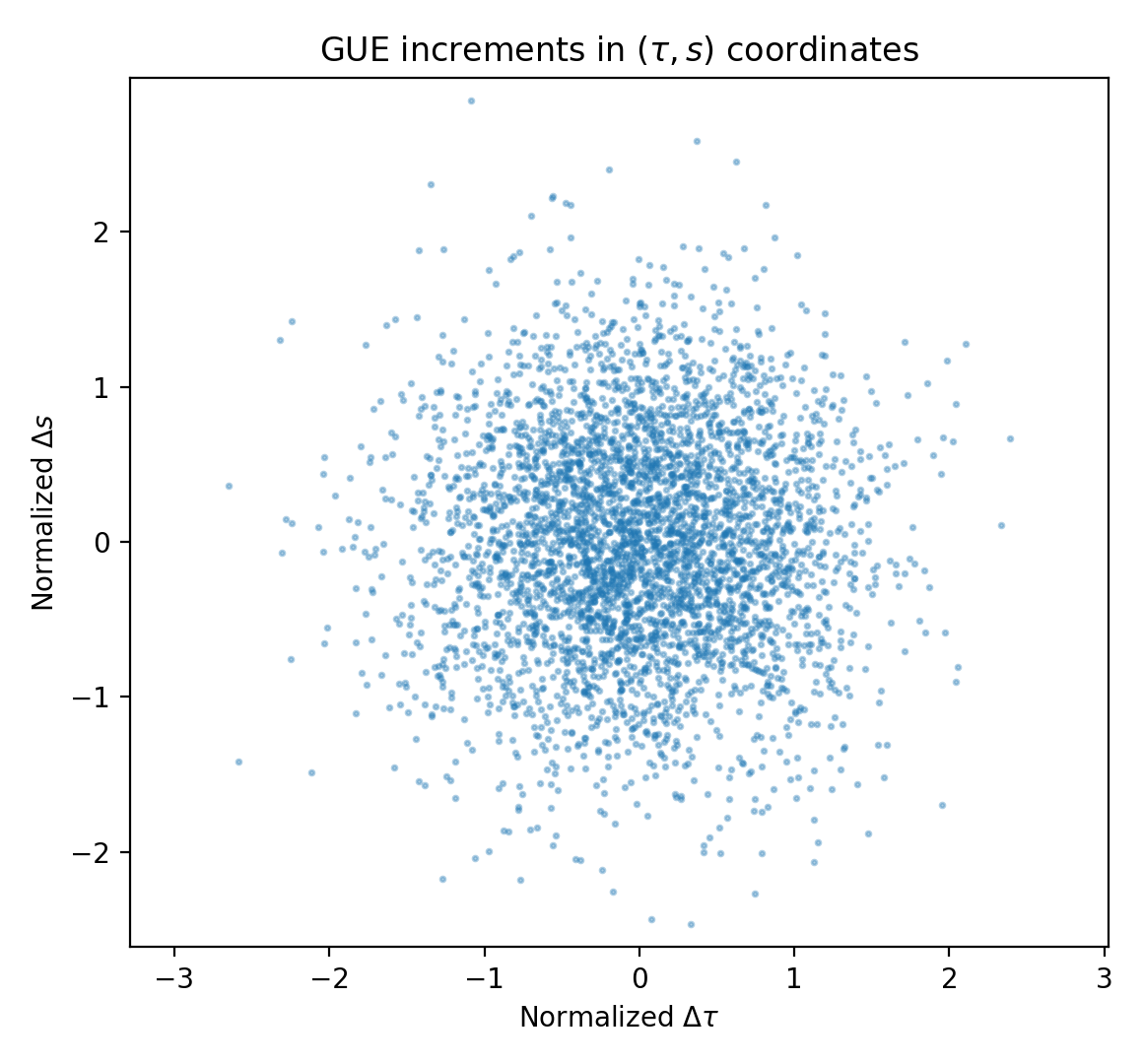}
\caption{Normalized GUE-induced increments in the \((\tau,s)\)-coordinates at
\((\tau,s)=(0,0)\). The circular cloud shows that the recorded-position
direction \(\tau\) and the localization direction \(s\) are orthogonal and have
uncorrelated Gaussian increments after normalization by the induced metric.}
\label{fig:tau-s-scatter}
\end{figure}

Figure~\ref{fig:tau-s-scatter} shows the normalized increments at
\((\tau,s)=(0,0)\). The approximately circular scatter plot is the expected
local signature of isotropic GUE-induced motion in the orthogonal
\((\tau,s)\)-coordinates.

\subsection{Calibration on the classical submanifold}

The Brownian scale on the classical submanifold is fixed by the transition
weight between localized Gaussian representatives. For equal-width Gaussians,
\begin{equation}
\label{calibration}
\cos^2\rho(g_{a,\sigma},g_{b,\sigma})
=
\exp\left[-\frac{(a-b)^2}{4\sigma^2}\right].
\end{equation}
Thus the Fubini--Study distance on \(M_1^\sigma\) reproduces the Gaussian
dependence of the classical normal law. This calibrates the state-space
diffusion scale by matching its restriction to \(M_1^\sigma\) with the
Brownian diffusion coefficient in the classical position coordinate.

\begin{figure}[h]
\centering
\includegraphics[width=0.76\textwidth]{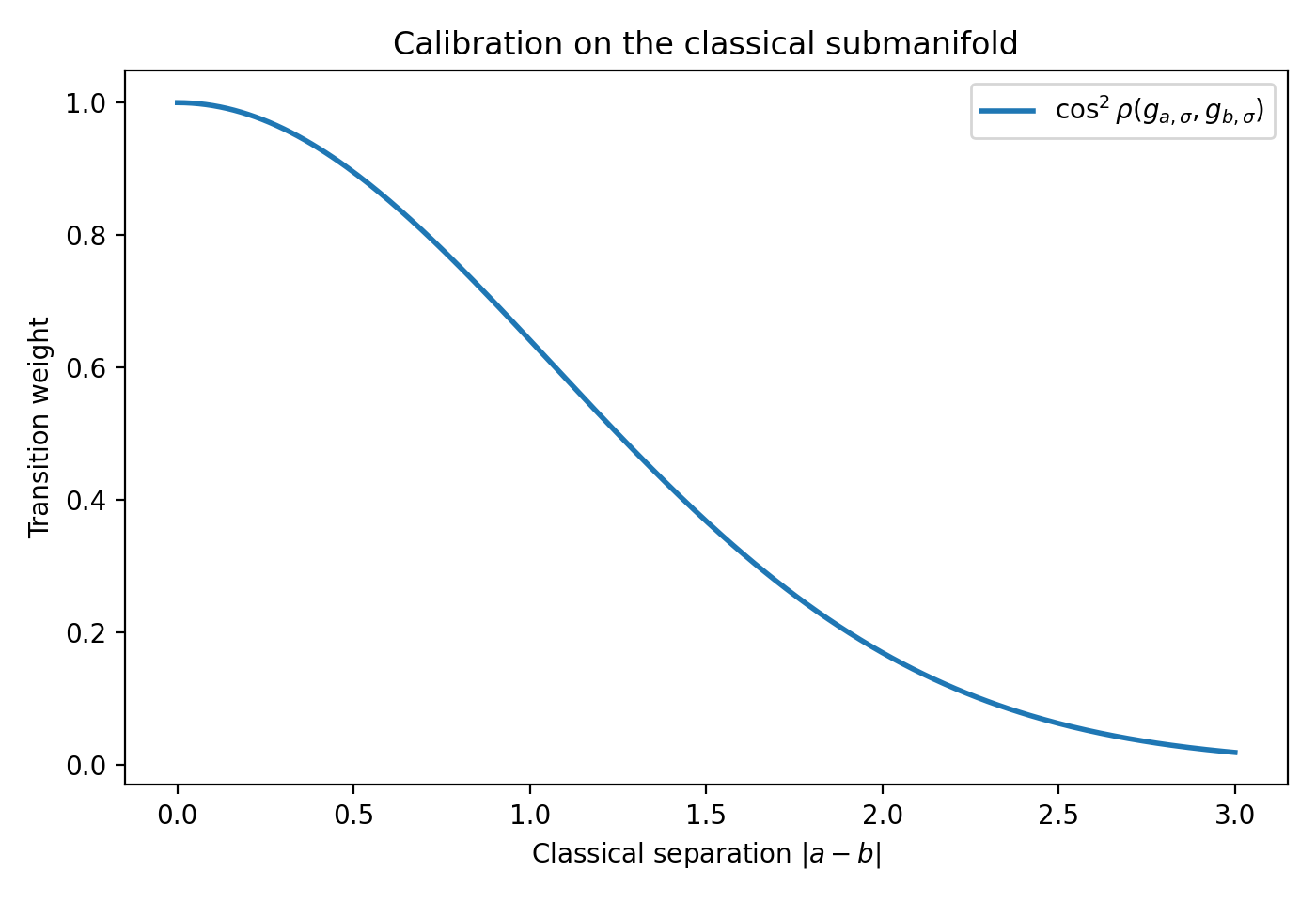}
\caption{Calibration on the classical submanifold. For equal-width Gaussian
representatives, the distance-dependent weight
\(\cos^2\rho(g_{a,\sigma},g_{b,\sigma})\) has the Gaussian form required by the
normal law for classical measurement errors.}
\label{fig:normal-calibration}
\end{figure}

Figure~\ref{fig:normal-calibration} shows the corresponding transition weight
as a function of the classical separation \(|a-b|\).

\subsection{Path weights and detector classes}
\label{subsec:calibration-and-weights}

The calibration~\eqref{calibration} is introduced on the classical submanifold
\(M_1^\sigma\). There, the projected \({\bf (RM)}\)-induced diffusion becomes
ordinary Brownian motion in the classical coordinate, and the transition
weights agree with the Brownian path weights associated with detector bins.

For states not lying on \(M_1^\sigma\), the same calibrated transition law
extends to detector-defined equivalence classes in projective state space. By
unitary invariance of the \({\bf (RM)}\) dynamics, transition probabilities can
depend only on Fubini--Study distance. Since the law has already been fixed on
the Gaussian submanifold \(M_1^\sigma\), its extension to arbitrary states is
therefore determined by the corresponding distances to the detector classes.

This is analogous to a classical Brownian measurement. If detectors are placed
in specified regions of \(\mathbb R^3\), and a Brownian particle starts at a
given point, the probability associated with a detector is the total weight of
Brownian paths that reach, or are recorded in, the corresponding region. This
weight is determined by the Brownian process, the geometry of the detector
regions, and the initial position of the particle. In the same way, the
detector-cell weights used below are aggregate weights of \({\bf (RM)}\)
state-space paths whose endpoints lie in the corresponding detector-defined
equivalence classes.

This extension is also needed because, for an initial state
\(\psi\notin M_1^\sigma\), the leaves of the foliation by fixed detector
coordinates \((\tau,s)\) carry no canonical volume measure. Thus detector
probabilities cannot be defined by uniformly counting paths through the leaves.
Instead, the calibrated \({\bf (RM)}\) path measure assigns weights to sets of
state-space paths, and the weight of a detector class is the aggregate weight of
all paths reaching that class.

For a detector class \(\mathcal C_j\), the relevant geometric quantity is the
Fubini--Study distance from the state to that class,
\[
\rho(\psi,\mathcal C_j)
=
\inf_{\chi\in\mathcal C_j}\rho(\psi,\chi).
\]
When the detector classes are represented by mutually orthogonal subspaces,
with projectors \(P_j\), this distance satisfies
\[
\cos^2\rho(\psi,\mathcal C_j)
=
\|P_j\psi\|^2.
\]

Thus the unique extension of the calibration~\eqref{calibration} yields
\[
W_j(\psi)=\|P_j\psi\|^2.
\]
In the position representation, for a finite-resolution detector bin \(I_j\),
this becomes
\[
W_j(\psi)=\|P_{I_j}\psi\|^2
=
\int_{I_j}|\psi(x)|^2\,dx.
\]

To reiterate, the detector-cell weight is determined geometrically: it depends
on the Fubini--Study distance from the state to the cell, viewed as the
corresponding detector-defined equivalence class. More precisely, this weight
represents the aggregate contribution of all {\bf (RM)} state-space paths
that reach the cell after a fixed number of steps; the farther the measured
state is from the cell, the smaller this aggregate contribution is.

\section{Born-rule frequencies from a reduction-coordinate walk}
\label{sec:born-frequency-simulation}

\subsection{Purpose of the simulation}

The preceding section fixed the detector-cell weights as aggregate path weights
associated with finite-resolution equivalence classes. We now test the corresponding record
process in the simplest two-outcome case. The effective reduction coordinate is
an unbiased one-dimensional walk on an interval, with the two endpoints
representing the two detector-defined outcome classes. Once the projected state
enters one of these classes, the corresponding outcome is recorded.

\subsection{Reduction coordinate and hitting probability}

Suppose that the two detector-defined outcome classes are represented by the
endpoints of an interval
\[
[a,b].
\]
An unbiased one-dimensional walk in the reduction coordinate \(\tau\), starting
at
\[
\tau_0\in[a,b],
\]
has absorbing endpoints \(a\) and \(b\). The probability of reaching \(b\)
before \(a\) is the affine coordinate of the starting point:
\[
\mathbb P(\text{hit }b)
=
\frac{\tau_0-a}{b-a}.
\]
This is the standard gambler's-ruin hitting probability.

For the spin-type two-outcome model, we take
\[
[a,b]=[-1,1].
\]
Let the endpoint classes correspond to the eigenstates \(|1\rangle\) and
\(|0\rangle\), with eigenvalues \(+1\) and \(-1\). For
\[
\psi=\alpha |1\rangle+\beta |0\rangle,
\qquad
|\alpha|^2+|\beta|^2=1,
\]
the initial reduction coordinate is
\[
\tau_0=\mu_z
=
\langle \psi,\widehat z\psi\rangle
=
|\alpha|^2-|\beta|^2
=
2|\alpha|^2-1.
\]
Therefore
\[
\mathbb P(\text{hit }+1)
=
\frac{\mu_z+1}{2}
=
|\alpha|^2,
\]
and
\[
\mathbb P(\text{hit }-1)
=
\frac{1-\mu_z}{2}
=
|\beta|^2.
\]
Thus the two-outcome Born frequencies are obtained as hitting probabilities of
an unbiased walk in the physical reduction coordinate.

For numerical convenience, the simulation below uses the affine coordinate
\[
p=\frac{\tau-a}{b-a}.
\]
In the normalization \([a,b]=[-1,1]\), this is
\[
p=\frac{\mu_z+1}{2}.
\]
The simulation verifies the resulting frequency law numerically.

\subsection{Numerical set-up}

The simulated walk on \([0,1]\) is the rescaled version of the
unbiased walk in the physical coordinate \(\tau=\mu_z\). The endpoint \(p=1\)
represents the outcome \(|1\rangle\), while \(p=0\) represents the outcome
\(|0\rangle\). 

The interval \([0,1]\) was discretized into
\[
N=100
\]
subintervals. For each chosen value of \(p_0\), the walk begins at
\[
k_0=\operatorname{round}(Np_0),
\qquad
p_0\simeq \frac{k_0}{N}.
\]
At each step,
\[
k\longmapsto k+1
\quad\text{or}\quad
k\longmapsto k-1
\]
with equal probability, until the walk reaches one of the absorbing endpoints
\[
k=0
\qquad\text{or}\qquad
k=N.
\]
The outcome is recorded as \(1\) if the walk reaches \(N\), and as \(0\) if it
reaches \(0\).

For each value of \(p_0\), we simulated
\[
50{,}000
\]
independent paths. The tested Born probabilities were
\[
p_0=0.05,\ 0.10,\ 0.20,\ 0.35,\ 0.50,\ 0.65,\ 0.80,\ 0.90,\ 0.95.
\]

\subsection{Results}

The simulated frequencies are shown in Table~\ref{tab:born-frequency-summary}.
They agree with the Born probabilities within sampling error.

\begin{table}[h]
\centering
\caption{Born-rule frequencies from the unbiased reduction-coordinate walk.
The simulated frequency is the fraction of paths absorbed at endpoint \(1\).}
\label{tab:born-frequency-summary}
\begin{tabular}{cccccc}
\toprule
\(p_0=|\alpha|^2\) & Paths & Frequency at \(1\) & Std. error &
Abs. error & Mean absorption steps \\
\midrule
0.05 & \(50{,}000\) & 0.0501 & 0.0010 & 0.0001 & 482.3 \\
0.10 & \(50{,}000\) & 0.1011 & 0.0013 & 0.0011 & 905.0 \\
0.20 & \(50{,}000\) & 0.2052 & 0.0018 & 0.0052 & 1612.7 \\
0.35 & \(50{,}000\) & 0.3560 & 0.0021 & 0.0060 & 2261.7 \\
0.50 & \(50{,}000\) & 0.4985 & 0.0022 & 0.0015 & 2486.0 \\
0.65 & \(50{,}000\) & 0.6465 & 0.0021 & 0.0035 & 2285.6 \\
0.80 & \(50{,}000\) & 0.8014 & 0.0018 & 0.0014 & 1599.2 \\
0.90 & \(50{,}000\) & 0.9014 & 0.0013 & 0.0014 & 892.9 \\
0.95 & \(50{,}000\) & 0.9490 & 0.0010 & 0.0010 & 477.2 \\
\bottomrule
\end{tabular}
\end{table}

\begin{figure}[h]
\centering
\includegraphics[width=0.76\textwidth]{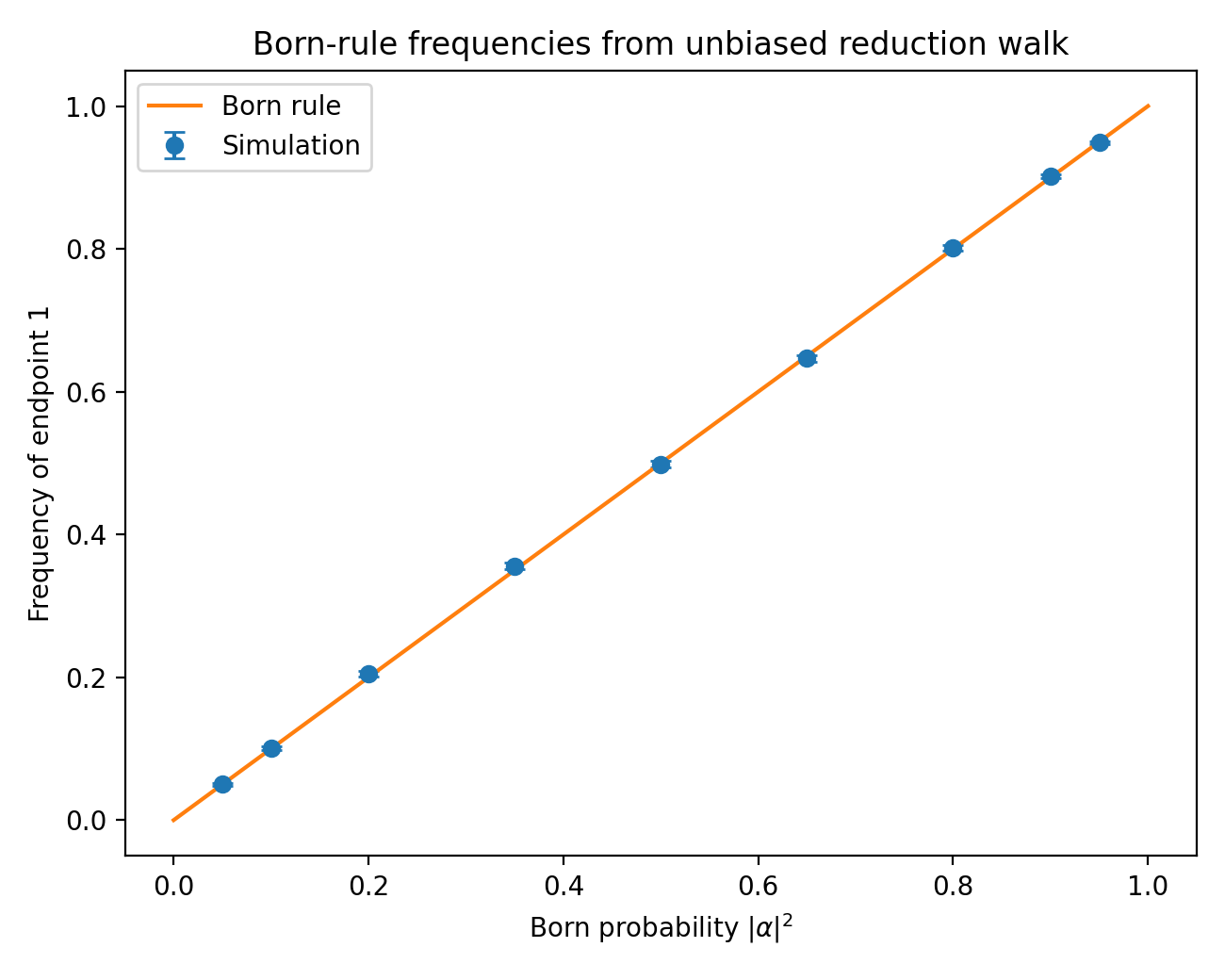}
\caption{Simulated endpoint frequencies for the unbiased reduction-coordinate
walk. The diagonal line is the Born-rule prediction \(P_1=|\alpha|^2\).}
\label{fig:born-frequency-plot}
\end{figure}

Figure~\ref{fig:born-frequency-plot} compares the simulated frequencies with the
Born-rule prediction. The points lie on the diagonal
\[
P_{\mathrm{sim}}=|\alpha|^2
\]
within the statistical error bars.

\begin{figure}[h]
\centering
\includegraphics[width=0.78\textwidth]{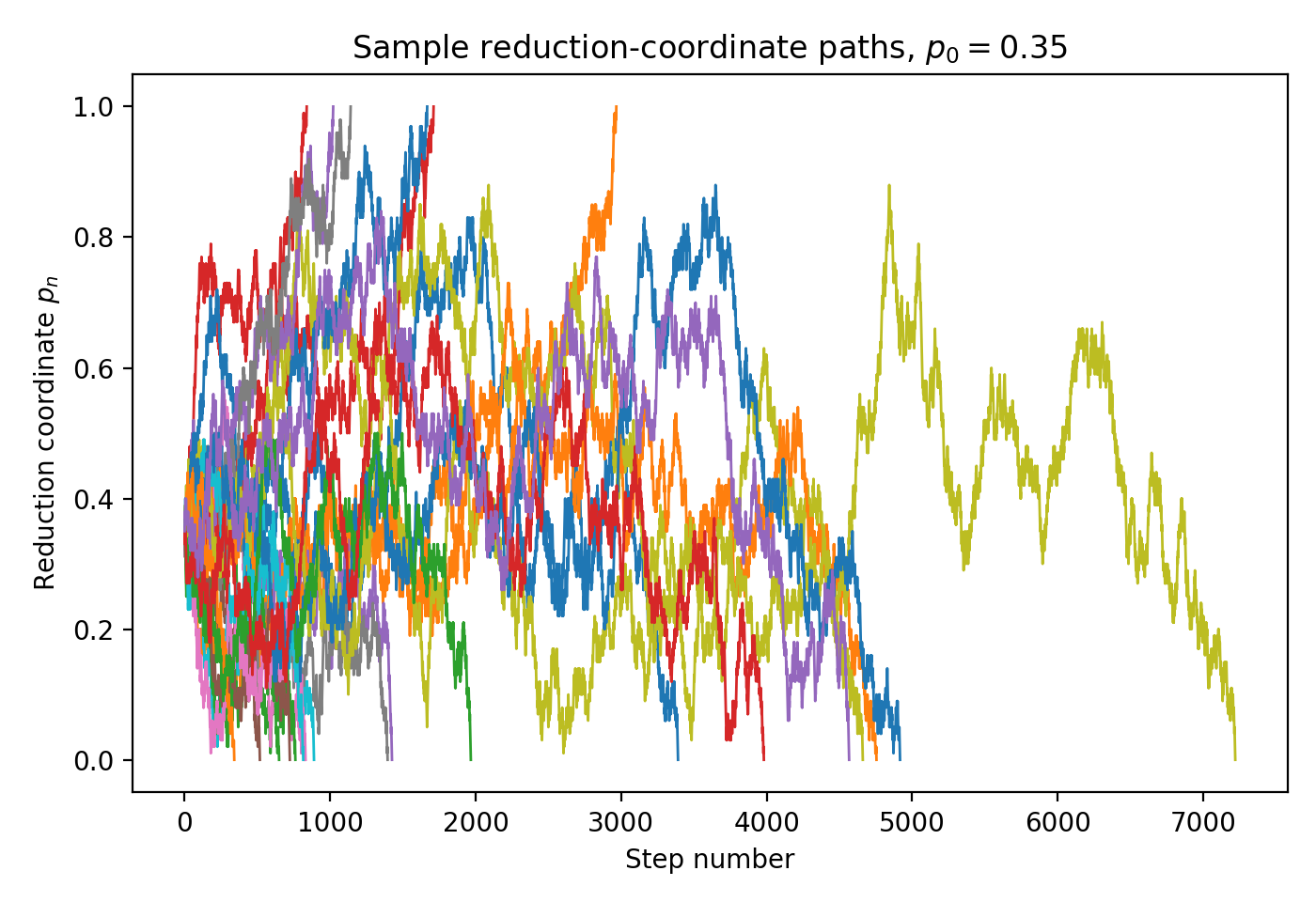}
\caption{Sample paths of the reduction coordinate \(p_n\) for \(p_0=0.35\).
The endpoints \(0\) and \(1\) represent the two detector-defined outcome
classes.}
\label{fig:sample-reduction-paths}
\end{figure}

Figure~\ref{fig:sample-reduction-paths} shows representative reduction paths
for the initial value
\[
p_0=0.35.
\]
Each path fluctuates until it reaches one of the two absorbing endpoints. The
frequency of absorption at \(1\) over many such paths is approximately \(0.35\),
as predicted by the Born rule.

\subsection{Apparent absorption in the Euclidean detector coordinate}

The normal-distribution and absorbing-boundary descriptions are not different
mechanisms. They are two projections of the same state-space diffusion. In the
interior of the represented family of outcome classes, the projection to the
relevant reduction coordinate is locally regular. Therefore the
Fubini--Study-normalized state-space diffusion appears as ordinary Brownian
motion in the reduced coordinate. Near an endpoint, however, the Euclidean
detector coordinate no longer resolves the continuing state-space motion: many
state-space directions are assigned to the same endpoint equivalence class.
Thus the endpoint appears absorbing in the reduced coordinate, although the
underlying state-space walk continues.

Equivalently, one may either wait until the projected process first reaches a
detector cell, or fix a sufficiently large number \(n\) of \({\bf (RM)}\) steps
and identify the detector class occupied by the state-space path at step \(n\).
For large enough \(n\), the probability that no record has yet formed is
negligible, and the fixed-\(n\) and first-passage descriptions give the same
outcome weights. The absorbing-boundary picture is therefore a reduced
description of record formation, not an additional collapse postulate.

This point is especially important in numerical simulations. A finite-dimensional approximation of \(L_2(\mathbb R)\) is necessarily obtained
by first replacing it with \(L_2([a,b])\) for some finite interval \([a,b]\),
and then discretizing. Consequently, the represented classical submanifold is also
restricted to \([a,b]\). In the reduced Euclidean coordinate, the endpoints
\(a\) and \(b\) are therefore boundaries of the representation: the coordinate
cannot move outside the interval because no components outside \([a,b]\) are
present in the finite-dimensional Hilbert space. Thus the endpoints are
absorbing for the projected detector coordinate. The underlying state-space
motion still continues inside the Hilbert space, as the state may become
squeezed or evolve into a narrow superposition, but its Euclidean classical
projection cannot pass beyond the represented interval.

To illustrate this in Euclidean detector coordinates, consider a finite
detector interval
\[
[a,b]
\]
in the Euclidean coordinate \(\tau\). We continue to represent states by
Gaussian packets, but require the packet to remain inside the represented
interval. Thus, if
\[
\delta=\sigma e^s
\]
is the width of the Gaussian, then near the endpoints the width must satisfy
\[
\delta(\tau)\leq \min\{\tau-a,\ b-\tau\}.
\]
Consequently, as \(\tau\) approaches either endpoint, the Gaussian must become
more localized. Near \(b\), for example,
\[
\delta(\tau)\lesssim b-\tau,
\]
and therefore
\[
s(\tau)
=
\ln\frac{\delta(\tau)}{\sigma}
\sim
\ln\frac{b-\tau}{\sigma}.
\]
It follows that
\[
\frac{ds}{d\tau}
\sim
-\frac{1}{b-\tau},
\]
which diverges as \(\tau\to b\).

Now suppose that the projected walk makes steps of fixed size in the
\((\tau,s)\)-metric,
\[
d\ell^2=d\tau^2+ds^2.
\]
Then a step of fixed metric size \(\Delta\ell\) has Euclidean \(\tau\)-component
approximately
\[
|\Delta\tau|
\approx
\frac{\Delta\ell}{\sqrt{1+(ds/d\tau)^2}}.
\]
Therefore
\[
|\Delta\tau|\longrightarrow0
\qquad
\text{as}
\qquad
\tau\to a \text{ or } b.
\]
Thus the walk does not stop in the intrinsic \((\tau,s)\)-metric, but its
Euclidean detector coordinate freezes at the endpoint. This is the Euclidean
appearance of absorption.

\begin{figure}[!htbp]
\centering
\includegraphics[width=0.78\textwidth]{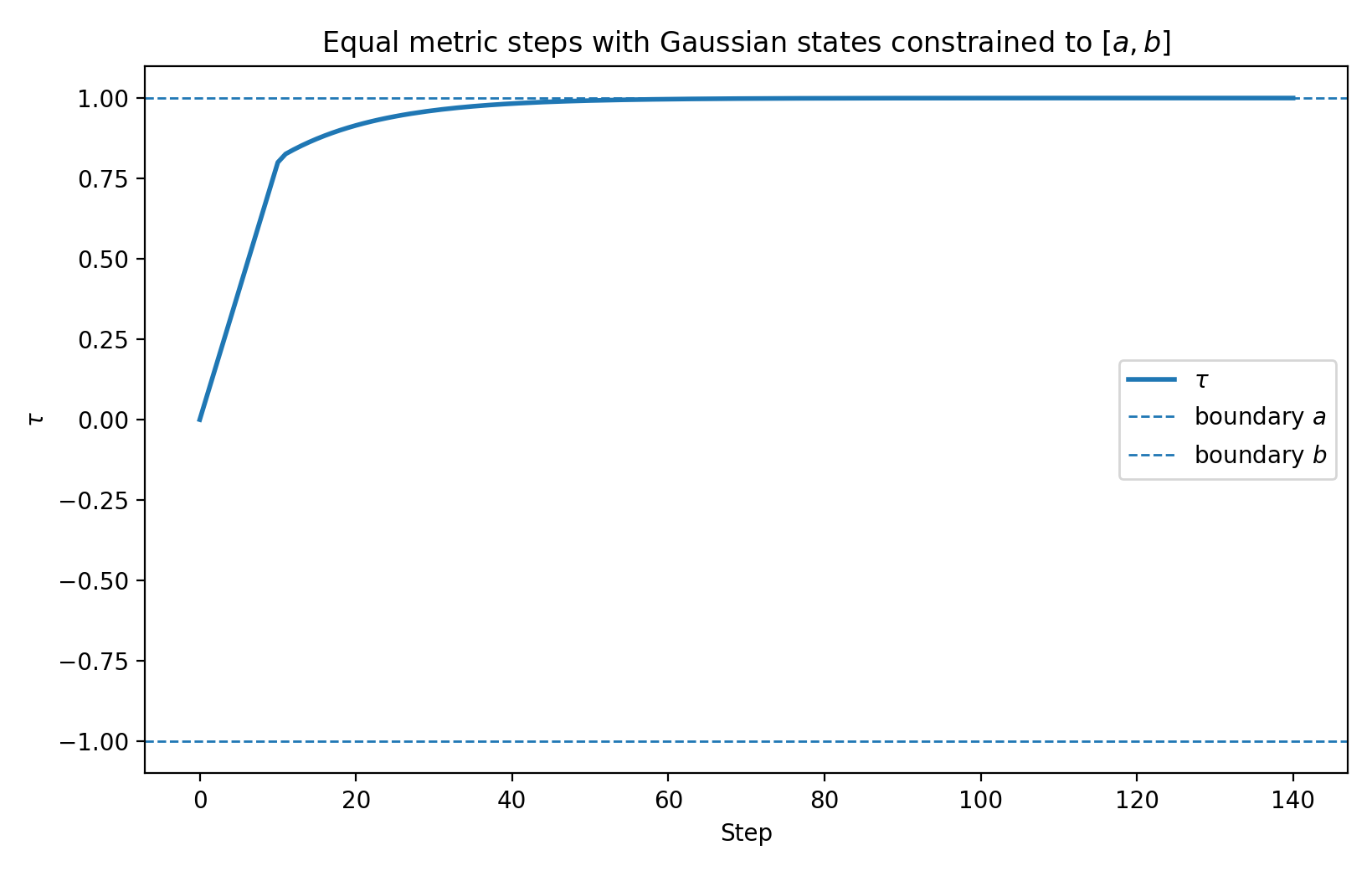}
\caption{Euclidean detector coordinate for Gaussian states constrained to
remain inside the interval \([a,b]\). Equal steps in the \((\tau,s)\)-metric
produce decreasing Euclidean \(\tau\)-increments near the endpoint.}
\label{fig:gaussian-constrained-tau-freezing}
\end{figure}

Figure~\ref{fig:gaussian-constrained-tau-freezing} shows a simulated path made
of equal steps in the \((\tau,s)\)-metric while the Gaussian width is forced to
shrink near the endpoint. The Euclidean coordinate \(\tau\) approaches the
boundary and then changes only by smaller and smaller amounts.

\begin{figure}[!htbp]
\centering
\includegraphics[width=0.78\textwidth]{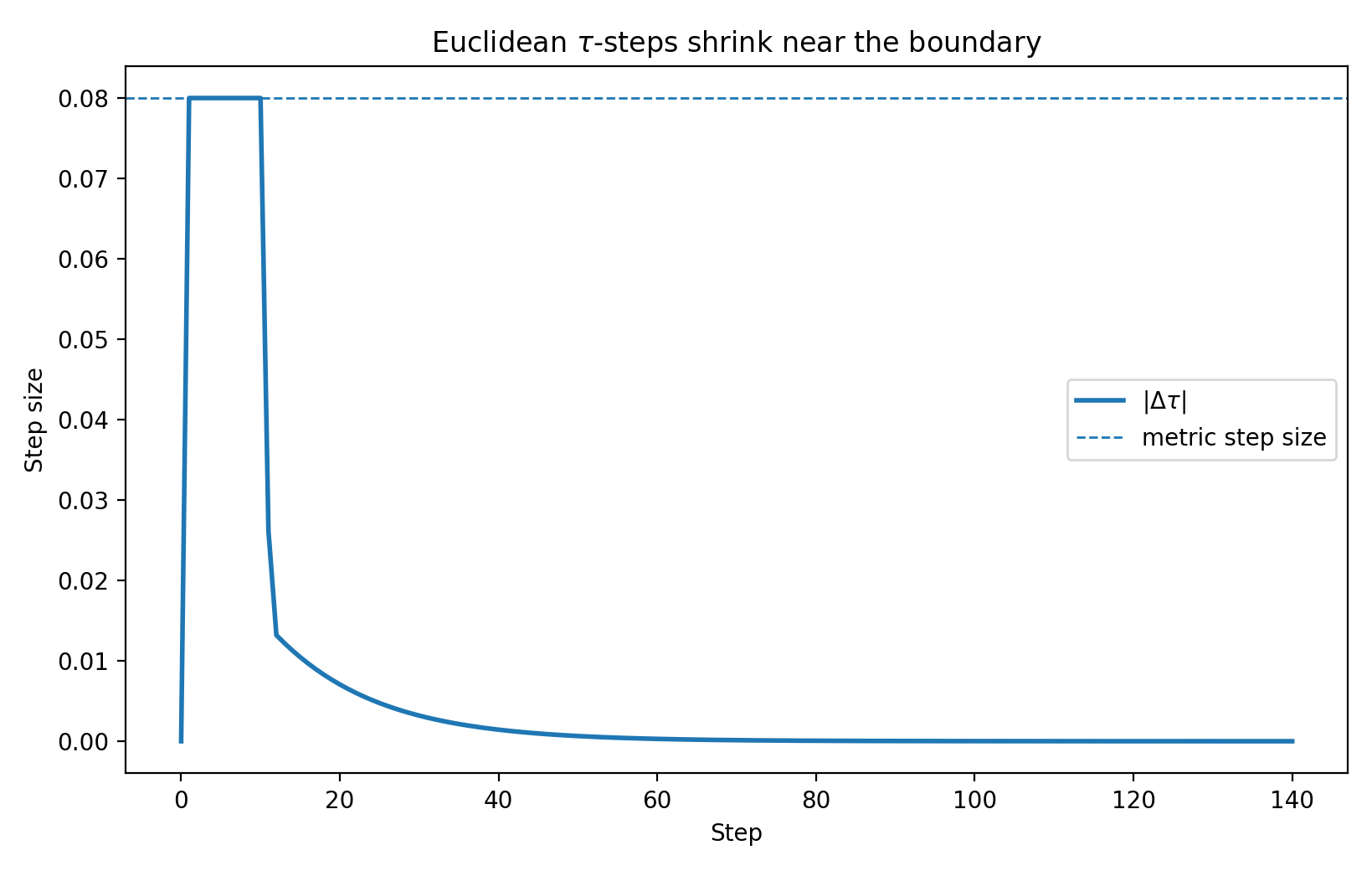}
\caption{Euclidean step size \(|\Delta\tau|\) for equal steps in the
\((\tau,s)\)-metric. The \(\tau\)-component decreases near the endpoint because
motion is increasingly converted into narrowing of the Gaussian packet.}
\label{fig:gaussian-constrained-step-size}
\end{figure}

Figure~\ref{fig:gaussian-constrained-step-size} shows the corresponding
Euclidean step size \(|\Delta\tau|\). Although the metric step size is fixed,
the \(\tau\)-component of the step tends to zero as the endpoint is approached.

\begin{figure}[!htbp]
\centering
\includegraphics[width=0.78\textwidth]{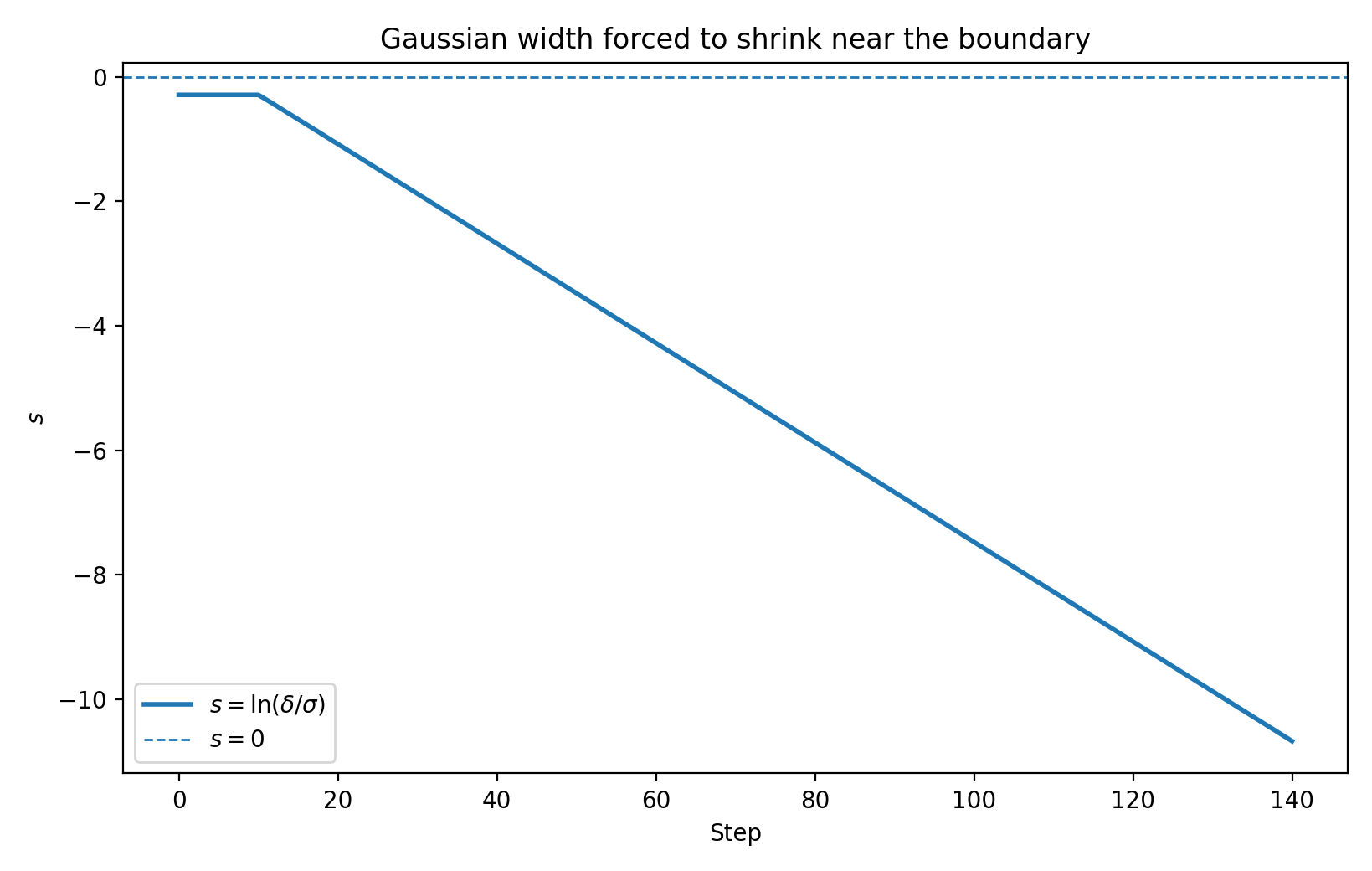}
\caption{Localization coordinate for the same constrained Gaussian path. Near
the endpoint, the packet must narrow in order to remain inside the detector
interval, forcing \(s\) to become more negative.}
\label{fig:gaussian-constrained-s}
\end{figure}

Figure~\ref{fig:gaussian-constrained-s} shows the localization coordinate
along the same path. As \(\tau\) approaches the endpoint, the Gaussian width is
forced to decrease, so \(s=\ln(\delta/\sigma)\) remains nonpositive and becomes
more negative.


This shows that apparent absorption is a property of the Euclidean detector
projection. The underlying state-space motion may continue, but once the path
approaches an endpoint class, the finite interval constraint forces the
Euclidean \(\tau\)-motion to freeze and the localization coordinate to remain
in the recorded sector.

\subsection{Zeno stability of a recorded equivalence class}
\label{subsec:zeno-stability}

The same mechanism gives a numerical form of the Zeno effect. Suppose that a
state has already entered a detector-defined position class, for example the
right endpoint class in a two-detector model. The subsequent \({\bf (RM)}\)
motion in projective state space need not stop. However, the question relevant
to a macroscopic record is not whether the exact state ray continues to move,
but whether the state leaves the recorded finite-resolution equivalence class.

In the Euclidean detector coordinate, the endpoint class is represented by the
finite interval boundary. As shown above, a Gaussian representative constrained
to remain inside \([a,b]\) must become narrower as its center approaches the
endpoint. Near \(b\), \(\Delta \tau \to 0\), and hence the projected Euclidean motion
freezes at the endpoint even though the state-space path continues.

This is the Zeno-type stabilization of the record. Repeated environmental
registration assigns the state to the same finite-resolution equivalence class
whenever the localization condition \(s<0\) is satisfied. After the record has
formed, continuing \({\bf (RM)}\) motion may squeeze the state further or move
it within directions not resolved by the detector, but the Euclidean
position class remains unchanged with overwhelming probability. In this sense,
the record is stable not because the state ray is frozen, but because the
projection to detector-defined equivalence classes becomes effectively
constant.

The effect is already visible in
Figures~\ref{fig:gaussian-constrained-step-size} and
\ref{fig:gaussian-constrained-s}. The Euclidean step size
\(|\Delta\tau|\) decreases as the endpoint class is approached, while the
localization coordinate remains in the recorded sector and becomes increasingly
negative. Thus a state that has entered a recorded position class remains in
that class under the continuing projected \({\bf (RM)}\) motion, giving the
equivalence-class form of the Zeno effect.

\subsection{Interpretation}

The simulation above illustrates the simplest stochastic mechanism behind the Born
rule in the \({\bf (RM)}\) framework. The coordinate
\[
p=|\alpha|^2
\]
is a martingale under the unbiased reduction walk. Therefore the probability of
absorption at the endpoint \(1\) is the initial value \(p_0\). Consequently,
\[
\mathbb P(\text{outcome }1)=|\alpha|^2,
\qquad
\mathbb P(\text{outcome }0)=|\beta|^2.
\]

The simulation it is the finite-dimensional two-outcome representation of the
state-space diffusion mechanism. The stochastic walk is unbiased; the
asymmetry of the outcome probabilities comes entirely from the initial
state-space position, encoded in the coordinate \(p_0=|\alpha|^2\). Thus the
Born weights arise as hitting probabilities for detector-defined outcome
classes.

\section{Double-slit screen records from projected \({\bf (RM)}\) diffusion}
\label{sec:double-slit-rm-records}

\subsection{Purpose of the simulation}

We now apply the detector-coordinate and path-weight construction of the
preceding sections to the double-slit experiment. The purpose of the simulation
is to show how the interference pattern is formed from finite-resolution
detector records on the final screen.

The screen is divided into bins
\[
I_1,\ldots,I_n,
\]
and each bin represents a detector-defined equivalence class. As explained in Section~\ref{subsec:calibration-and-weights}, the calibrated
weight associated with the \(j\)-th bin is the aggregate weight of
\({\bf (RM)}\)-generated paths from the state to that bin. For the screen
bins considered here, this weight is determined by projecting the state
arriving at the screen onto the bin. Thus, if \(\psi_T\) is the state arriving at the screen,
then
\[
W_j(\psi_T)=\|P_{I_j}\psi_T\|^2
=
\int_{I_j}|\psi_T(x)|^2\,dx,
\]
where \(P_{I_j}\) is the projection associated with the screen bin \(I_j\).

\subsection{Projected record process}

In the simulations, the random Hamiltonian motion is used through its projected
effect on the detector coordinates. This is the same projected process used in
Subsection~\ref{projectedGUE}. At each short time step, a GUE Hamiltonian
generates a tangent vector
\[
v=-iH\psi
\]
in projective state space. After removal of the vertical phase component, this
tangent vector is projected onto the Fubini--Study normal plane spanned by the
detector-coordinate directions, here represented by \(\tau\) and \(s\). This
projection gives the reduced random walk in the detector coordinates. The
preceding simulations showed that the resulting normalized increments are
Gaussian and isotropic in the induced Fubini--Study metric.

For the screen simulation we use this reduced record process rather than
displaying every microscopic GUE path in the full Hilbert space. The calibrated
aggregate path weights determine the relative weights of the detector-defined
screen classes, and screen records are generated by sampling from these
weights.

\subsection{Coherent and which-slit records}

The state arriving at the final screen depends on whether a which-slit record
has already been formed. If no which-slit record has formed, the arriving state
is the coherent two-slit state \(\psi_T\). Since \(\psi_T\) is a coherent
superposition of the two propagated slit components, the screen-bin weights
contain interference terms.

If a which-slit record has formed at the slit screen, the run is first assigned
to one of two detector-defined equivalence classes: the left-slit class or the
right-slit class. Conditional on the left-slit record, the state arriving at the
final screen is \(\psi_L\), and the screen-bin weights are
\[
W_j^{L}
=
\int_{I_j}|\psi_L(x)|^2\,dx.
\]
Conditional on the right-slit record, the state arriving at the final screen is
\(\psi_R\), and the screen-bin weights are
\[
W_j^{R}
=
\int_{I_j}|\psi_R(x)|^2\,dx.
\]
Thus the which-slit case is simulated by first selecting the slit record and
then sampling the final screen record from the corresponding conditional
distribution. The two alternatives are combined as recorded alternatives, not
as amplitudes, and therefore no interference terms appear.

The difference between the coherent and which-slit cases is therefore not in
the final screen-recording rule, which is the same in both cases. It is in the
state arriving at the screen: a coherent superposition in the first case and
conditional single-slit states in the second.

\subsection{Formation of the screen patterns}

\begin{figure}[h]
\centering
\includegraphics[width=0.82\textwidth]{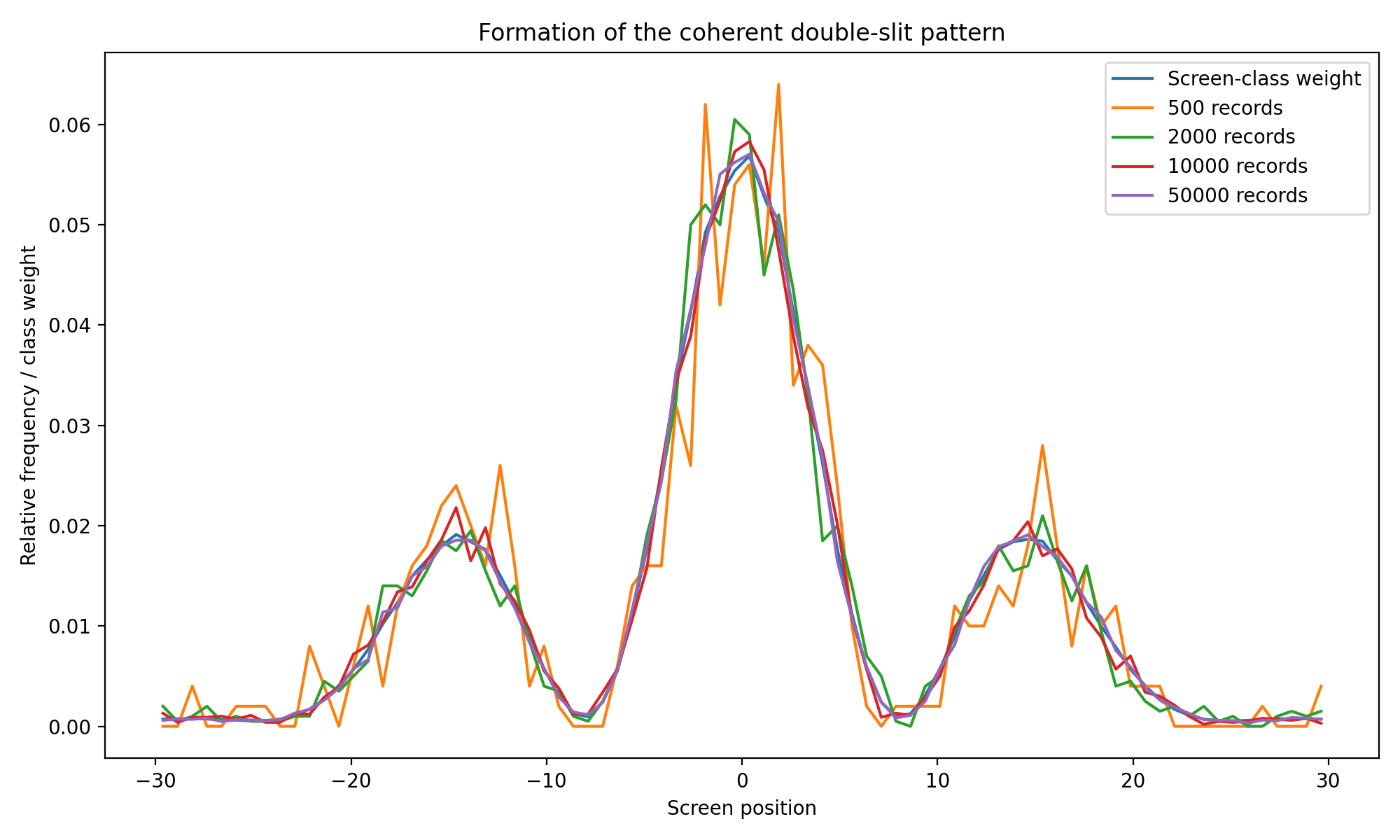}
\caption{Formation of the coherent double-slit pattern. The reference curve is
the screen-class endpoint weight \(W_j\), obtained from the projected
\({\bf (RM)}\) diffusion. The other curves show cumulative record frequencies
after \(500\), \(2000\), \(10000\), and \(50000\) records.}
\label{fig:coherent-build-up}
\end{figure}

The screen was divided into
\[
n=80
\]
finite-resolution classes on the interval \([-30,30]\). Repeated screen records
were generated using the endpoint weights described above.

Figure~\ref{fig:coherent-build-up} shows the formation of the coherent
double-slit pattern after
\[
500,\quad 2000,\quad 10000,\quad 50000
\]
records. The reference curve is the screen-class weight \(W_j\), and the other
curves are cumulative record frequencies. As the number of records increases,
the histogram approaches the interference pattern.

\begin{figure}[h]
\centering
\includegraphics[width=0.82\textwidth]{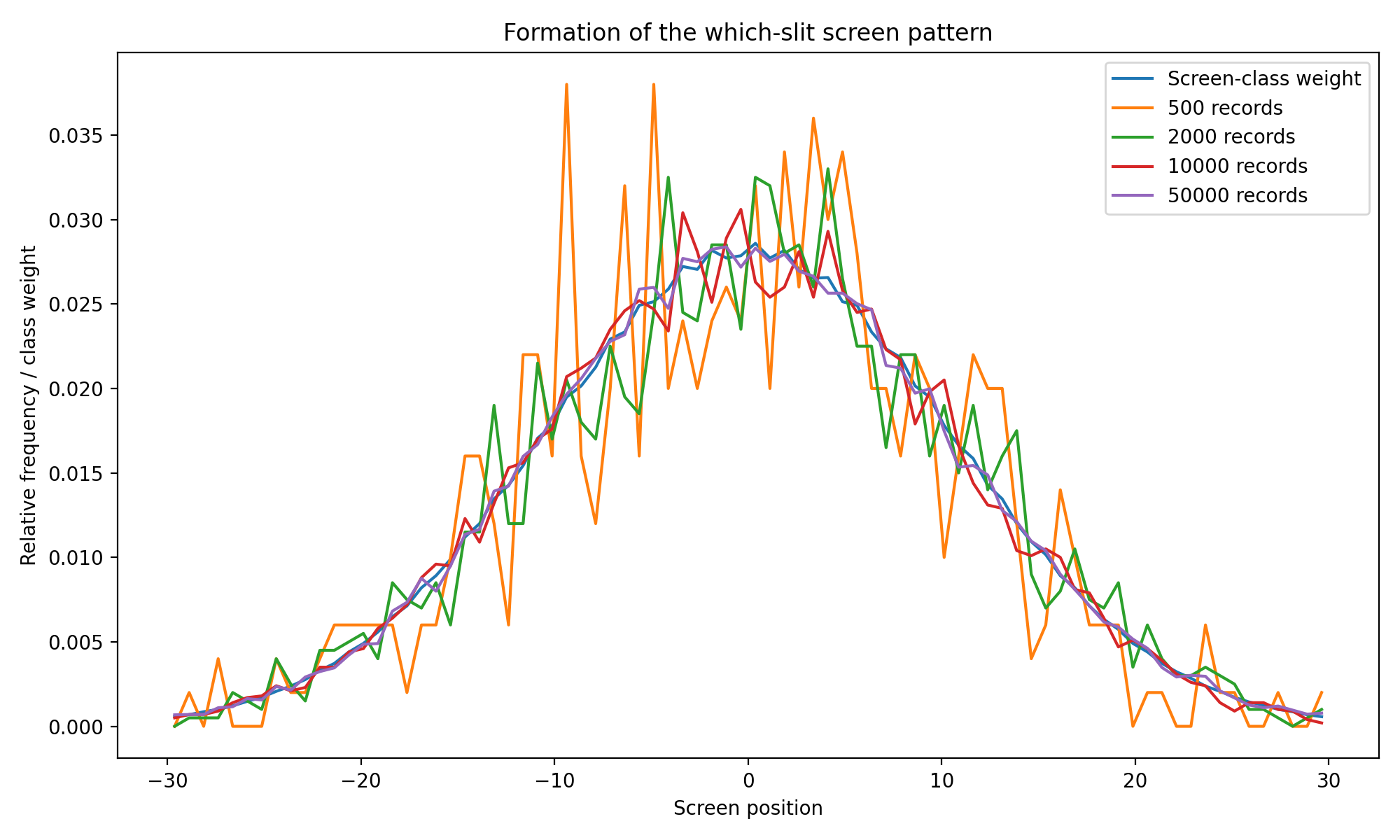}
\caption{Formation of the screen pattern when a which-slit record has already
been formed. The record frequencies converge to the endpoint weights for the
conditional slit records, and the interference pattern is absent.}
\label{fig:which-slit-build-up}
\end{figure}

Figure~\ref{fig:which-slit-build-up} shows the corresponding case in which a
which-slit record is formed before the particle reaches the final screen. The
accumulated records converge to the noninterference screen distribution.

The convergence of the cumulative record frequencies is summarized in
Table~\ref{tab:double-slit-build-up-errors}.

\begin{table}[h]
\centering
\caption{Convergence of cumulative screen-record frequencies to the
finite-resolution screen-class endpoint weights.}
\label{tab:double-slit-build-up-errors}
\begin{tabular}{llcc}
\toprule
Case & Records & \(L^1\)-error & Maximum bin error \\
\midrule
Coherent & \(500\)   & 0.2608 & 0.0153 \\
Coherent & \(2000\)  & 0.1199 & 0.0087 \\
Coherent & \(10000\) & 0.0627 & 0.0027 \\
Coherent & \(50000\) & 0.0279 & 0.0022 \\
\midrule
Which-slit & \(500\)   & 0.2822 & 0.0185 \\
Which-slit & \(2000\)  & 0.1413 & 0.0066 \\
Which-slit & \(10000\) & 0.0596 & 0.0032 \\
Which-slit & \(50000\) & 0.0275 & 0.0013 \\
\bottomrule
\end{tabular}
\end{table}

\subsection{Interpretation}

Thus the difference between the two regimes lies in the state that reaches the
screen, rather than in the final screen measurement rule. The projected
{\bf (RM)} mechanism records finite-resolution screen classes in both
cases; interference is present or absent according to whether the state
arriving at the screen is coherent or incoherent. Equivalently, the aggregate
weights of {\bf (RM)} paths from these states to the detector bins are
different in the two cases.


\section{Stroboscopic Newtonian motion}
\label{sec:stroboscopic-simulation}

\subsection{Purpose of the simulation}

The companion theoretical paper argues that macroscopic Newtonian motion arises
as a stroboscopic process. Between environmental interactions, the tangent
component of the Schr\"odinger flow gives Newtonian drift on the classical
phase-space submanifold. The \({\bf (RM)}\) interaction produces stochastic
state-space displacements, and environmental monitoring repeatedly returns the
state to the localized sector. A classical record is produced only when the
state lies in this localized sector.

The purpose of the present simulation is to illustrate this mechanism at the
level of recorded positions. We consider a one-dimensional Newtonian system and
a sequence of candidate recording times
\[
t_k=k\Delta t.
\]
At each such time, the state has a projected position coordinate and a
localization coordinate
\[
s_k=\ln(\delta_k/\sigma),
\]
where \(\delta_k\) is the position uncertainty and \(\sigma\) is the detector
resolution. The localized sector is
\[
s_k<0.
\]
Only candidate times satisfying this condition are counted as classical
records.

Thus the recorded positions are
\[
a_k^{\rm rec}=a_N(t_k)+\eta_k
\qquad
\text{for those }k\text{ such that }s_k<0,
\]
where \(a_N(t_k)\) is the Newtonian position and \(\eta_k\) is a normal random
variable with variance \(D_a\Delta t\):
\[
\eta_k\sim N(0,D_a\Delta t).
\]

The localization coordinate is modeled by a Gaussian transverse walk
\[
s_{k+1}=s_k+\xi_k,
\qquad
\xi_k\sim N(0,2D_s\Delta t),
\]
where \(D_s\) is written in the standard Brownian diffusion convention.
A classical record is produced when
\[
s_{k+1}<0,
\]
after which the process is renewed from a localized representative
\(s_{\rm rec}<0\).

The stochastic variables \(\eta_k\) and \(s_k\) are not introduced as
additional classical noise sources. They represent the tangential and
transverse coordinate displacements induced by the accumulated {\bf (RM)}
motion between returns to the localized sector. By the Brownian-restriction
simulation of Section~\ref{sec:brownian-restriction} and the subsequent
\((\tau,s)\)-coordinate simulations, the \(\tau\)- and \(s\)-components of the
{\bf (RM)}-induced displacement on the corresponding Gaussian surface are
independent Gaussian variables.

\subsection{Numerical set-up}

For definiteness, we use a dimensionless harmonic oscillator with
\[
M=1,
\qquad
\omega=1.
\]
The Newtonian solution is
\[
a_N(t)
=
a_0\cos(\omega t)
+
\frac{p_0}{M\omega}\sin(\omega t).
\]
We take
\[
a_0=1,
\qquad
p_0=0.35.
\]
The candidate recording times are
\[
t_k=k\Delta t,
\qquad
\Delta t=0.05,
\]
over the time interval
\[
0\leq t\leq 20.
\]
This gives \(401\) candidate recording times.

The detector resolution for the position record is
\[
\sigma=0.10,
\]
and the induced position-diffusion coefficient is
\[
D_a=0.002.
\]
Therefore
\[
D_a\Delta t
=
0.0001,
\qquad
\frac{D_a\Delta t}{\sigma^2}
=
0.01\ll1.
\]
Thus the tangential stochastic displacement between returns is small compared
with the detector resolution.

As discussed above, we use
\[
s_{k+1}=s_k+\xi_k,
\qquad
\xi_k\sim N(0,2D_s\Delta t),
\]
with a classical record produced when
\[
s_{k+1}<0.
\]
When this condition is satisfied, the state is assigned to the corresponding
localized equivalence class, and the next interval begins from a localized
representative \(s_{\rm rec}<0\).

In the simulation we use
\[
D_s=0.25,
\qquad
\sqrt{2D_s\Delta t}=0.1581,
\qquad
s_{\rm rec}=-0.5.
\]
Thus the transverse motion is an unbiased Gaussian walk between records, and
localization enters through conditioning on the event \(s<0\), followed by
renewal from a localized representative. We generated
\[
5000
\]
independent trajectories.

\subsection{Conditioned stroboscopic records}

\begin{figure}[h]
\centering
\includegraphics[width=0.78\textwidth]{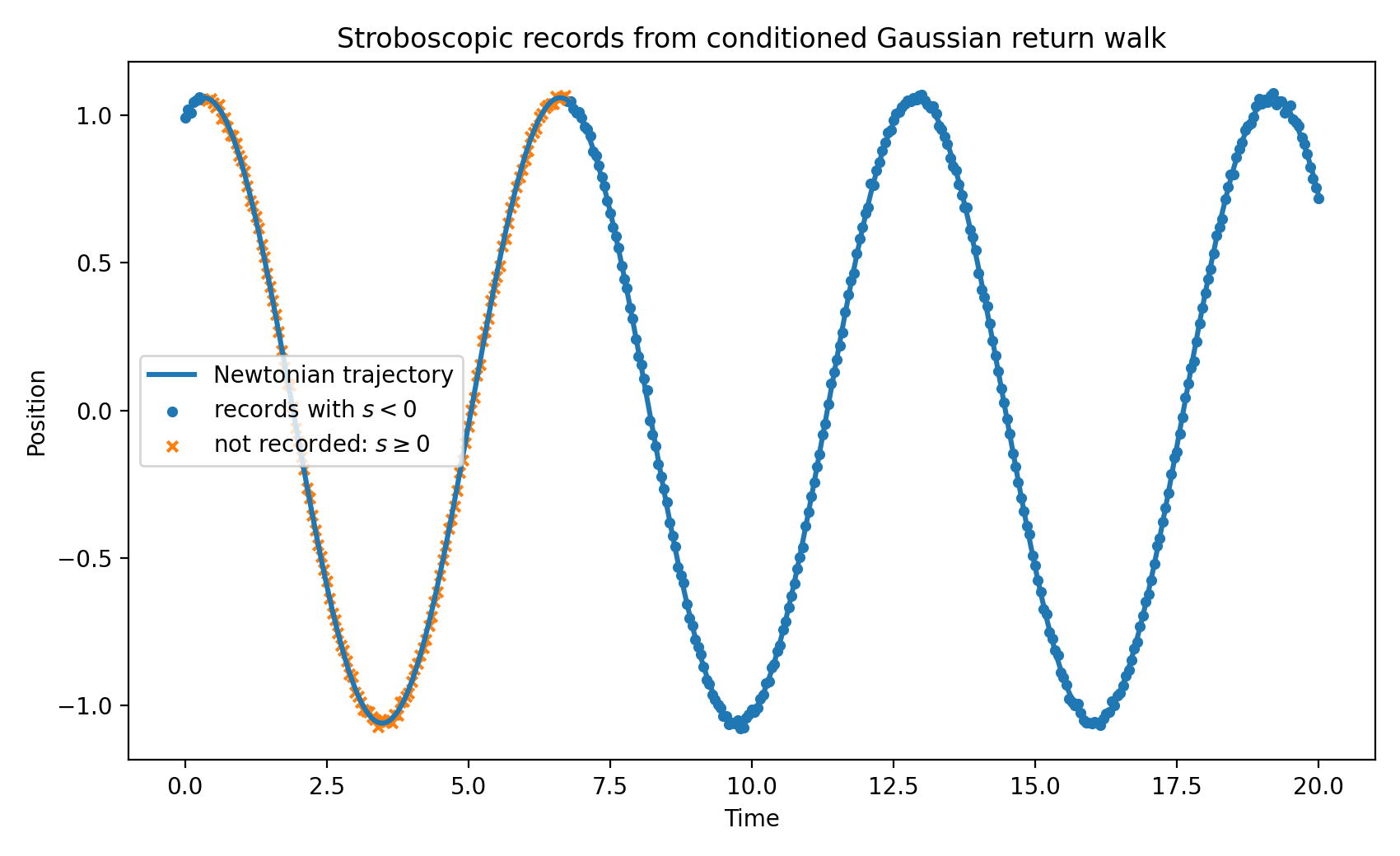}
\caption{Stroboscopic position records generated by a conditioned Gaussian
return walk in the localization coordinate. The solid curve is the Newtonian
trajectory. Points with \(s<0\) are counted as records; points with \(s\geq0\)
are not counted as classical records.}
\label{fig:sim4c-conditioned-records}
\end{figure}

Figure~\ref{fig:sim4c-conditioned-records} shows one representative trajectory.
The solid curve is the Newtonian trajectory. Points marked as records satisfy
\(s_k<0\). Candidate times with \(s_k\geq0\) are shown separately and are not
counted as classical position records.

\begin{figure}[h]
\centering
\includegraphics[width=0.78\textwidth]{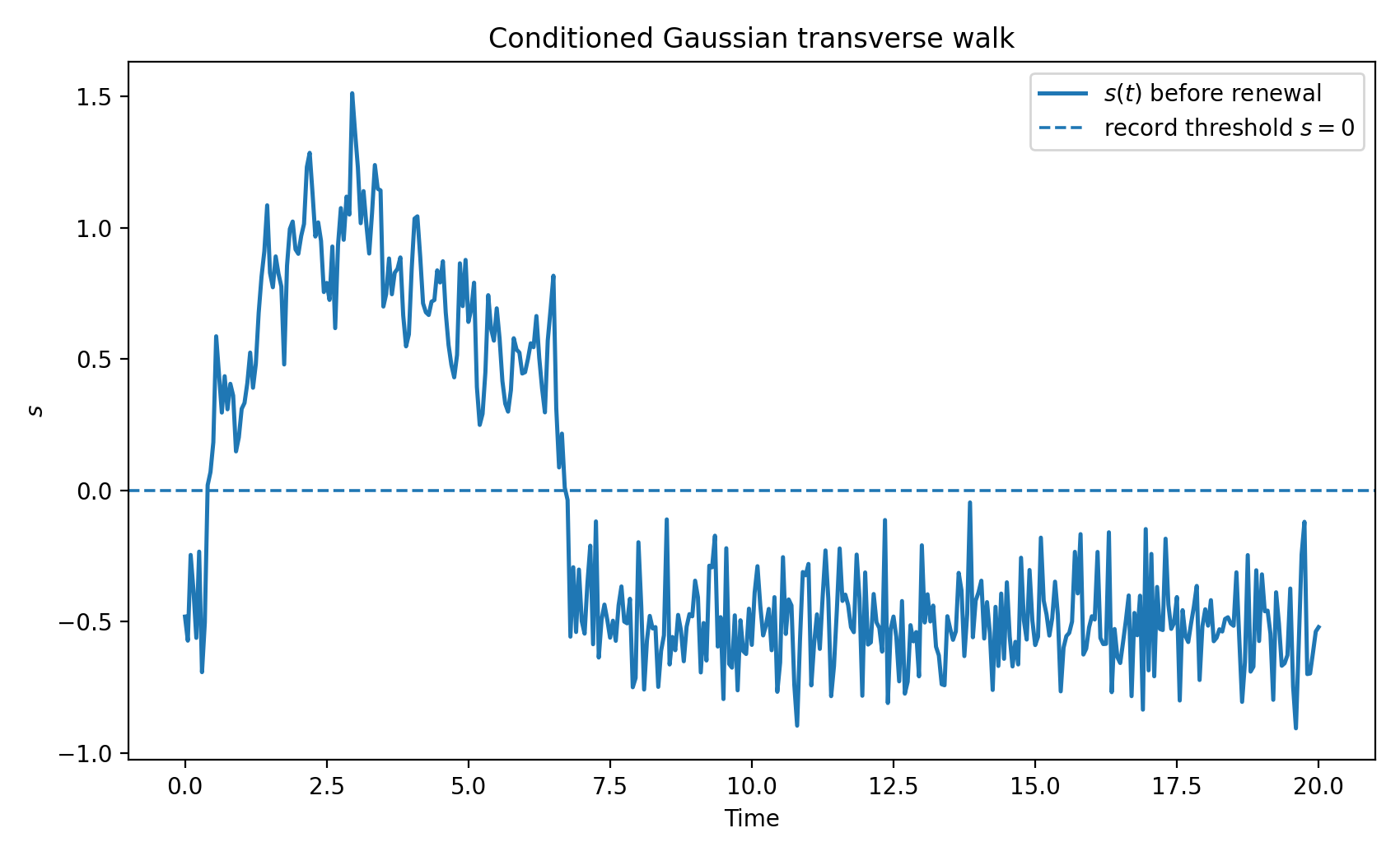}
\caption{Conditioned Gaussian transverse walk for a sample trajectory. A
classical position record is counted only when \(s<0\); after such a record the
next interval begins from a localized representative.}
\label{fig:sim4c-s-coordinate}
\end{figure}

Figure~\ref{fig:sim4c-s-coordinate} shows the corresponding localization
coordinate \(s_k\) for the same sample trajectory. The dashed line marks the
recording threshold \(s=0\). When the path satisfies \(s<0\), a record is made
and the transverse coordinate is renewed from the localized representative
\(s_{\rm rec}=-0.5\).

\begin{figure}[h]
\centering
\includegraphics[width=0.78\textwidth]{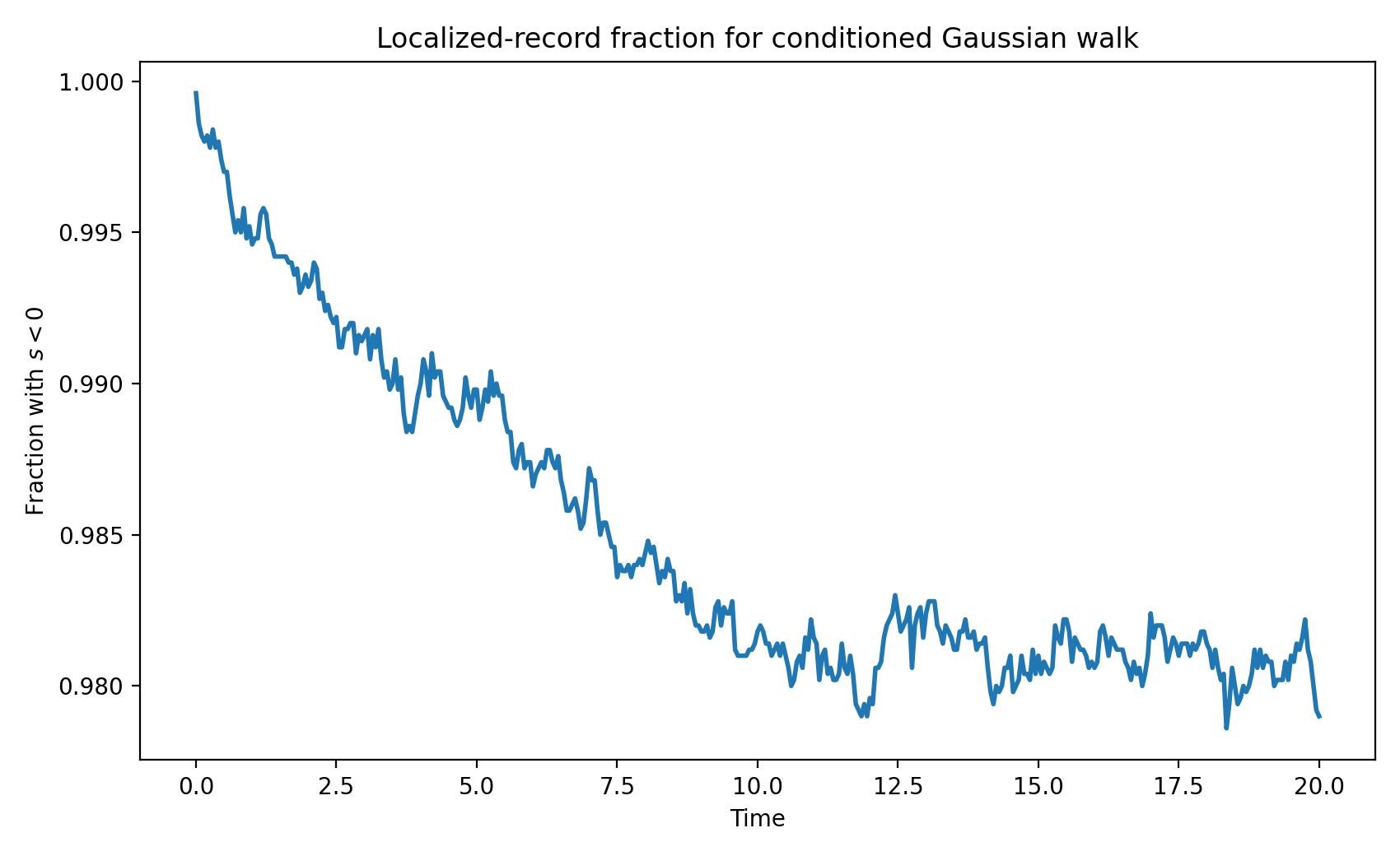}
\caption{Fraction of trajectories in the localized sector \(s<0\) as a
function of time for the conditioned Gaussian return walk.}
\label{fig:sim4c-record-fraction}
\end{figure}

The fraction of trajectories satisfying \(s_k<0\) at each candidate time is
shown in Figure~\ref{fig:sim4c-record-fraction}. In this run the overall
fraction of localized candidate points was
\[
0.9849.
\]
Thus most candidate times produce localized records, as required for
macroscopic Newtonian behavior.

\subsection{Newtonian consistency on recorded points}

The deviation from Newtonian motion was computed only over the recorded set
\[
\{(i,k):s_k^{(i)}<0\}.
\]
For these recorded points,
\[
a_k^{\rm rec}-a_N(t_k)=\eta_k.
\]
The RMS deviation over all recorded points was
\[
0.010004,
\]
which agrees with the imposed tangential scale
\[
\sqrt{D_a\Delta t}=0.01.
\]
The mean absolute deviation was
\[
0.007983,
\]
and the maximum absolute deviation observed in the simulation was
\[
0.050327.
\]

\begin{figure}[h]
\centering
\includegraphics[width=0.78\textwidth]{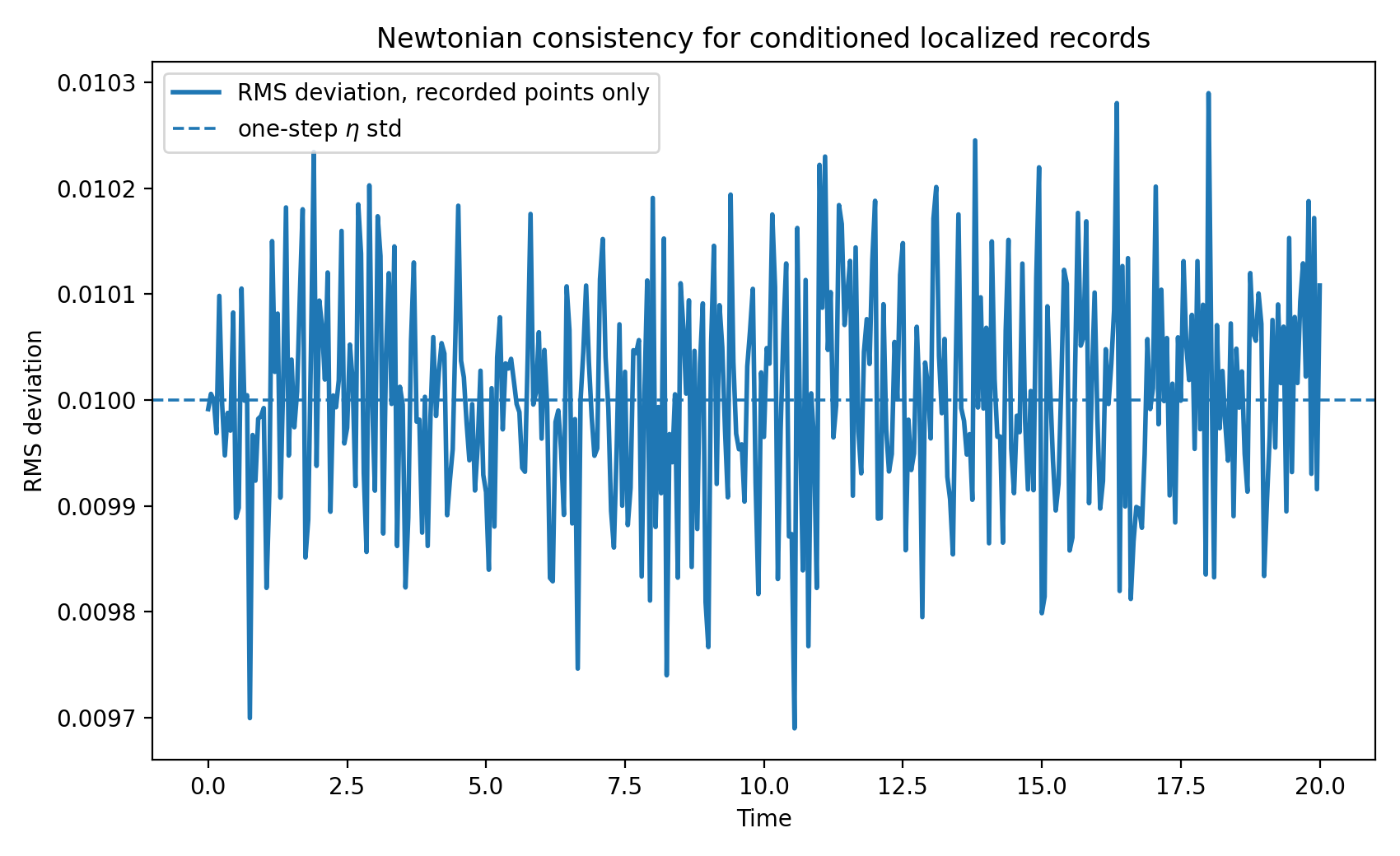}
\caption{RMS deviation of recorded positions from the Newtonian trajectory,
computed only over localized records satisfying \(s<0\). The dashed line marks
the one-step standard deviation \(\sqrt{D_a\Delta t}=0.01\).}
\label{fig:sim4c-rms-conditioned}
\end{figure}

Figure~\ref{fig:sim4c-rms-conditioned} shows the RMS deviation as a function of
time, computed only from trajectories satisfying \(s_k<0\) at that time.

The numerical parameters and recorded-position statistics are summarized in
Table~\ref{tab:sim4c-summary}.

\begin{table}[h]
\centering
\caption{Parameters and recorded-position deviations in the conditioned
stroboscopic Newtonian simulation.}
\label{tab:sim4c-summary}
\begin{tabular}{lc}
\toprule
Quantity & Value \\
\midrule
Mass \(M\) & \(1\) \\
Oscillator frequency \(\omega\) & \(1\) \\
Initial position \(a_0\) & \(1\) \\
Initial momentum \(p_0\) & \(0.35\) \\
Candidate recording interval \(\Delta t\) & \(0.05\) \\
Number of candidate times & \(401\) \\
Number of simulated trajectories & \(5000\) \\
Resolution \(\sigma\) & \(0.10\) \\
Position-diffusion coefficient \(D_a\) & \(0.002\) \\
\(\sqrt{D_a\Delta t}\) & \(0.0100\) \\
\(D_a\Delta t/\sigma^2\) & \(0.01\) \\
Localization diffusion coefficient \(D_s\) & \(0.25\) \\
\(\sqrt{2D_s\Delta t}\) & \(0.1581\) \\
Renewal value \(s_{\rm rec}\) & \(-0.5\) \\
Fraction of candidate points with \(s<0\) & \(0.9849\) \\
RMS deviation, recorded points only & \(0.010004\) \\
Mean absolute deviation, recorded points only & \(0.007983\) \\
Maximum absolute deviation, recorded points only & \(0.050327\) \\
\bottomrule
\end{tabular}
\end{table}

\subsection{Return-gap statistics}

The intervals during which
\[
s_k\geq0
\]
represent failures of localization at candidate recording times. A return gap
is a maximal consecutive interval of candidate times without a record, ending
when the trajectory next satisfies \(s<0\).

\begin{figure}[h]
\centering
\includegraphics[width=0.78\textwidth]{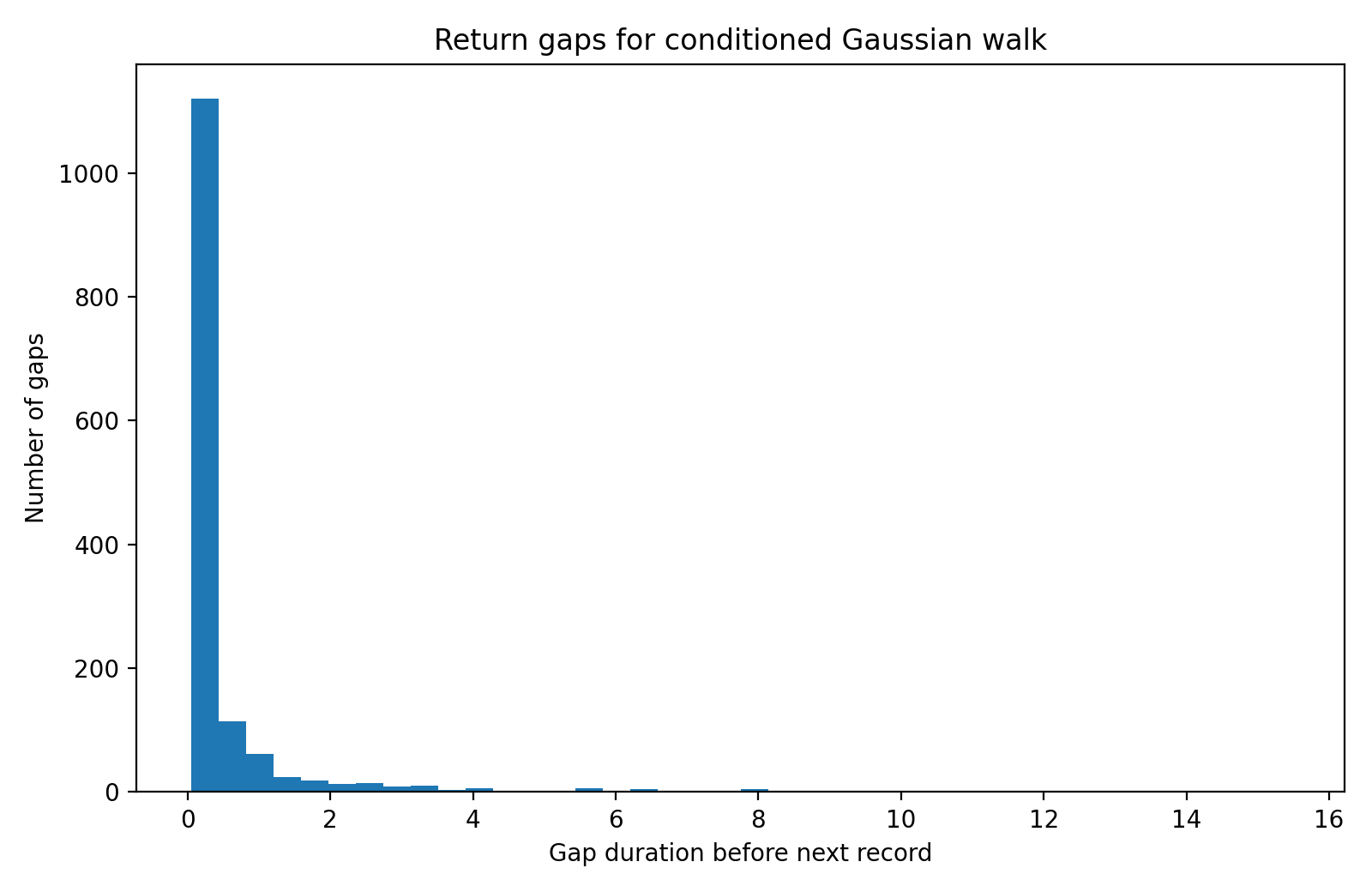}
\caption{Distribution of return gaps for the conditioned Gaussian transverse
walk. A gap is an interval with \(s\geq0\) before the next return to \(s<0\).}
\label{fig:sim4c-return-gap-histogram}
\end{figure}

The histogram of return-gap durations is shown in
Figure~\ref{fig:sim4c-return-gap-histogram}. The median gap is short, while
rare longer gaps occur because the transverse motion between records is an
unbiased Gaussian walk.

The return-gap statistics are summarized in
Table~\ref{tab:sim4c-return-gap-summary}. In the simulation there were
\[
1430
\]
return gaps. The mean gap length was
\[
10.95
\]
steps, corresponding to a mean gap duration of approximately
\[
0.547.
\]
The median gap length was only \(2\) steps. The \(90\%\) and \(99\%\) quantiles
were \(23\) and \(160.84\) steps, respectively.

\begin{table}[h]
\centering
\caption{Return-gap statistics for intervals without localized records.}
\label{tab:sim4c-return-gap-summary}
\begin{tabular}{lc}
\toprule
Quantity & Value \\
\midrule
Number of return gaps & \(1430\) \\
Mean gap length & \(10.95\) steps \\
Median gap length & \(2\) steps \\
\(90\%\) quantile & \(23\) steps \\
\(99\%\) quantile & \(160.84\) steps \\
Maximum gap length & \(309\) steps \\
Mean gap duration & \(0.547\) \\
Median gap duration & \(0.100\) \\
\bottomrule
\end{tabular}
\end{table}

\subsection{Parameter sweep for the stroboscopic Newtonian regime}
\label{subsec:stroboscopic-parameter-sweep}

The preceding simulation used one set of parameters satisfying
\[
\frac{D_a\Delta t}{\sigma^2}\ll1.
\]
We now vary this dimensionless ratio in order to test the robustness of the
stroboscopic Newtonian regime. Define
\[
\varepsilon
=
\frac{D_a\Delta t}{\sigma^2}.
\]
Since the recorded tangential displacement satisfies
\[
\eta_k\sim N(0,D_a\Delta t),
\]
the expected scaling is
\[
\frac{\mathrm{RMS}}{\sigma}
\approx
\sqrt{\varepsilon}.
\]
Thus the condition for Newtonian records to remain sharp on the detector
resolution scale is
\[
\varepsilon\ll1.
\]

We used the same harmonic oscillator and candidate recording times as in the
previous subsection:
\[
M=1,\qquad
\omega=1,\qquad
a_0=1,\qquad
p_0=0.35,
\]
\[
\Delta t=0.05,
\qquad
0\leq t\leq20,
\]
with
\[
\sigma=0.10.
\]
For each value of \(\varepsilon\), the diffusion coefficient was chosen as
\[
D_a=\frac{\varepsilon\sigma^2}{\Delta t}.
\]
The localization coordinate was generated by the same conditioned Gaussian
return process, with
\[
D_s=0.25,
\qquad
\sqrt{2D_s\Delta t}=0.1581,
\qquad
s_{\rm rec}=-0.5.
\]
Only points satisfying \(s_k<0\) were counted as records. The sweep used
\(2500\) independent trajectories.

\begin{figure}[h]
\centering
\includegraphics[width=0.78\textwidth]{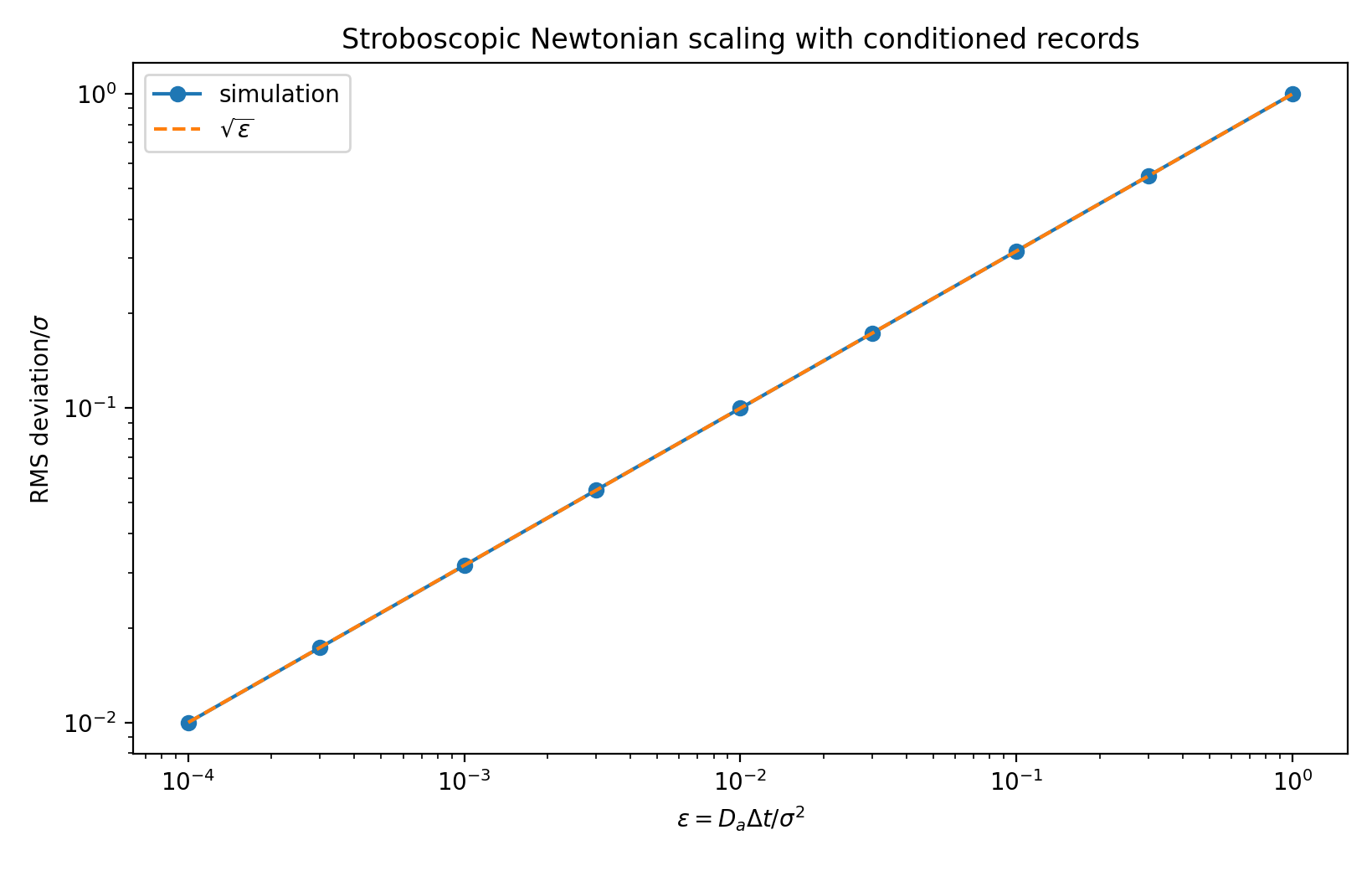}
\caption{Parameter sweep for the conditioned stroboscopic Newtonian regime. The
normalized RMS deviation of recorded positions satisfies
\(\mathrm{RMS}/\sigma\approx\sqrt{\varepsilon}\), where
\(\varepsilon=D_a\Delta t/\sigma^2\).}
\label{fig:sim9c-epsilon-sweep-rms}
\end{figure}

Figure~\ref{fig:sim9c-epsilon-sweep-rms} shows the normalized RMS deviation of
the recorded positions from the Newtonian trajectory as a function of
\(\varepsilon\). The numerical results agree with the expected
\(\sqrt{\varepsilon}\) scaling.

Selected numerical values are shown in
Table~\ref{tab:sim9c-epsilon-sweep}.

\begin{table}[h]
\centering
\caption{Scaling of recorded-position deviations with
\(\varepsilon=D_a\Delta t/\sigma^2\) for the conditioned return process.}
\label{tab:sim9c-epsilon-sweep}
\begin{tabular}{ccc}
\toprule
\(\varepsilon\) & Simulated \(\mathrm{RMS}/\sigma\) & \(\sqrt{\varepsilon}\) \\
\midrule
\(10^{-4}\) & \(0.009995\) & \(0.010000\) \\
\(10^{-3}\) & \(0.031584\) & \(0.031623\) \\
\(10^{-2}\) & \(0.099952\) & \(0.100000\) \\
\(10^{-1}\) & \(0.316389\) & \(0.316228\) \\
\(1\)       & \(0.999496\) & \(1.000000\) \\
\bottomrule
\end{tabular}
\end{table}

\begin{figure}[h]
\centering
\includegraphics[width=0.78\textwidth]{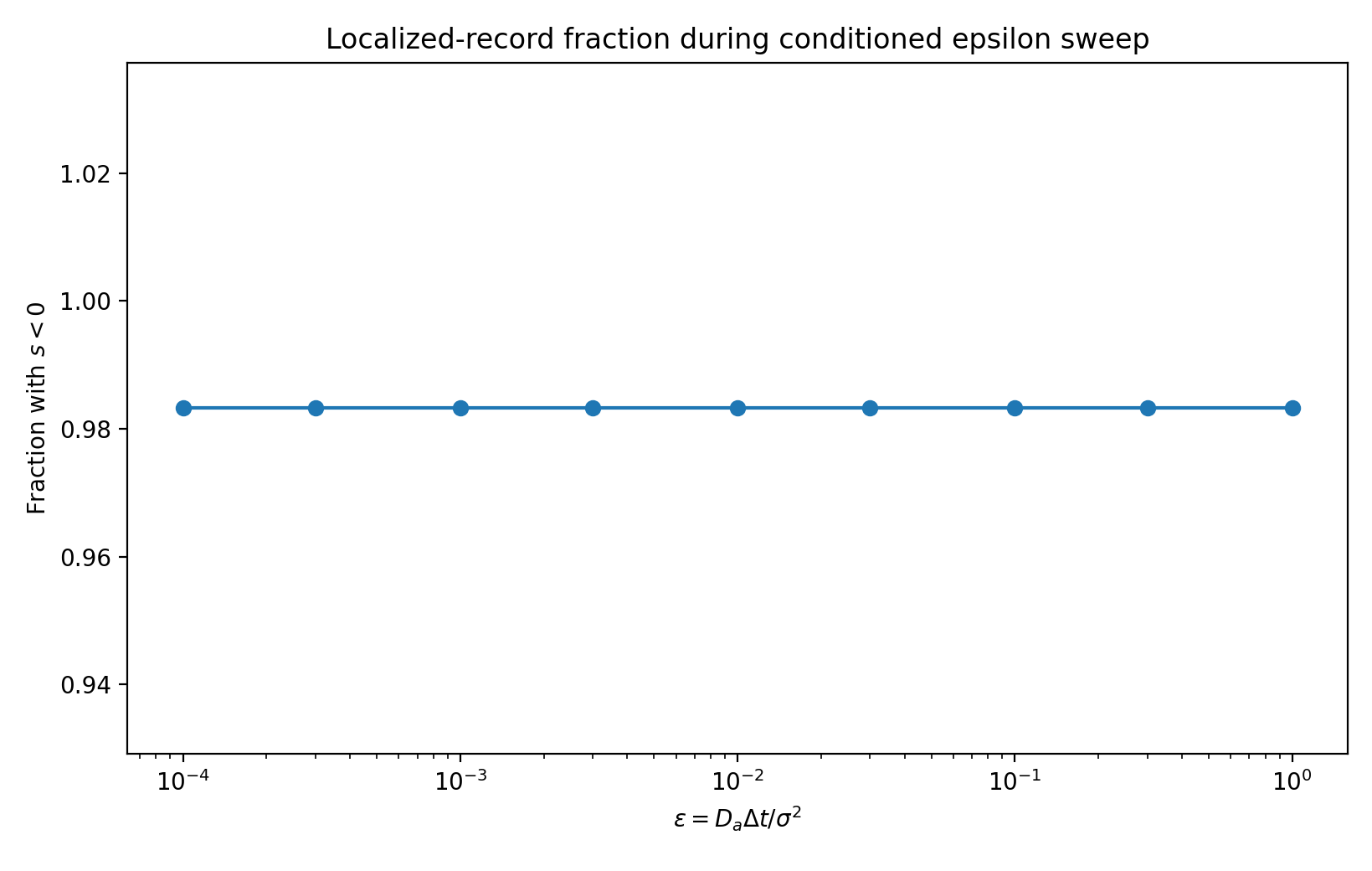}
\caption{Localized-record fraction during the \(\varepsilon\)-sweep for the
conditioned Gaussian return process. The fraction is essentially independent of
\(\varepsilon\).}
\label{fig:sim9c-epsilon-record-fraction}
\end{figure}

In the same sweep, the fraction of candidate times satisfying \(s<0\) was
approximately
\[
0.9832.
\]
Figure~\ref{fig:sim9c-epsilon-record-fraction} shows that this fraction is
independent of \(\varepsilon\), as expected, since \(\varepsilon\) controls the
tangential spread while the conditioned \(s\)-process controls return to the
localized sector.

\begin{figure}[h]
\centering
\includegraphics[width=0.78\textwidth]{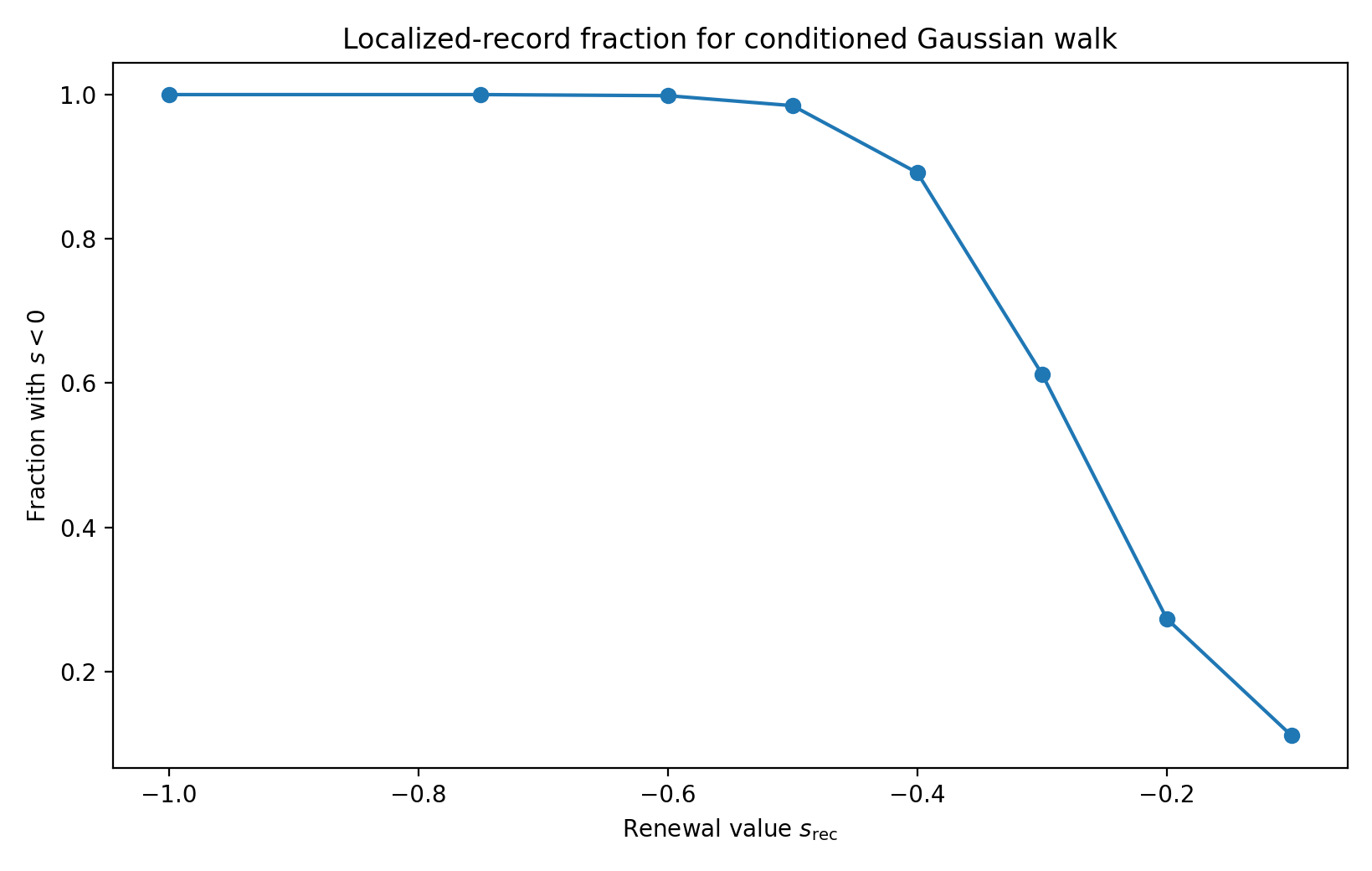}
\caption{Localized-record fraction as a function of the renewal value
\(s_{\rm rec}\). More negative renewal values produce more frequent localized
records.}
\label{fig:sim9c-record-fraction-srec}
\end{figure}

We also varied the renewal value \(s_{\rm rec}\), keeping
\[
D_s=0.25,
\qquad
\sqrt{2D_s\Delta t}=0.1581.
\]
More negative values of \(s_{\rm rec}\) correspond to stronger renewal into the
localized sector after a record. The result is shown in
Figure~\ref{fig:sim9c-record-fraction-srec}.

\begin{figure}[h]
\centering
\includegraphics[width=0.78\textwidth]{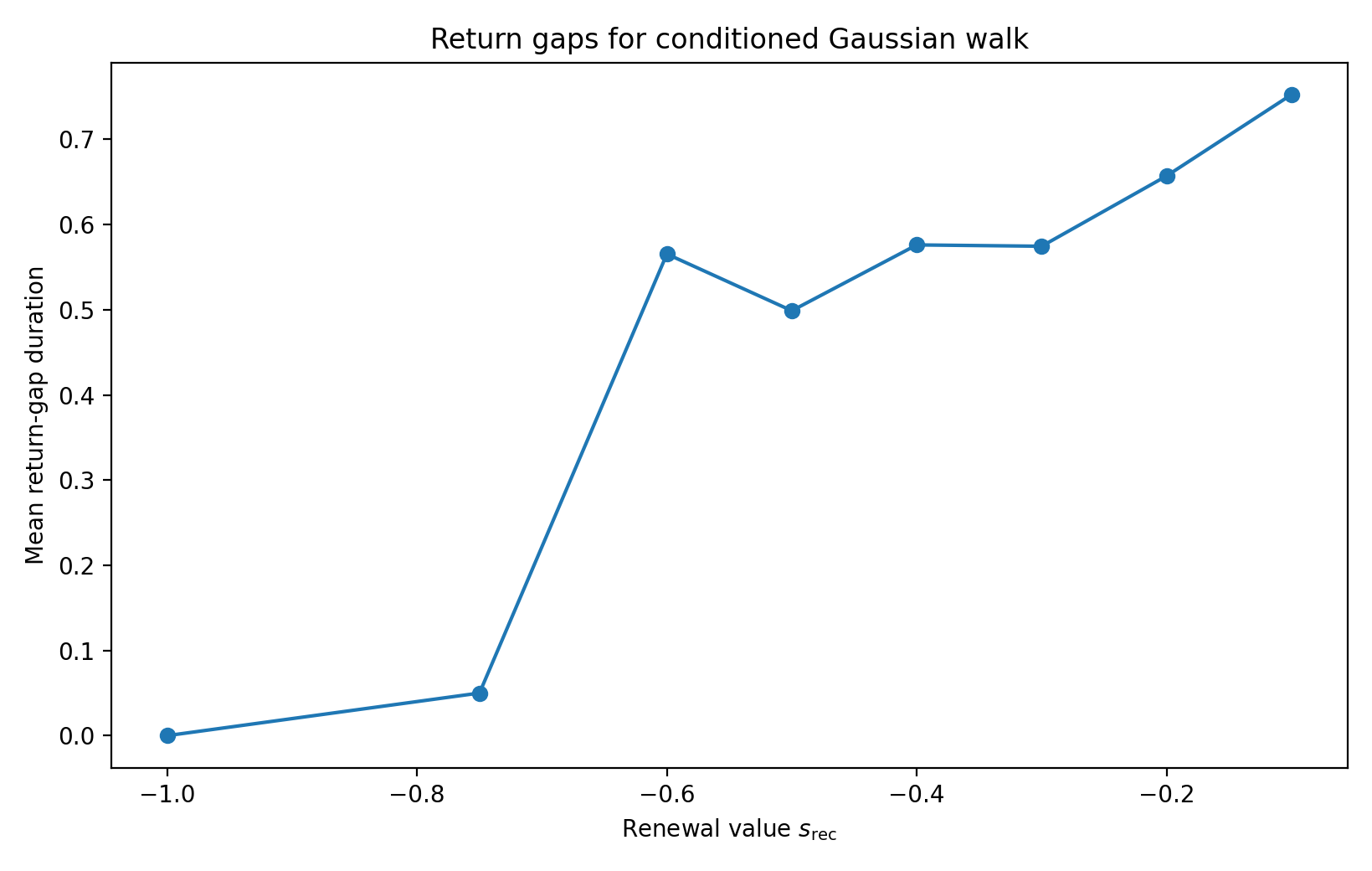}
\caption{Mean return-gap duration as a function of the renewal value
\(s_{\rm rec}\).}
\label{fig:sim9c-return-gap-srec}
\end{figure}

The return-gap duration also depends on \(s_{\rm rec}\). Figure~\ref{fig:sim9c-return-gap-srec}
shows the mean return-gap duration as a function of the renewal value.

Selected values are summarized in
Table~\ref{tab:sim9c-localization-sweep}.

\begin{table}[h]
\centering
\caption{Localized-record fraction and return-gap duration as functions of the
renewal value \(s_{\rm rec}\).}
\label{tab:sim9c-localization-sweep}
\begin{tabular}{ccc}
\toprule
\(s_{\rm rec}\) & Fraction with \(s<0\) & Mean return-gap duration \\
\midrule
\(-1.00\) & \(1.000000\) & \(0.000\) \\
\(-0.75\) & \(0.999999\) & \(0.050\) \\
\(-0.60\) & \(0.998486\) & \(0.565\) \\
\(-0.50\) & \(0.984658\) & \(0.499\) \\
\(-0.40\) & \(0.891651\) & \(0.576\) \\
\(-0.30\) & \(0.612388\) & \(0.574\) \\
\(-0.10\) & \(0.111176\) & \(0.753\) \\
\bottomrule
\end{tabular}
\end{table}

These parameter sweeps identify the two independent conditions for
stroboscopic Newtonian behavior. First, the tangential displacement between
returns must be small on the detector scale:
\[
\frac{D_a\Delta t}{\sigma^2}\ll1.
\]
Second, the conditioned transverse process must produce localized records with
high probability:
\[
\mathbb P(s<0)\approx1.
\]
When both conditions hold, the sequence of recorded positions remains sharply
concentrated around the Newtonian trajectory. When the first condition fails,
the recorded positions are no longer sharp on the detector scale. When the
second condition fails, many candidate times do not produce localized classical
records. Thus the simulation identifies a parameter regime, rather than a
fine-tuned example, in which the \({\bf (RM)}\) dynamics yields stroboscopic
Newtonian motion.

\subsection{Interpretation}

The simulation confirms the effective macroscopic regime described in the
companion theoretical paper. The free Schr\"odinger dynamics supplies the
Newtonian tangent drift. The \({\bf (RM)}\) interaction supplies a small
tangential displacement \(\eta_k\), and the localization coordinate \(s_k\)
determines whether a classical record is actually produced.

The important point is that Newtonian consistency is tested only on the
recorded subset satisfying
\[
s_k<0.
\]
On this subset, the recorded positions remain concentrated around the Newtonian
trajectory with RMS deviation
\[
\sqrt{D_a\Delta t}=0.01,
\]
well below the detector resolution \(\sigma=0.10\). At the same time, the
return-gap statistics show that excursions outside the localized sector are
rare for the chosen macroscopic parameters, although the unbiased transverse
walk can produce occasional longer gaps.

The parameter sweep shows that the relevant dimensionless control parameter for
the tangential accuracy of the records is
\[
\varepsilon=\frac{D_a\Delta t}{\sigma^2}.
\]
When \(\varepsilon\ll1\), the stochastic tangential displacement accumulated
between successive localized records is small on the detector scale, and the
records form a sharp stroboscopic Newtonian trajectory. This is the regime
appropriate to macroscopic track formation, such as bubble- or cloud-chamber
records: the individual detector events have finite spatial resolution, while
the accumulated stochastic displacement between neighboring records is small on
that scale.

The same simulation also separates the two requirements for macroscopic
classicality. The smallness of \(\varepsilon\) controls the tangential accuracy
of recorded positions, while the condition \(s<0\) controls whether the state is
sufficiently localized to produce a classical record at all. Thus Newtonian
motion requires both
\[
\frac{D_a\Delta t}{\sigma^2}\ll1
\]
and frequent returns to the localized sector.

Thus the numerical model illustrates the macroscopic limit of the
\({\bf (RM)}\) framework:
\[
\text{Schr\"odinger tangent drift}
+
\text{\({\bf (RM)}\)-induced tangential diffusion}
+
\text{conditioning on }s<0
\]
produces a sequence of classical records concentrated around a Newtonian
trajectory.

This also clarifies the relation to the Born-rule regime. In the macroscopic
Newtonian regime, the state remains near a single classical branch and the
\({\bf (RM)}\)-induced diffusion appears as small normal measurement error. In
a microscopic measurement, or when the state is not confined to one classical
branch, the relevant coarse-grained objects are instead detector-defined
equivalence classes. The same state-space diffusion then yields outcome
weights for those classes, giving the Born-rule frequencies.

\section{Isotropy and homogeneity in tensor-product state space}
\label{sec:tensor-product-isotropy-homogeneity}

\subsection{Purpose of the simulation}

Before using the tensor-product Brownian calibration to derive the aggregate
weights of paths reaching detector classes, we first check that the underlying
\({\bf (RM)}\)-induced state-space diffusion remains isotropic and homogeneous
in tensor-product Hilbert space.
This is important because the Born-weight construction must apply not only to
product states on
\[
M_1^{\sigma_1}\otimes M_1^{\sigma_2},
\]
but also to arbitrary superpositions and entangled states.

The finite-dimensional test is performed in
\[
\mathcal H
=
\mathbb C^{10}\otimes\mathbb C^{10},
\qquad
\dim_{\mathbb C}\mathcal H=100.
\]
For a normalized state \(\psi\in\mathcal H\), a GUE-induced infinitesimal
projective displacement is
\[
\delta\psi_\perp
=
-iH\psi
-
\psi\langle\psi,-iH\psi\rangle,
\qquad
H\in{\rm GUE}(100).
\]
Homogeneity means that the distribution of
\[
\|\delta\psi_\perp\|^2
\]
does not depend on the initial state \(\psi\). Isotropy means that the
variance of the projection of \(\delta\psi_\perp\) onto tangent directions is
independent of the chosen tangent direction.

\subsection{Test states}

We tested four representative states in the tensor-product Hilbert space:
\[
\psi_1=|0\rangle\otimes |0\rangle,
\]
a product basis state;
\[
\psi_2=u\otimes v,
\]
a product superposition state;
\[
\psi_3=\frac{1}{\sqrt2}
\left(|0\rangle\otimes|0\rangle+|1\rangle\otimes|1\rangle\right),
\]
a two-branch entangled state; and a randomly chosen entangled state
\[
\psi_4=\psi_{\rm rand}.
\]
For each state, \(20000\) independent GUE-induced projective increments were
sampled.

\subsection{Homogeneity of the step-size distribution}

The step-size distributions are summarized in
Table~\ref{tab:tensor-product-homogeneity-summary}. The means and quantile
ranges are essentially the same for product and entangled states.

\begin{table}[h]
\centering
\caption{Homogeneity of the GUE-induced projective step size in
\(\mathbb C^{10}\otimes\mathbb C^{10}\).}
\label{tab:tensor-product-homogeneity-summary}
\begin{tabular}{lccc}
\toprule
State & Mean \(\|\delta\psi_\perp\|^2\) & \(5\%\) quantile & \(95\%\) quantile \\
\midrule
Product basis & \(99.054\) & \(83.385\) & \(116.109\) \\
Product superposition & \(99.094\) & \(83.380\) & \(115.780\) \\
Two-branch entangled & \(99.083\) & \(83.089\) & \(116.052\) \\
Random entangled & \(98.978\) & \(83.166\) & \(115.696\) \\
\bottomrule
\end{tabular}
\end{table}

\begin{figure}[h]
\centering
\includegraphics[width=0.78\textwidth]{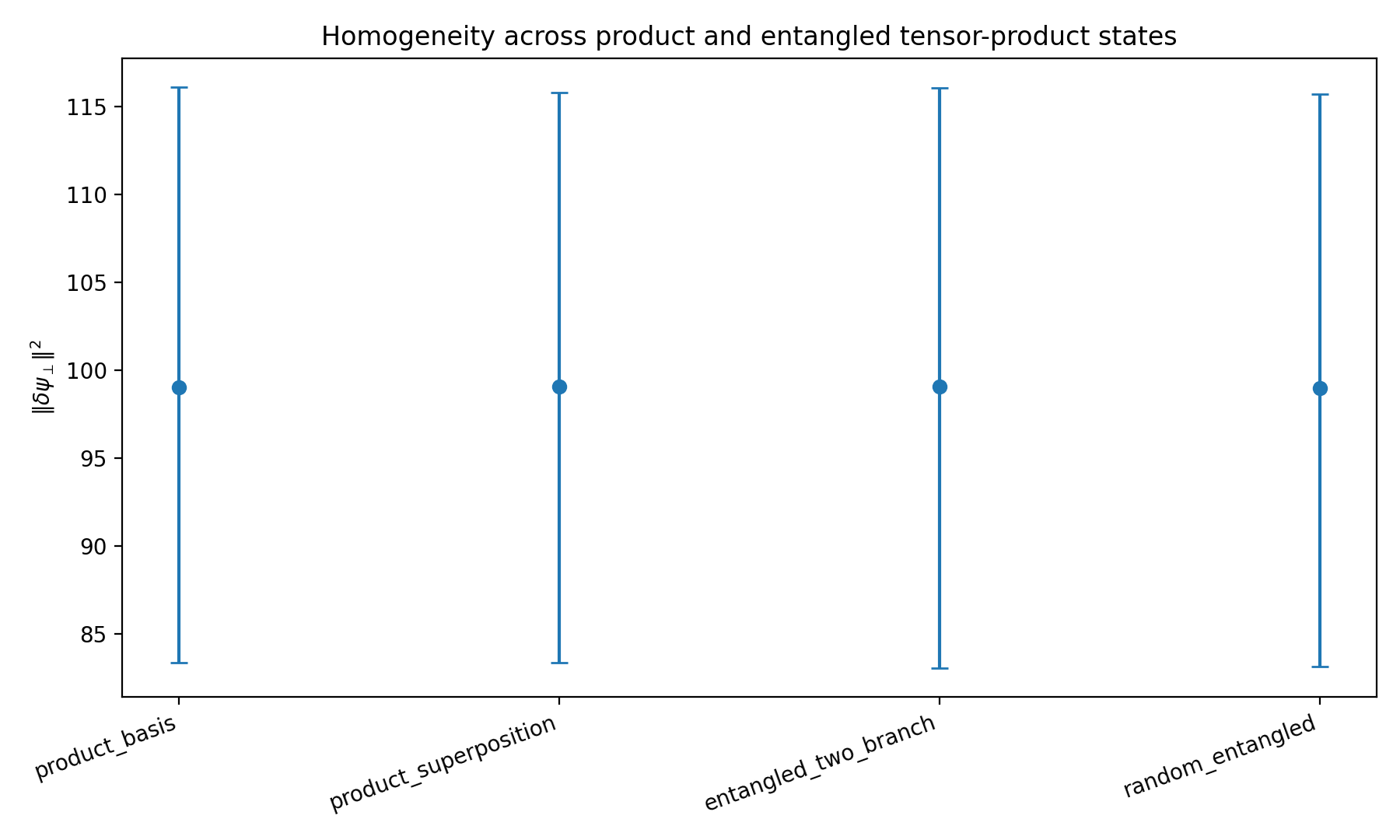}
\caption{Homogeneity of GUE-induced projective step sizes across product and
entangled states in \(\mathbb C^{10}\otimes\mathbb C^{10}\).}
\label{fig:tensor-product-homogeneity-norms}
\end{figure}

Figure~\ref{fig:tensor-product-homogeneity-norms} shows the same data
graphically.

We also compared the step-size distribution at the product basis state with
the distributions at the other states using two-sample
Kolmogorov--Smirnov tests. The results are shown in
Table~\ref{tab:tensor-product-homogeneity-ks}. No statistically significant
difference is observed.

\begin{table}[h]
\centering
\caption{Kolmogorov--Smirnov tests comparing the step-size distribution at the
product basis state with the distributions at the other tensor-product states.}
\label{tab:tensor-product-homogeneity-ks}
\begin{tabular}{lcc}
\toprule
Comparison & KS statistic & \(p\)-value \\
\midrule
Product basis vs. product superposition & \(0.00830\) & \(0.494\) \\
Product basis vs. two-branch entangled & \(0.00805\) & \(0.533\) \\
Product basis vs. random entangled & \(0.00645\) & \(0.797\) \\
\bottomrule
\end{tabular}
\end{table}

\subsection{Isotropy in tangent directions}

To test isotropy, we selected random tangent directions at each of the four
states and computed the variance of the projection of
\(\delta\psi_\perp\) onto those directions. For an isotropic complex Gaussian
tangent distribution, the real and imaginary components of these projections
should have equal variance, independent of direction and independent of the
base state.

\begin{figure}[h]
\centering
\includegraphics[width=0.78\textwidth]{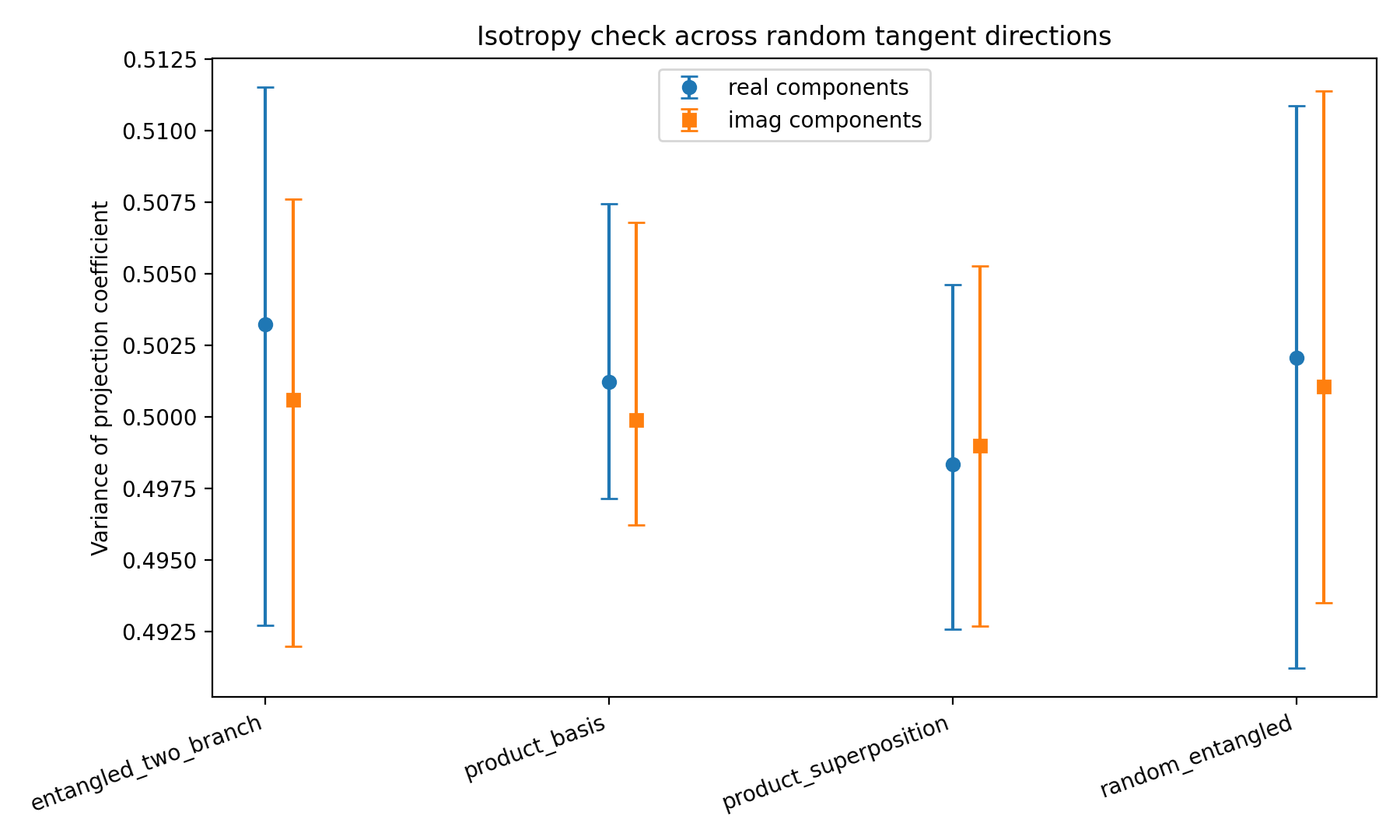}
\caption{Isotropy check in tensor-product state space. For each product or
entangled base state, random tangent directions were sampled and the variances
of the real and imaginary projection components were computed. The variances
are independent of both the tangent direction and the base state.}
\label{fig:tensor-product-direction-variances}
\end{figure}

The results are shown in
Figure~\ref{fig:tensor-product-direction-variances}. The variances are all
approximately \(0.5\), with only sampling-level fluctuations.

Table~\ref{tab:tensor-product-direction-variance-summary} gives the mean
variance over the sampled tangent directions.

\begin{table}[h]
\centering
\caption{Mean variance of projection components over random tangent directions.}
\label{tab:tensor-product-direction-variance-summary}
\begin{tabular}{lcc}
\toprule
State & Mean variance, real part & Mean variance, imaginary part \\
\midrule
Product basis & \(0.5012\) & \(0.4999\) \\
Product superposition & \(0.4983\) & \(0.4990\) \\
Two-branch entangled & \(0.5032\) & \(0.5006\) \\
Random entangled & \(0.5021\) & \(0.5011\) \\
\bottomrule
\end{tabular}
\end{table}

\subsection{Interpretation}

The simulation confirms that the GUE-induced projective diffusion is not
special to product Gaussian states. It is homogeneous across product and
entangled states in the tensor-product Hilbert space, and it is isotropic in
the tangent directions at those states. This justifies using the same
state-space diffusion when passing from the Brownian calibration on
\[
M_1^{\sigma_1}\otimes M_1^{\sigma_2}
\]
to detector-cell weights for arbitrary two-particle states.

Thus the Born-weight construction used below is not tied to states lying on
the classical product submanifold. The Brownian calibration fixes the scale on
the classical sector, while GUE homogeneity and isotropy extend the same
diffusion to arbitrary superpositions and entangled states in the full
tensor-product projective space.

\section{Tensor-product Brownian restriction and the device limit}
\label{sec:tensor-product-device}

\subsection{Purpose of the simulation}

The preceding simulations tested the one-particle Brownian restriction and the
two-outcome reduction-coordinate model. We now test the corresponding
tensor-product situation. The goal is to verify that, for a product of two
localized Gaussian states, the projected \({\bf (RM)}\)-induced motion gives
independent Brownian increments in the two classical position coordinates.

This is the first step toward a particle--device model. We begin with two
ordinary particles, with localization widths \(\sigma_1\) and \(\sigma_2\),
without assuming that either one is a macroscopic device. After verifying the
independence of the two projected Brownian motions, we let \(\sigma_2\) become
small. In this limit the second coordinate becomes stable in Euclidean space,
which is the expected behavior of a sharply localized measuring device.

\subsection{Product Gaussian states and tangent directions}

We consider the product Gaussian state
\[
\Psi_{a,b}(x,y)
=
g_{a,\sigma_1}(x)\,g_{b,\sigma_2}(y),
\]
where
\[
g_{a,\sigma_1}(x)
=
\left(\frac{1}{2\pi\sigma_1^2}\right)^{1/4}
\exp\left[-\frac{(x-a)^2}{4\sigma_1^2}\right],
\]
and similarly for \(g_{b,\sigma_2}(y)\). The corresponding product classical
submanifold is
\[
M_1^{\sigma_1}\otimes M_1^{\sigma_2}.
\]

The two tangent directions associated with translations of the two particles
are
\[
T_1
=
\partial_a g_{a,\sigma_1}\otimes g_{b,\sigma_2},
\]
and
\[
T_2
=
g_{a,\sigma_1}\otimes \partial_b g_{b,\sigma_2}.
\]
The induced Fubini--Study metric on these tangent directions is diagonal:
\[
G_{11}
=
\|T_1\|^2
=
\frac{1}{4\sigma_1^2},
\qquad
G_{22}
=
\|T_2\|^2
=
\frac{1}{4\sigma_2^2},
\qquad
G_{12}=0.
\]
Thus the two translation directions are orthogonal in the induced
Fubini--Study metric.

At a fixed projective state, the local GUE-induced tangent distribution is
isotropic in projective Hilbert space. Therefore its projections onto the two
orthogonal tangent directions \(T_1,T_2\) should be independent Gaussian
variables when measured in the Fubini--Study metric. Equivalently, the
normalized increments
\[
\sqrt{G_{11}}\,\Delta a_1,
\qquad
\sqrt{G_{22}}\,\Delta a_2
\]
should be independent standard normal variables, up to the common overall
diffusion scale.

\subsection{Independent Brownian components}

We first take
\[
\sigma_1=0.75,
\qquad
\sigma_2=1.0.
\]
The metric coefficients are therefore
\[
G_{11}=0.444444,
\qquad
G_{22}=0.25.
\]
The simulated projected increments were sampled from the local GUE tangent
distribution and then projected onto \(T_1\) and \(T_2\). The results are shown
in Table~\ref{tab:tensor-product-increment-summary}.

\begin{table}[h]
\centering
\caption{Projected tensor-product increments for
\(\sigma_1=0.75\), \(\sigma_2=1.0\). The variables
\(z_i=\sqrt{G_{ii}}\Delta a_i\) are the Fubini--Study-normalized increments.}
\label{tab:tensor-product-increment-summary}
\begin{tabular}{lclc}
\toprule
\multicolumn{2}{c}{Metric data} &
\multicolumn{2}{c}{Increment statistics} \\
\cmidrule(r){1-2}
\cmidrule(l){3-4}
Quantity & Value & Quantity & Value \\
\midrule
Samples & \(50000\) &
\(\operatorname{Std}(\Delta a_1)\) & \(1.5001\) \\
\(G_{11}\) & \(0.444444\) &
\(\operatorname{Std}(\Delta a_2)\) & \(1.9925\) \\
\(G_{22}\) & \(0.25\) &
\(\operatorname{Std}(z_1)\) & \(1.0001\) \\
\(G_{12}\) & \(0\) &
\(\operatorname{Std}(z_2)\) & \(0.9962\) \\
& &
\(\operatorname{Corr}(z_1,z_2)\) & \(-0.0012\) \\
\bottomrule
\end{tabular}
\end{table}

\begin{figure}[h]
\centering
\includegraphics[width=0.68\textwidth]{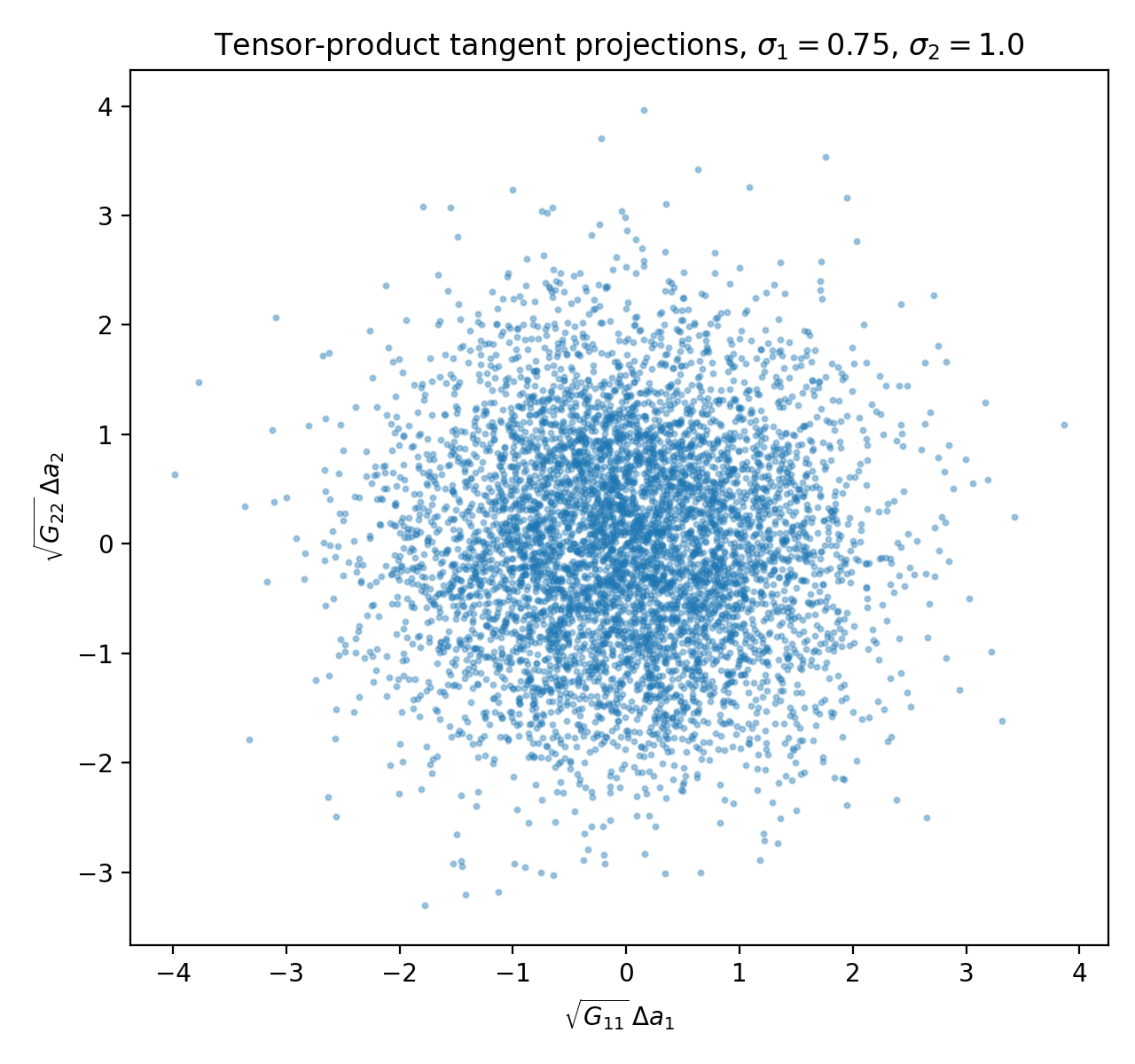}
\caption{Fubini--Study-normalized projected increments for the product
Gaussian state. The two coordinates represent the translation directions of the
two particles on \(M_1^{\sigma_1}\otimes M_1^{\sigma_2}\). The near-circular
scatter plot shows that the projected increments are independent Gaussian
variables in the induced metric.}
\label{fig:tensor-product-increment-scatter}
\end{figure}

Figure~\ref{fig:tensor-product-increment-scatter} shows the scatter plot of the
Fubini--Study-normalized increments
\[
z_1=\sqrt{G_{11}}\Delta a_1,
\qquad
z_2=\sqrt{G_{22}}\Delta a_2.
\]
The approximately circular cloud and near-zero correlation confirm that the
projected tensor-product motion gives independent Brownian increments in the
two classical coordinates.

In Euclidean coordinates, the increments are not normalized by the metric.
Since
\[
G_{ii}=\frac{1}{4\sigma_i^2},
\]
a fixed Fubini--Study-normalized step corresponds to a Euclidean displacement
whose standard deviation is proportional to \(\sigma_i\):
\[
\operatorname{Std}(\Delta a_i)
\propto
\sigma_i.
\]

\subsection{The device limit}

We next keep
\[
\sigma_1=0.75
\]
fixed and vary \(\sigma_2\). The second particle is interpreted as becoming
more device-like as \(\sigma_2\) decreases, because its Euclidean position
coordinate becomes more sharply localized. The Fubini--Study-normalized
increment remains of order one, but the Euclidean increment \(\Delta a_2\)
shrinks proportionally to \(\sigma_2\).

\begin{figure}[h]
\centering
\includegraphics[width=0.78\textwidth]{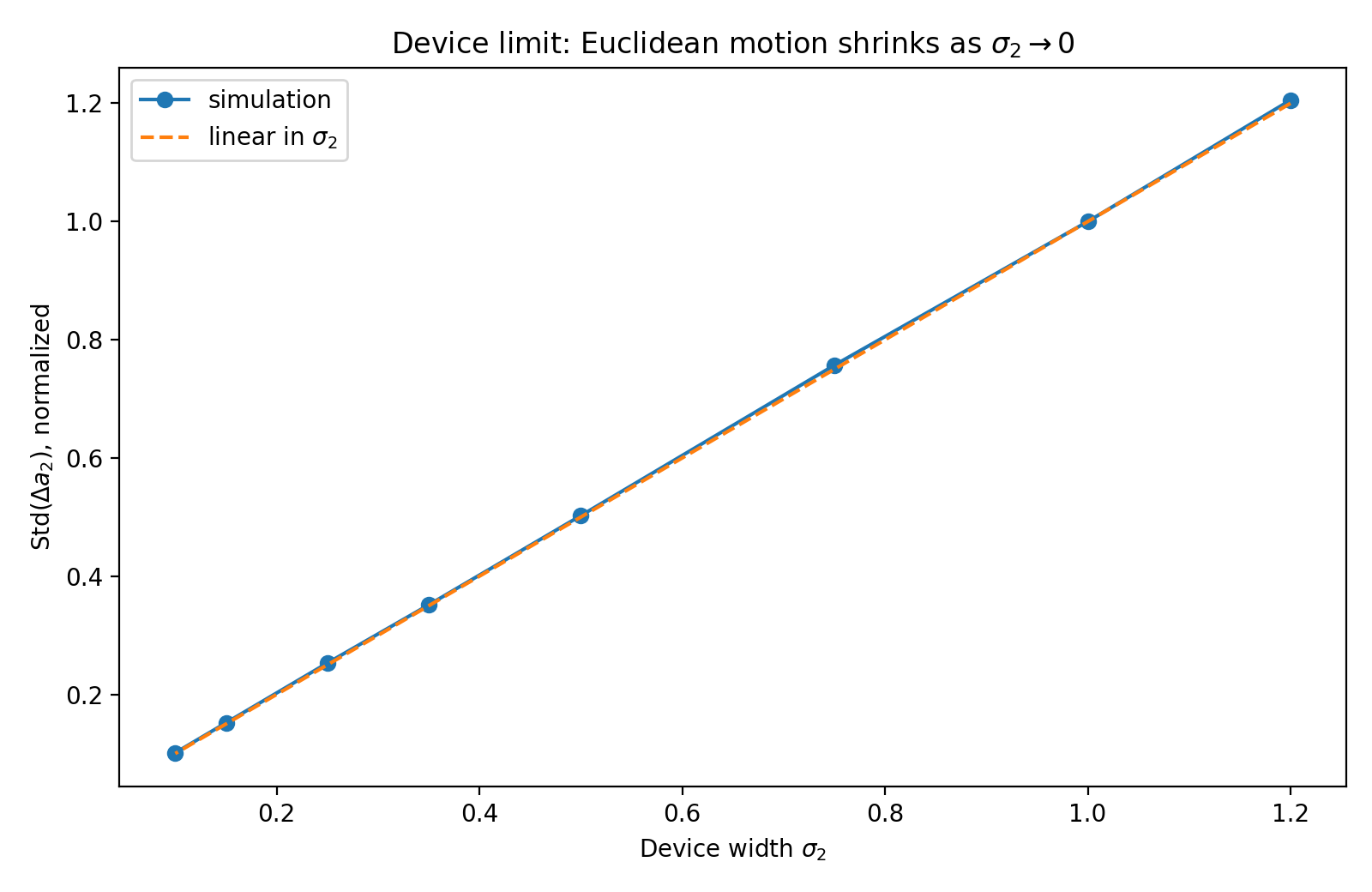}
\caption{Device limit for the second coordinate. The Euclidean standard
deviation of the projected increment \(\Delta a_2\) is proportional to
\(\sigma_2\). Thus, as \(\sigma_2\to0\), the second coordinate becomes stable
in Euclidean space, although the Fubini--Study-normalized increment remains of
order one.}
\label{fig:sigma2-device-limit}
\end{figure}

Figure~\ref{fig:sigma2-device-limit} shows the numerical scaling of
\(\operatorname{Std}(\Delta a_2)\) as a function of \(\sigma_2\). The values
are normalized by the value at \(\sigma_2=1\). The dashed line shows the
expected linear dependence on \(\sigma_2\).

Selected values are shown in Table~\ref{tab:sigma2-device-limit}.

\begin{table}[h]
\centering
\caption{Scaling of the Euclidean increment of the second coordinate as
\(\sigma_2\) decreases. The standard deviation is normalized by its value at
\(\sigma_2=1\).}
\label{tab:sigma2-device-limit}
\begin{tabular}{ccc}
\toprule
\(\sigma_2\) & Normalized \(\operatorname{Std}(\Delta a_2)\) &
\(\sigma_2/\sigma_{\rm ref}\) \\
\midrule
\(1.20\) & \(1.2054\) & \(1.20\) \\
\(1.00\) & \(1.0000\) & \(1.00\) \\
\(0.75\) & \(0.7568\) & \(0.75\) \\
\(0.50\) & \(0.5024\) & \(0.50\) \\
\(0.35\) & \(0.3515\) & \(0.35\) \\
\(0.25\) & \(0.2530\) & \(0.25\) \\
\(0.15\) & \(0.1519\) & \(0.15\) \\
\(0.10\) & \(0.1012\) & \(0.10\) \\
\bottomrule
\end{tabular}
\end{table}

Thus the same Fubini--Study-normalized \({\bf (RM)}\)-induced motion has
different Euclidean manifestations depending on the localization scale. For two
ordinary particles with comparable \(\sigma_1,\sigma_2\), the projected motion
gives two independent Brownian motions. When \(\sigma_2\) becomes very small,
the second coordinate becomes stable in Euclidean space and can be interpreted
as a sharply localized device coordinate.

\subsection{Interpretation}

The simulation confirms the tensor-product version of the Brownian calibration.
The product Gaussian manifold
\[
M_1^{\sigma_1}\otimes M_1^{\sigma_2}
\]
has two orthogonal translation directions, and the projected GUE-induced
motion gives independent Gaussian increments in those directions when measured
with the induced Fubini--Study metric. This is the two-particle analogue of the
one-particle Brownian restriction tested earlier.

The device limit arises from the metric scaling. A small value of
\(\sigma_2\) makes the second coordinate sharply localized in Euclidean space.
Since
\[
\operatorname{Std}(\Delta a_2)\propto \sigma_2,
\]
the Euclidean motion of that coordinate becomes small as
\(\sigma_2\to0\). Thus a sharply localized second particle can serve as a
model for a measuring device or pointer coordinate.

This prepares the next simulation, in which a measurement is applied to one
coordinate of an entangled two-particle state. The relevant detector weights are the aggregate weights of {\bf (RM)}
paths reaching the corresponding detector classes. They are obtained from the
same tensor-product calibration. For detector cells \(I_j\) in the measured
coordinate,
\[
W_j
=
\|(P_{I_j}\otimes I)\Psi\|^2.
\]
The conditional state of the second particle after outcome \(I_j\) is obtained
by restricting the joint state to that detector class and normalizing.

\section{Two-particle Born weights and conditional records}
\label{sec:two-particle-born-weights}

\subsection{Purpose of the simulation}

The preceding tensor-product simulations showed that the \({\bf (RM)}\)-induced
state-space diffusion remains homogeneous and isotropic in tensor-product
Hilbert space, and that its restriction to the product Gaussian manifold gives
independent Brownian motion in the two position coordinates. We now use this calibration to assign aggregate path weights, equivalently
detector-cell weights, for two-particle states.

The goal of the present simulation is twofold. First, we verify that for a
product state the detector-cell weights factor into the product of the
one-particle weights. Second, we consider an entangled two-particle state and
show that measuring the position of one particle gives the Born weights for the
measured particle and the corresponding conditional state of the second
particle.

The detector acts only on the first coordinate. For a detector cell \(I_j\) in
the \(x\)-coordinate, the corresponding projection is
\[
P_{I_j}\otimes I.
\]
The weight of the outcome \(I_j\) is therefore
\[
W_j
=
\|(P_{I_j}\otimes I)\Psi\|^2
=
\int_{I_j}\int |\Psi(x,y)|^2\,dy\,dx.
\]
After the outcome \(I_j\) is recorded, the conditional two-particle state is
\[
\Psi_j
=
\frac{(P_{I_j}\otimes I)\Psi}{\|(P_{I_j}\otimes I)\Psi\|}.
\]
The corresponding conditional state of the second particle is obtained by
tracing out, or equivalently integrating over, the measured coordinate within
the recorded cell.

\subsection{Product-state factorization}

We first consider a product state
\[
\Psi(x,y)=\psi_1(x)\psi_2(y).
\]
Let the detector cells in the two coordinates be
\[
I_j,\qquad J_k.
\]
The joint detector-cell weight is
\[
W_{jk}
=
\int_{I_j\times J_k}|\Psi(x,y)|^2\,dx\,dy.
\]
For a product state this must factor:
\[
W_{jk}
=
\left(\int_{I_j}|\psi_1(x)|^2\,dx\right)
\left(\int_{J_k}|\psi_2(y)|^2\,dy\right)
=
W_j^{(1)}W_k^{(2)}.
\]

The numerical simulation used Gaussian factors
\[
\psi_1(x)=g_{-1,0.8}(x),
\qquad
\psi_2(y)=g_{1.2,1.1}(y),
\]
on a grid
\[
x,y\in[-6,6],
\]
with eight detector bins in each coordinate. The joint weights were computed
directly from the two-dimensional density and compared with the factorized
weights obtained from the one-dimensional marginals.

\begin{figure}[h]
\centering
\includegraphics[width=0.78\textwidth]{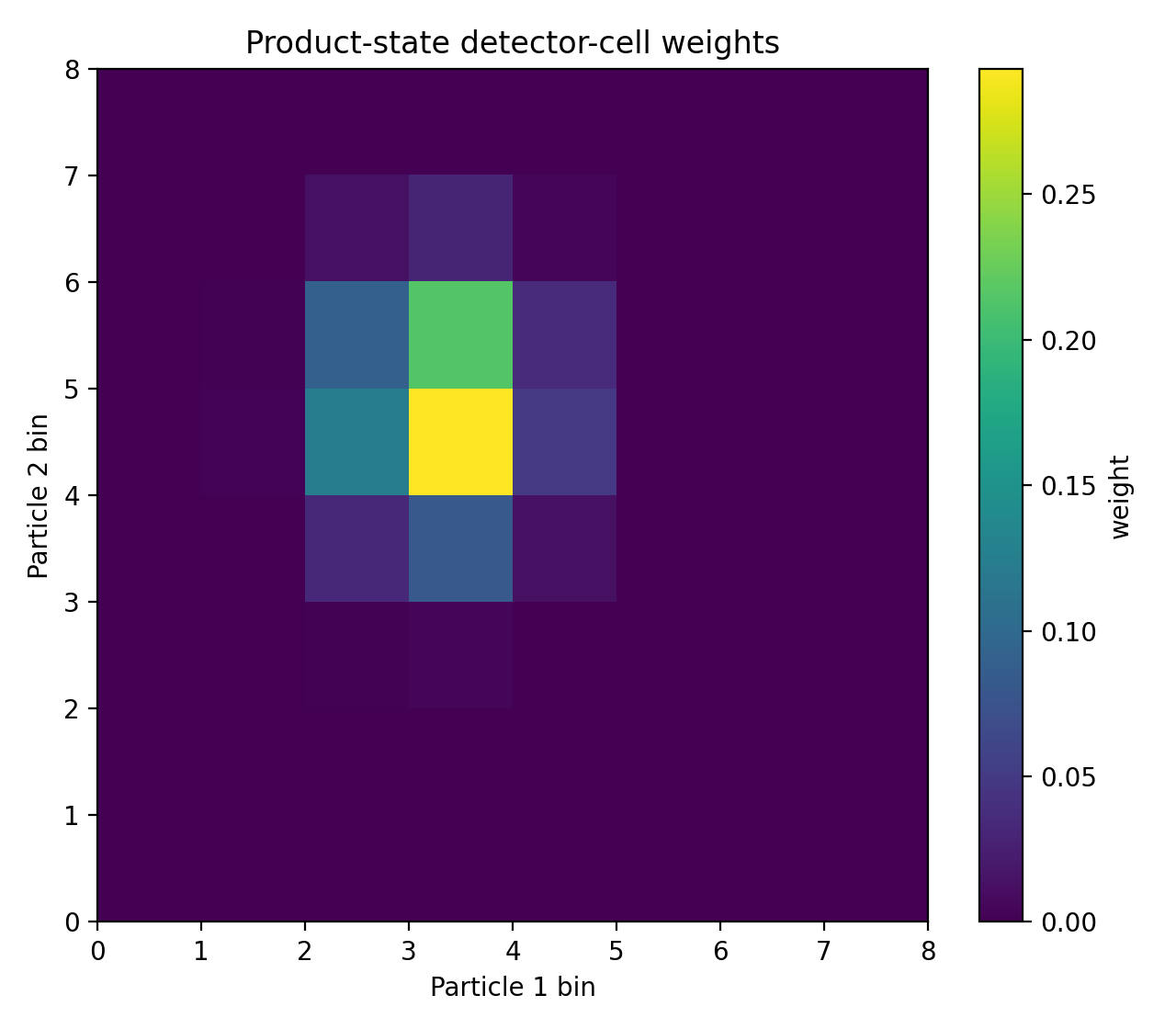}
\caption{Detector-cell weights for a product two-particle state. The direct
two-dimensional weights agree with the factorized product of the two
one-particle marginal weights.}
\label{fig:two-particle-product-weights}
\end{figure}

Figure~\ref{fig:two-particle-product-weights} compares the direct and
factorized joint weights. The two agree to numerical precision.

The maximum absolute difference between the direct and factorized weights was
\[
2.78\times10^{-17},
\]
and the total \(L^1\) difference was
\[
8.62\times10^{-17}.
\]
This confirms that the tensor-product detector weights reduce to independent
one-particle Brownian weights when the state is a product state.

\subsection{Entangled two-branch state}

We next consider an entangled two-particle state of the form
\[
\Psi(x,y)
=
\sqrt{p_L}\,g_L(x)h_L(y)
+
\sqrt{p_R}\,g_R(x)h_R(y),
\]
where
\[
p_L=0.35,
\qquad
p_R=0.65.
\]
The packets \(g_L,g_R\) are localized in the measured coordinate \(x\), while
\(h_L,h_R\) are the corresponding conditional packets of the second particle.
In the simulation we used well-separated Gaussian packets centered at
\[
x=-2,\quad x=2,
\]
for the first particle, and
\[
y=-1.5,\quad y=1.5,
\]
for the second particle. The two detector cells for the first particle were
\[
I_L=(-\infty,0),
\qquad
I_R=[0,\infty).
\]

\begin{figure}[h]
\centering
\includegraphics[width=0.78\textwidth]{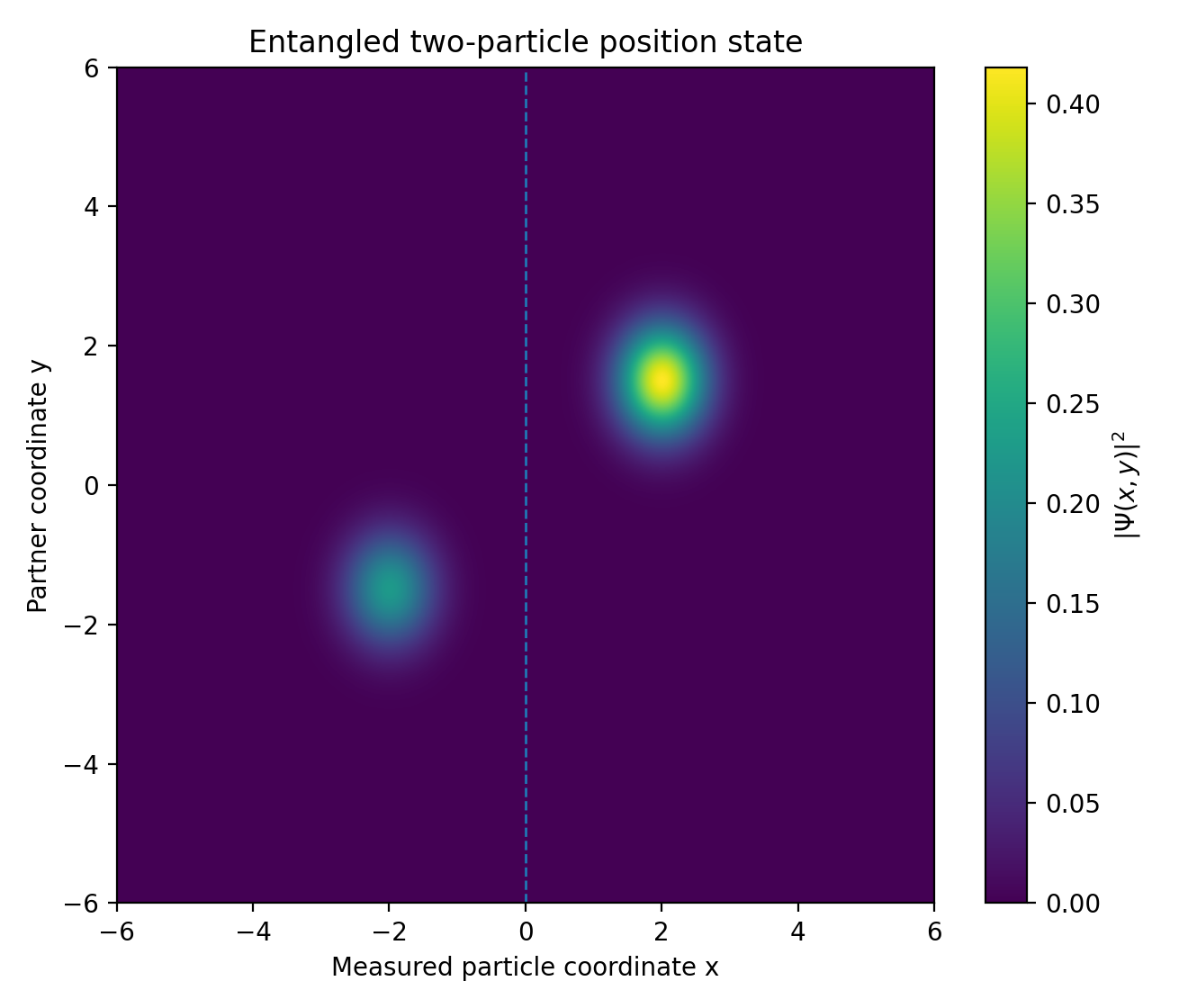}
\caption{Joint density of the entangled two-particle state. The two branches
correlate the measured particle position with the conditional position of the
second particle.}
\label{fig:two-particle-entangled-density}
\end{figure}

The joint probability density \(|\Psi(x,y)|^2\) is shown in
Figure~\ref{fig:two-particle-entangled-density}. The two branches are
correlated: the left packet of the measured particle is correlated with the
left packet of the second particle, and similarly for the right packet.

The detector weights for the first particle are
\[
W_L
=
\|(P_{I_L}\otimes I)\Psi\|^2,
\qquad
W_R
=
\|(P_{I_R}\otimes I)\Psi\|^2.
\]
Numerically we obtained
\[
W_L=0.350001,
\qquad
W_R=0.649999,
\]
in agreement with the branch weights \(p_L=0.35\) and \(p_R=0.65\).

\begin{figure}[h]
\centering
\includegraphics[width=0.72\textwidth]{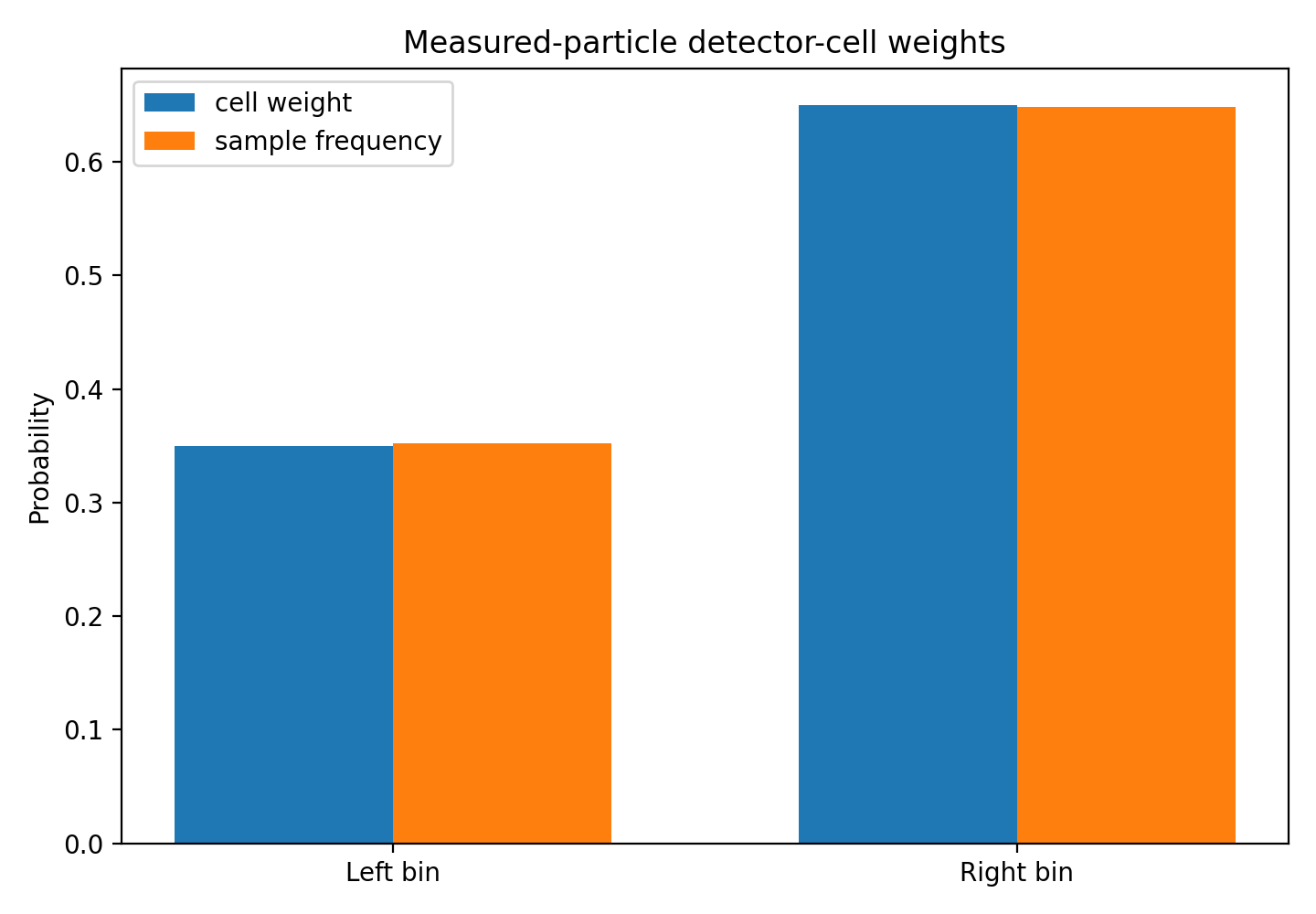}
\caption{Outcome frequencies for the measured particle in the entangled
two-branch state. The sampled frequencies agree with the detector-cell weights
\(W_L\) and \(W_R\).}
\label{fig:two-particle-entangled-frequencies}
\end{figure}

We then sampled
\[
50000
\]
detector records from these weights. The observed frequencies were
\[
f_L=0.3519,
\qquad
f_R=0.6481.
\]
These frequencies are shown in
Figure~\ref{fig:two-particle-entangled-frequencies}.

\subsection{Conditional state of the second particle}

After the left outcome is recorded, the conditional two-particle state is
\[
\Psi_L
=
\frac{(P_{I_L}\otimes I)\Psi}{\|(P_{I_L}\otimes I)\Psi\|}.
\]
The corresponding distribution of the second particle is obtained from
\[
\rho_{2|L}(y)
=
\int_{I_L}|\Psi_L(x,y)|^2\,dx.
\]
Similarly, after the right outcome,
\[
\rho_{2|R}(y)
=
\int_{I_R}|\Psi_R(x,y)|^2\,dx.
\]

\begin{figure}[h]
\centering
\includegraphics[width=0.78\textwidth]{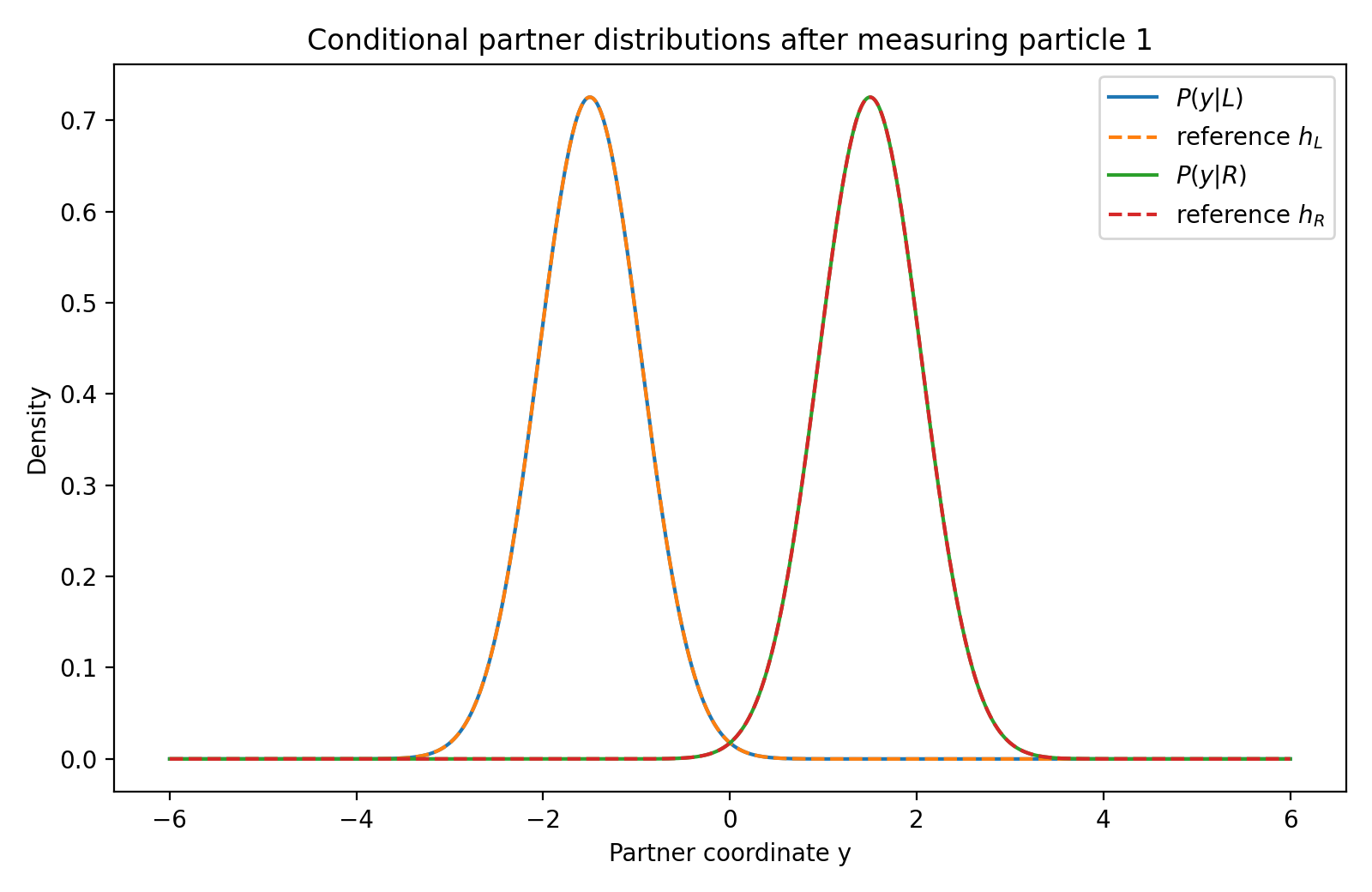}
\caption{Conditional distributions of the second particle after measuring the
first particle. The left outcome prepares the second particle in the packet
\(h_L\), while the right outcome prepares it in the packet \(h_R\), up to
finite-resolution numerical error.}
\label{fig:two-particle-conditional-partner}
\end{figure}

Figure~\ref{fig:two-particle-conditional-partner} compares these conditional
distributions with the expected packets \(h_L\) and \(h_R\). The agreement is
excellent. The \(L^1\) error for the left conditional distribution was
\[
1.9\times10^{-5},
\]
and for the right conditional distribution it was
\[
6.0\times10^{-6}.
\]

\subsection{Interpretation}

This simulation shows how the Born-weight construction extends from one
particle to two-particle tensor-product states. For product states, the
detector-cell weights factor into the product of the one-particle weights. For
entangled states, the weights do not factor, but are still given by the same
detector-cell rule
\[
W_j=\|(P_{I_j}\otimes I)\Psi\|^2.
\]
Thus measuring the position of the first particle gives the Born weights for
the measured coordinate, while the second particle is assigned the corresponding
conditional state.

The important point is that the \({\bf (RM)}\) dynamics acts in the full
tensor-product state space, but the detector record is defined by a
finite-resolution equivalence class in the measured coordinate. Once the first
particle is assigned to a detector cell \(I_j\), the partner state is not chosen
independently. It is the conditional state determined by the original
entangled state and the recorded detector class.

This is the position-measurement version of the usual entangled-pair
measurement rule. Spin measurements can be treated in the same framework after
the spin alternatives have been converted into spatially separated packets, as
in a Stern--Gerlach apparatus. The subsequent record is a position record, and
the same detector-cell weighting applies.

\section{Particle-device stability in the device limit}
\label{sec:particle-device-stability}

\subsection{Purpose of the simulation}

The preceding tensor-product simulations show that the \({\bf (RM)}\)-induced
diffusion is homogeneous and isotropic in tensor-product state space, and that
its restriction to the product Gaussian manifold gives independent Brownian
increments in the particle and device coordinates. We now test the consequence
that is most directly relevant to measurement: in the device limit, the
particle coordinate can move toward an outcome class with Born probabilities,
while the device coordinate remains in the same macroscopic equivalence class.

The point is not that the microscopic device ray is exactly fixed. The device
record is a finite-resolution equivalence class. The class is determined by a
position coordinate and a localization coordinate,
\[
(\tau_D,s_D).
\]
The device remains in the same record class when
\[
|\tau_D-\tau_{D,0}|\leq R_D,
\qquad
s_D<0,
\]
where \(R_D\) is the device resolution. Components corresponding to displaced
device classes are measured by the complementary weight
\[
W_{\rm out}^{(D)}
=
\|(I\otimes Q_D^\perp)\Psi\|^2,
\]
where \(Q_D\) is the projector onto the original device equivalence class. The
device limit requires
\[
W_{\rm out}^{(D)}\ll1.
\]

Thus the effective product-state description
\[
\Psi\sim \psi_p\otimes D_A
\]
is preserved at the level of equivalence classes: the particle state may move
toward one of its detector classes, while the device remains represented by the
same macroscopic position class \(D_A\).

\subsection{Four projected coordinates}

Near a product Gaussian state
\[
\Psi
=
g_{\tau_p,\delta_p}(x)\,
g_{\tau_D,\delta_D}(y),
\qquad
\delta_i=\sigma_i e^{s_i},
\]
the projected particle-device motion is described by four coordinates
\[
(\tau_p,s_p;\tau_D,s_D).
\]
The corresponding tangent directions are
\[
\partial_{\tau_p}\Psi,\quad
\partial_{s_p}\Psi,\quad
\partial_{\tau_D}\Psi,\quad
\partial_{s_D}\Psi.
\]
For product Gaussian states these directions are orthogonal in the induced
Fubini--Study metric, and the metric is
\[
G
=
\begin{pmatrix}
\dfrac{1}{4\delta_p^2} & 0 & 0 & 0\\[4pt]
0 & \dfrac12 & 0 & 0\\[4pt]
0 & 0 & \dfrac{1}{4\delta_D^2} & 0\\[4pt]
0 & 0 & 0 & \dfrac12
\end{pmatrix}.
\]
Therefore, in Fubini--Study-normalized coordinates,
\[
z_{\tau_p}=\sqrt{G_{\tau_p\tau_p}}\,\Delta\tau_p,
\qquad
z_{s_p}=\sqrt{G_{s_ps_p}}\,\Delta s_p,
\]
\[
z_{\tau_D}=\sqrt{G_{\tau_D\tau_D}}\,\Delta\tau_D,
\qquad
z_{s_D}=\sqrt{G_{s_Ds_D}}\,\Delta s_D,
\]
the projected \({\bf (RM)}\) increments should be independent Gaussian
variables.

\begin{figure}[h]
\centering
\includegraphics[width=0.70\textwidth]{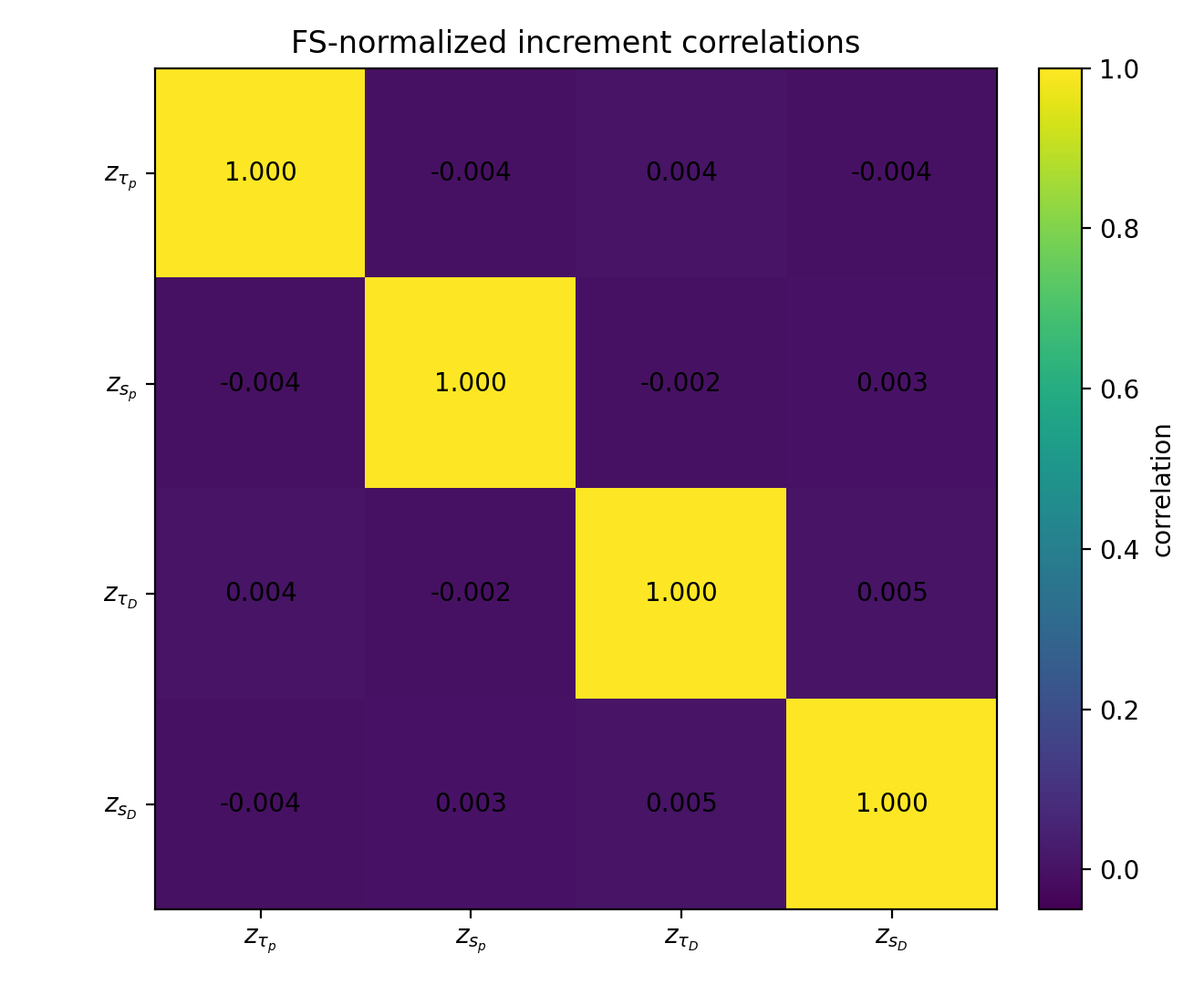}
\caption{Correlation matrix for the Fubini--Study-normalized increments
\((z_{\tau_p},z_{s_p},z_{\tau_D},z_{s_D})\). The off-diagonal correlations are
near zero, confirming independence of the four projected Gaussian increments.}
\label{fig:sim15-four-coordinate-correlations}
\end{figure}

We tested this with
\[
\delta_p=0.75,
\qquad
\delta_D=0.10,
\]
using \(80000\) samples. Figure~\ref{fig:sim15-four-coordinate-correlations}
shows the correlation matrix of the four normalized increments. The largest
off-diagonal correlation was
\[
0.0052.
\]
Thus the four projected coordinates are independent to numerical accuracy.

The corresponding standard deviations are summarized in
Table~\ref{tab:sim15-four-coordinate-summary}.

\begin{table}[h]
\centering
\caption{Projected increments in the four particle-device coordinates.}
\label{tab:sim15-four-coordinate-summary}
\begin{tabular}{lcc}
\toprule
Quantity & Euclidean standard deviation & FS-normalized standard deviation \\
\midrule
\(\Delta\tau_p\) & \(1.4972\) & \(0.9981\) \\
\(\Delta s_p\) & \(1.4182\) & \(1.0028\) \\
\(\Delta\tau_D\) & \(0.2011\) & \(1.0055\) \\
\(\Delta s_D\) & \(1.4172\) & \(1.0021\) \\
\bottomrule
\end{tabular}
\end{table}

The important feature is the difference between the normalized and Euclidean
device displacements. Since
\[
G_{\tau_D\tau_D}=\frac{1}{4\delta_D^2},
\]
a unit Fubini--Study-normalized step corresponds to
\[
\Delta\tau_D\sim 2\delta_D Z.
\]
Thus, as \(\delta_D\to0\), the Euclidean displacement of the device coordinate
is suppressed.

\subsection{Device equivalence class}

We now impose the device-class condition
\[
|\Delta\tau_D|\leq R_D,
\qquad
s_D<0.
\]
In the simulation
\[
R_D=1,
\qquad
s_{\rm rec}=-0.75,
\qquad
D_s=0.25,
\qquad
\Delta t=0.05.
\]
The \(s_D\)-coordinate is sampled from the Gaussian return step
\[
s_D=s_{\rm rec}+\xi_D,
\qquad
\xi_D\sim N(0,2D_s\Delta t).
\]
The condition \(s_D<0\) represents renewal into the localized sector, as in the
stroboscopic simulation.

Figure~\ref{fig:sim15-device-retention} shows the fraction of candidate steps
for which the device remains in its original equivalence class as a function
of
\[
\delta_D/R_D.
\]
As the device localization scale becomes small compared with the detector
resolution, the in-class fraction approaches one.

\begin{figure}[h]
\centering
\includegraphics[width=0.78\textwidth]{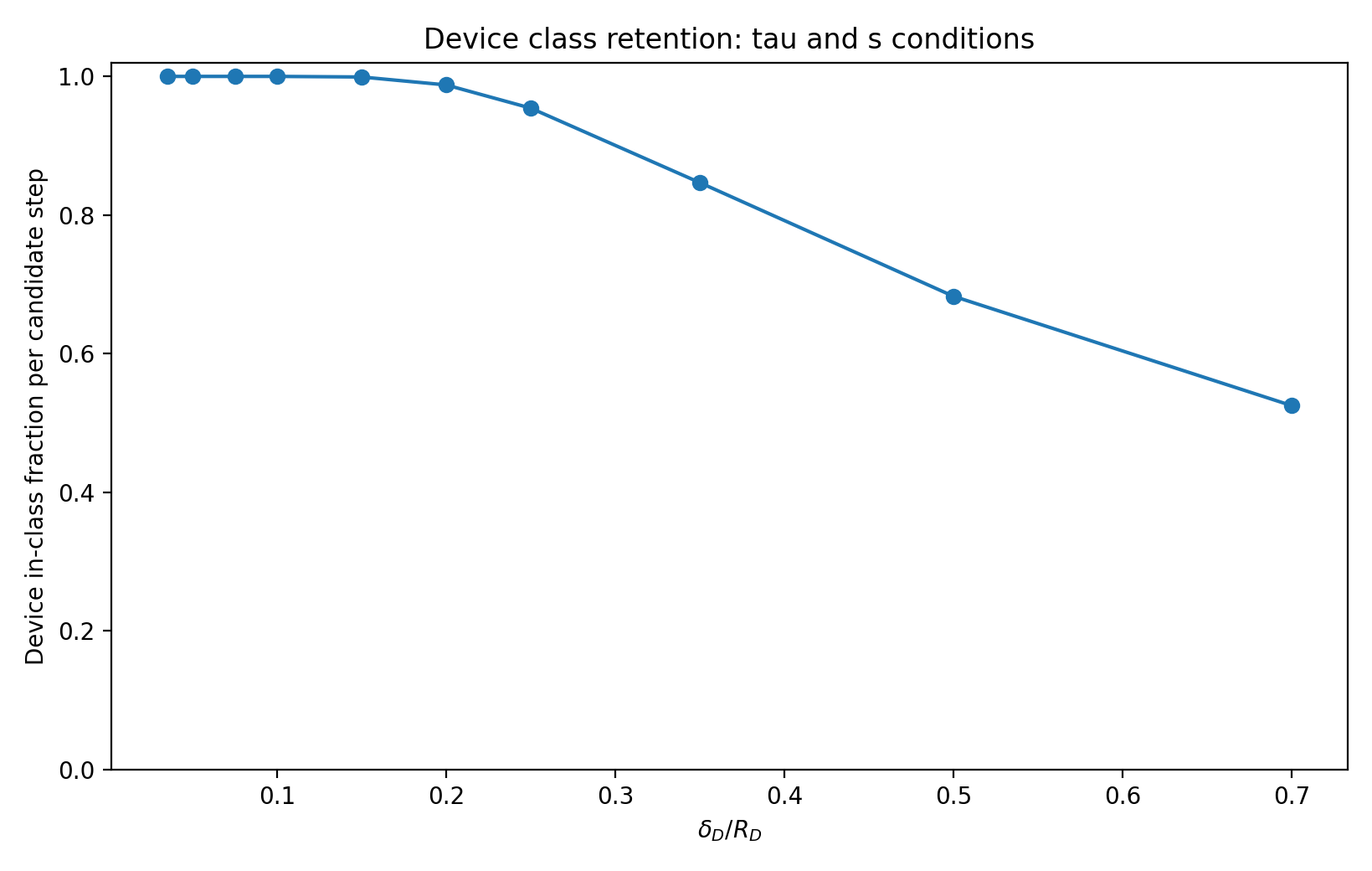}
\caption{Device class retention as a function of \(\delta_D/R_D\). The device
is counted as remaining in the same equivalence class when
\(|\Delta\tau_D|\leq R_D\) and \(s_D<0\).}
\label{fig:sim15-device-retention}
\end{figure}

Equivalently, Figure~\ref{fig:sim15-device-outside-weight} shows the
outside-class weight
\[
W_{\rm out}^{(D)}.
\]
The dashed curve is the Gaussian tail prediction for the \(\tau_D\)-condition,
and the dotted line is the small contribution from the \(s_D\)-condition. For
\(\delta_D/R_D=0.10\), the outside-class weight is already of order
\(10^{-6}\). For \(\delta_D/R_D=0.05\), the displaced-\(\tau_D\) contribution is
approximately
\[
1.5\times10^{-23},
\]
so the remaining outside weight is dominated by the rare failure of the
\(s_D<0\) condition.

\begin{figure}[h]
\centering
\includegraphics[width=0.78\textwidth]{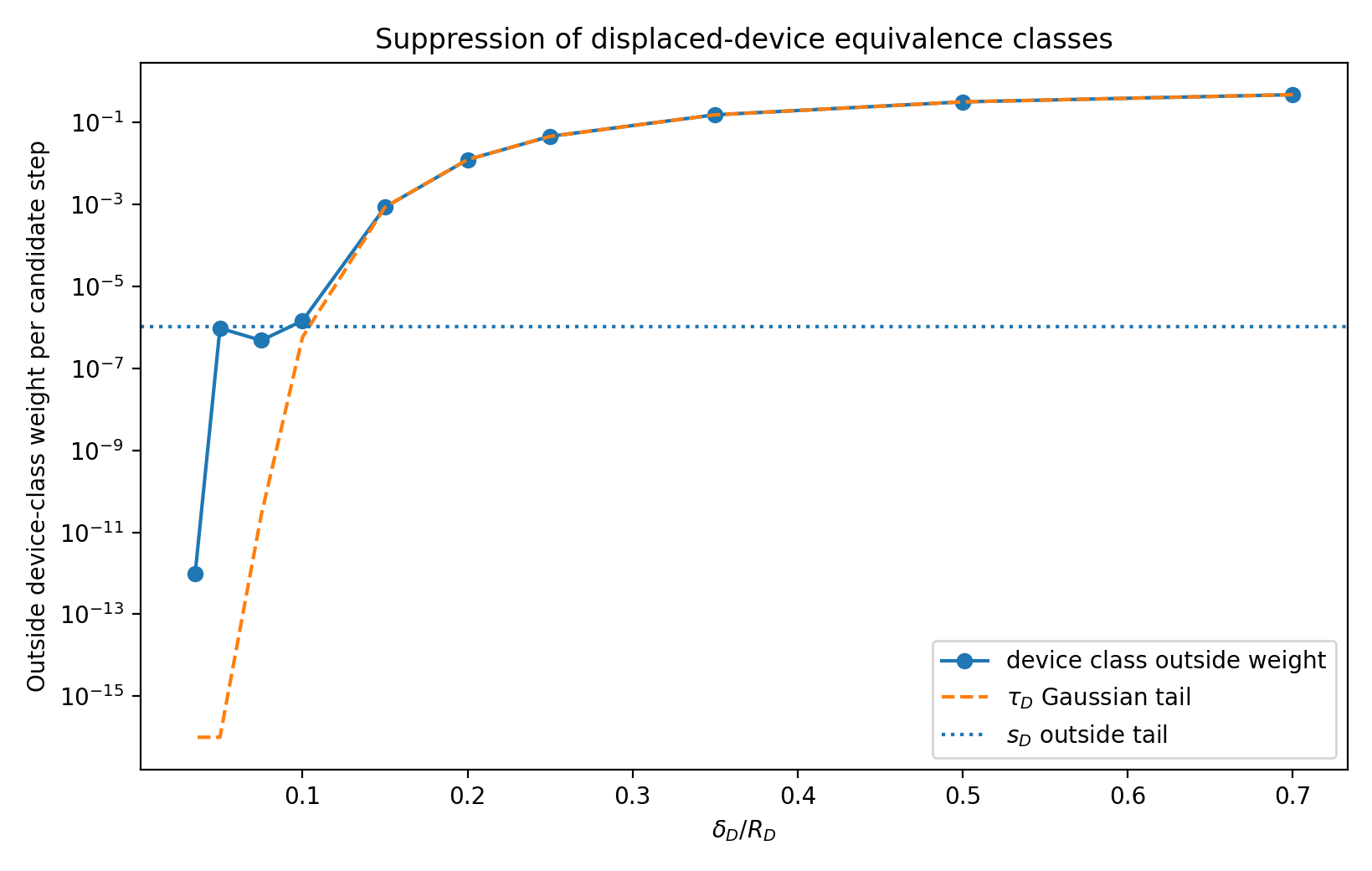}
\caption{Suppression of the outside-device-class weight. Components
corresponding to displaced device classes become negligible as
\(\delta_D/R_D\to0\).}
\label{fig:sim15-device-outside-weight}
\end{figure}

Selected values are shown in
Table~\ref{tab:sim15-device-stability}.

\begin{table}[h]
\centering
\caption{Device equivalence-class retention as \(\delta_D/R_D\) decreases.}
\label{tab:sim15-device-stability}
\begin{tabular}{ccc}
\toprule
\(\delta_D/R_D\) & In-class fraction & Outside-class weight \\
\midrule
\(0.70\) & \(0.5251\) & \(4.75\times10^{-1}\) \\
\(0.50\) & \(0.6829\) & \(3.17\times10^{-1}\) \\
\(0.35\) & \(0.8468\) & \(1.53\times10^{-1}\) \\
\(0.25\) & \(0.9545\) & \(4.55\times10^{-2}\) \\
\(0.20\) & \(0.9876\) & \(1.24\times10^{-2}\) \\
\(0.15\) & \(0.9991\) & \(8.61\times10^{-4}\) \\
\(0.10\) & \(0.999998\) & \(1.83\times10^{-6}\) \\
\(0.075\) & \(0.999999\) & \(1.10\times10^{-6}\) \\
\(0.050\) & \(0.999999\) & \(7.31\times10^{-7}\) \\
\bottomrule
\end{tabular}
\end{table}

Thus, in the device limit, the tensor-product state may contain small
components associated with displaced device classes, but their total weight is
vanishingly small. The device state remains fixed at the level of the
finite-resolution equivalence class.

\subsection{Particle reduction with a stable device class}

We next combine the device-class test with the particle reduction coordinate.
The particle coordinate is represented by an unbiased walk on \([0,1]\), with
absorbing endpoints corresponding to the two particle outcome classes. The
initial particle weight is denoted by \(p_0\). At each candidate step, the
device coordinates \((\tau_D,s_D)\) are also sampled and tested for membership
in the original device class.

For this test we used
\[
\delta_D/R_D=0.05.
\]
The device remained in its equivalence class with probability essentially one
at each candidate step. At the same time, the particle coordinate reached the
upper endpoint with frequencies agreeing with the Born prediction
\[
\mathbb P(\text{upper outcome})=p_0.
\]

Figure~\ref{fig:sim15-particle-born-stable-device} shows the outcome
frequencies for several values of \(p_0\).

\begin{figure}[h]
\centering
\includegraphics[width=0.72\textwidth]{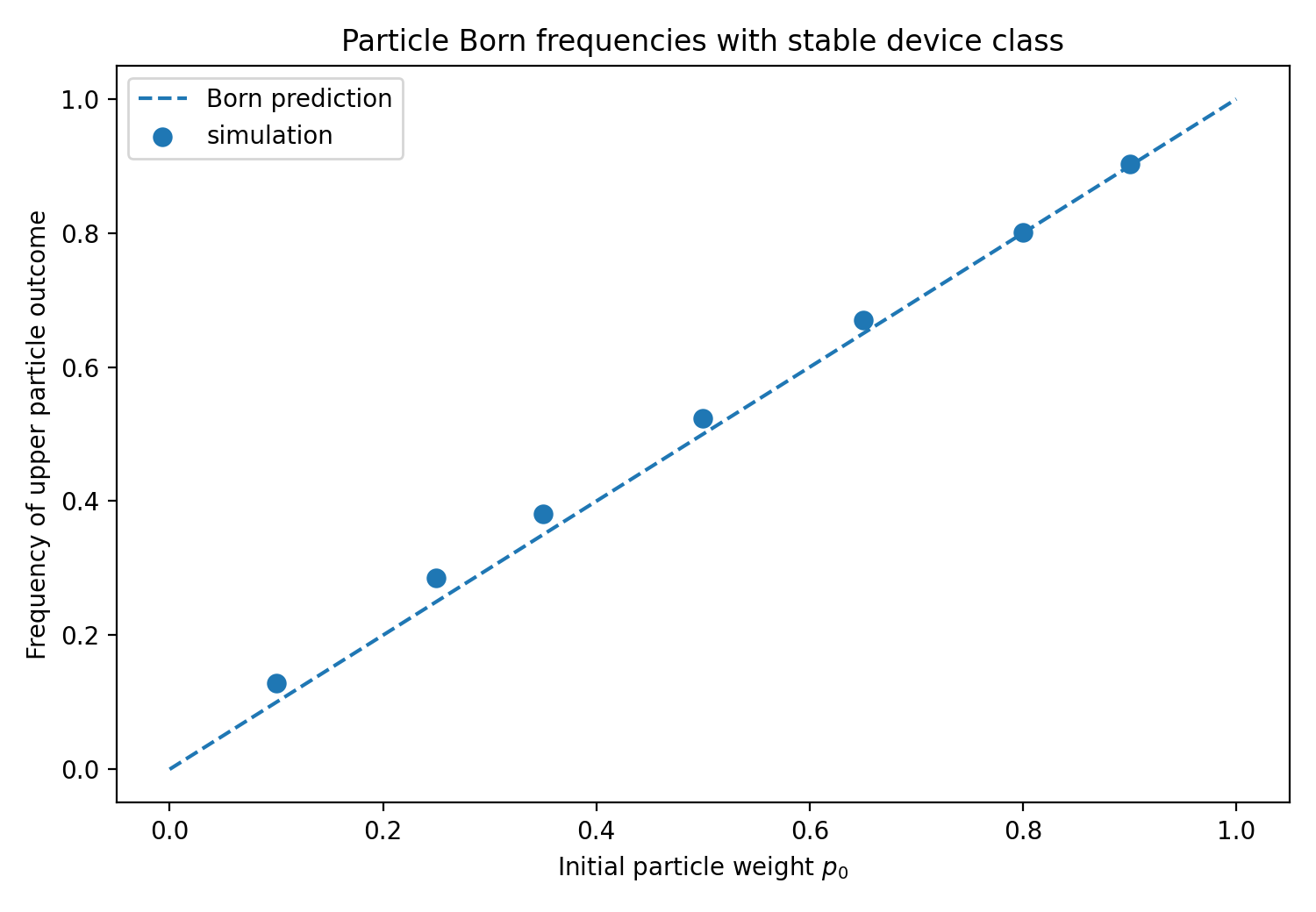}
\caption{Particle Born frequencies in the presence of a stable device
equivalence class. The particle coordinate reaches the upper outcome with
frequency \(p_0\), while the sharply localized device remains in its original
record class.}
\label{fig:sim15-particle-born-stable-device}
\end{figure}

The numerical values are summarized in
Table~\ref{tab:sim15-particle-born-stable-device}.

\begin{table}[h]
\centering
\caption{Particle outcome frequencies with stable device class
\((\delta_D/R_D=0.05)\).}
\label{tab:sim15-particle-born-stable-device}
\begin{tabular}{ccc}
\toprule
\(p_0\) & Frequency of upper outcome & Device outside-class weight \\
\midrule
\(0.10\) & \(0.10046\) & \(0\) \\
\(0.25\) & \(0.25296\) & \(1.86\times10^{-6}\) \\
\(0.35\) & \(0.34896\) & \(4.40\times10^{-7}\) \\
\(0.50\) & \(0.50106\) & \(6.00\times10^{-7}\) \\
\(0.65\) & \(0.64528\) & \(1.10\times10^{-6}\) \\
\(0.80\) & \(0.79930\) & \(1.87\times10^{-6}\) \\
\(0.90\) & \(0.90148\) & \(5.58\times10^{-7}\) \\
\bottomrule
\end{tabular}
\end{table}

\begin{figure}[h]
\centering
\includegraphics[width=0.78\textwidth]{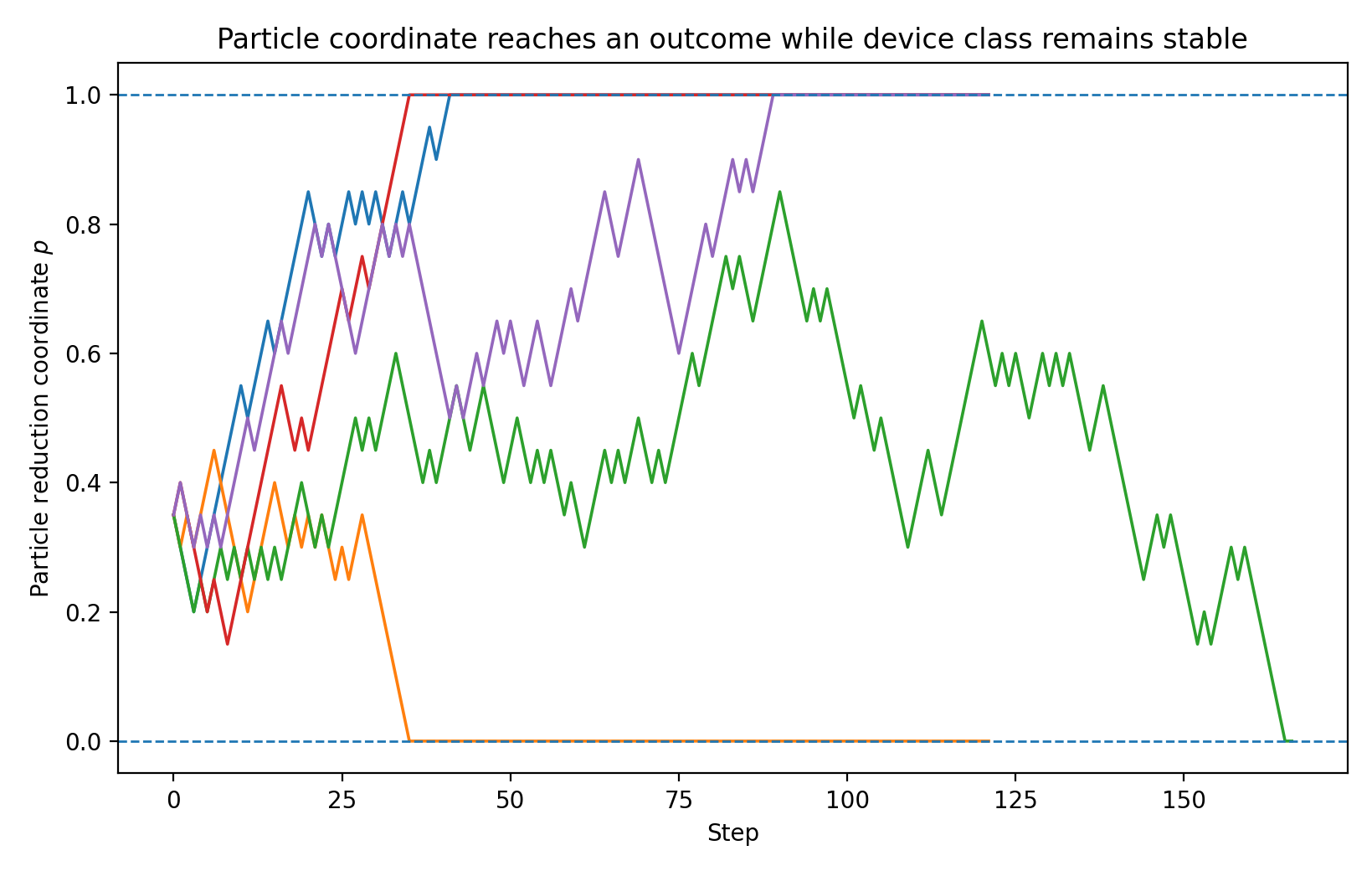}
\caption{Sample particle reduction-coordinate paths. The particle coordinate
continues to move until it reaches one of the two outcome classes.}
\label{fig:sim15-sample-particle-paths}
\end{figure}

Figure~\ref{fig:sim15-sample-particle-paths} shows sample particle reduction
paths. The particle coordinate continues to move until it reaches one of the
outcome classes.

\begin{figure}[h]
\centering
\includegraphics[width=0.78\textwidth]{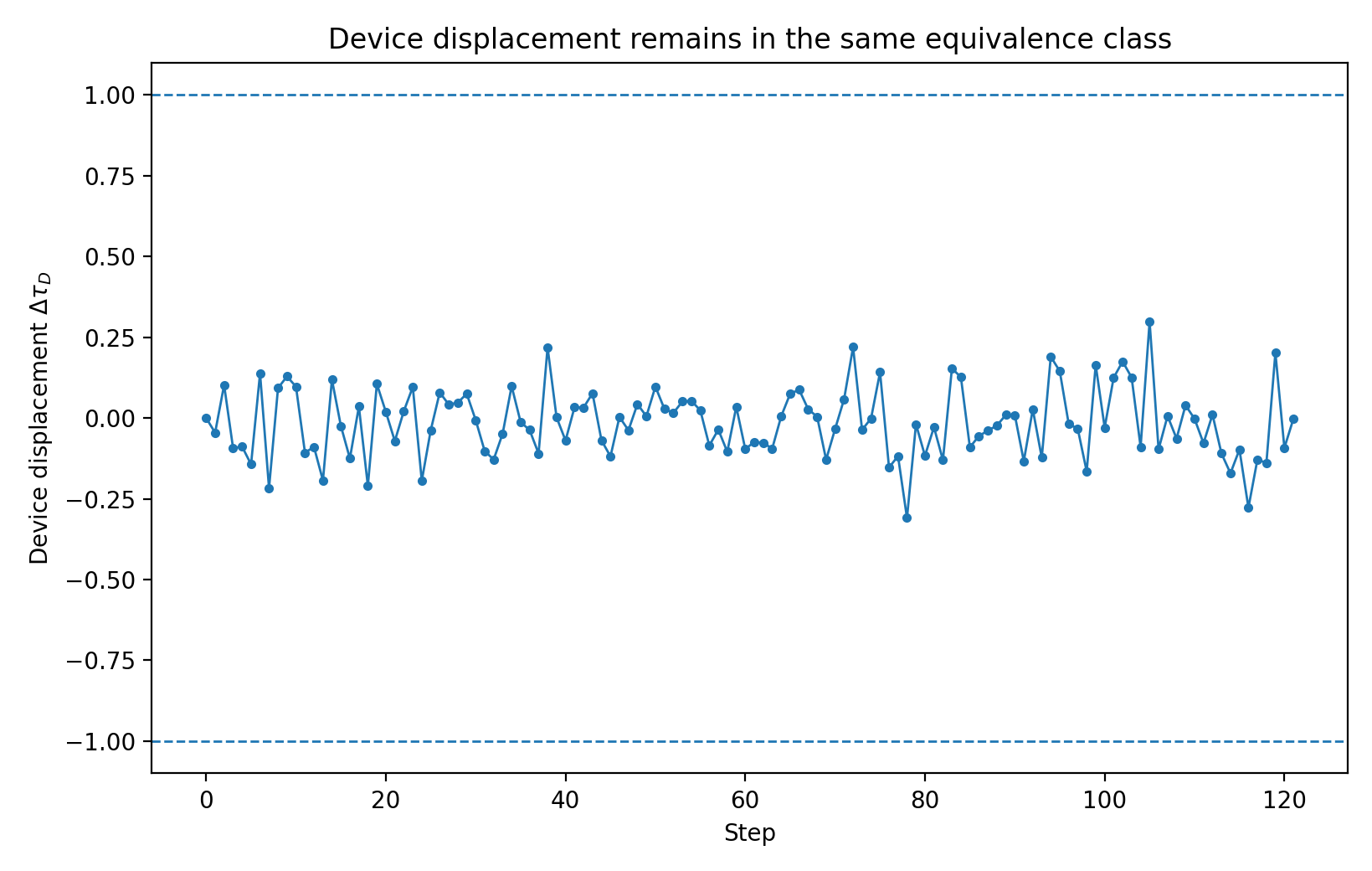}
\caption{Device displacement \(\Delta\tau_D\) along a sample path. The dashed
lines mark the device resolution interval.}
\label{fig:sim15-sample-device-tau}
\end{figure}

\begin{figure}[h]
\centering
\includegraphics[width=0.78\textwidth]{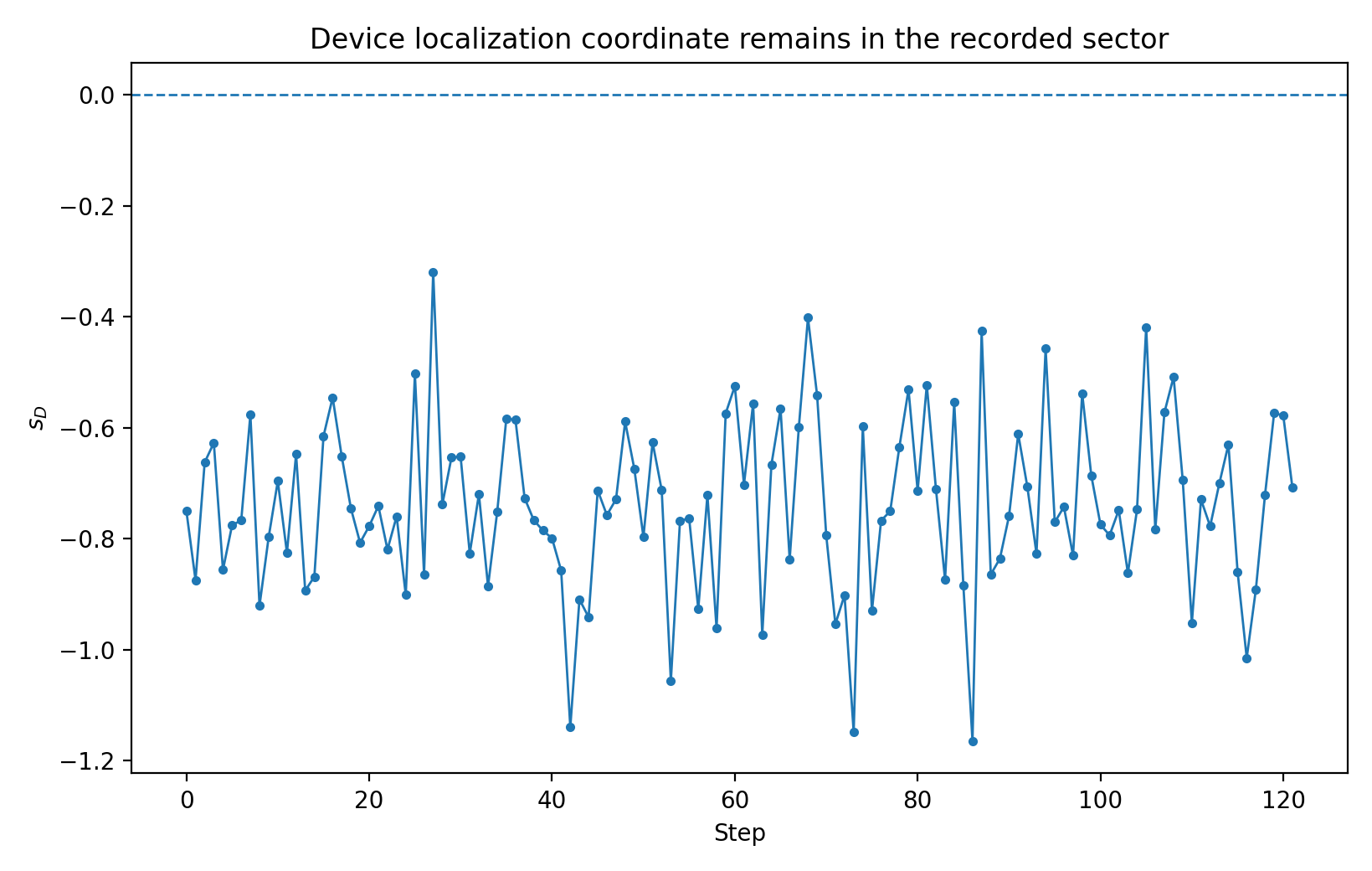}
\caption{Device localization coordinate \(s_D\) along the same sample path.
The condition \(s_D<0\) keeps the device in the recorded sector.}
\label{fig:sim15-sample-device-s}
\end{figure}

Figures~\ref{fig:sim15-sample-device-tau} and
\ref{fig:sim15-sample-device-s} show the corresponding device coordinates for
one sample path. The device displacement remains inside the resolution
interval, and the localization coordinate remains in the recorded sector.

\subsection{Interpretation}

The simulation demonstrates the particle-device separation required by the
measurement picture. The particle coordinate can undergo the \({\bf (RM)}\)
reduction process and reach its detector-defined outcome class with Born
frequencies. At the same time, the device coordinate remains in the same
macroscopic equivalence class when the device localization scale is small
compared with its resolution.

Thus the state of the particle-device system remains effectively a product
state at the level of recorded equivalence classes:
\[
\Psi
\sim
\psi_p\otimes D_A.
\]
The equivalence class represented by \(D_A\) may contain microscopic
superpositions and entangled components, but components corresponding to
different device position classes have vanishingly small total weight in the
device limit. This is the sense in which the device record is stable while the
particle state undergoes the stochastic reduction process.

This distinction is essential. The theory does not require the exact
microscopic device ray to be frozen. It requires only that the macroscopic
record, represented by a finite-resolution device equivalence class, remains
fixed with overwhelming probability. The simulation shows precisely this:
\[
\text{particle reduction with Born weights}
\quad+\quad
\text{stable device equivalence class}.
\]

\section{Sources of effective irreversibility in \({\bf (RM)}\) dynamics}
\label{sec:rm-irreversibility}

The \({\bf (RM)}\) dynamics is unitary for each realized sequence of
Hamiltonians. Thus the irreversibility considered in this section is not a
failure of microscopic unitary reversibility. If the full realized Hamiltonian
history is retained, the corresponding unitary path can be reversed exactly.
The effective arrow of time arises only after information about the realized
state-space path is not retained. There are three related, but distinct,
sources of this effective irreversibility.

\subsection{High-dimensional state-space irreversibility}

The first source is purely geometric. Unlike ordinary Brownian motion in
\(\mathbb R^3\), the \({\bf (RM)}\) process is a stochastic dynamics on
projective Hilbert space. Once the exact Hamiltonian history has been discarded,
a newly sampled GUE history almost never returns the state to a prescribed
small neighborhood of its initial ray. In high dimension this is a consequence
of concentration of volume in projective state space.

Let \(\psi_0\) be a fixed initial state in \(\mathbb C^N\), and let
\[
F(\psi,\psi_0)=|\langle \psi,\psi_0\rangle|^2
\]
be the fidelity with \(\psi_0\). In the Haar limit, the probability that a
random state lies in the fidelity neighborhood
\[
F(\psi,\psi_0)>1-\varepsilon
\]
is
\[
\mathbb P\big(F(\psi,\psi_0)>1-\varepsilon\big)
=
\varepsilon^{N-1}.
\]
Equivalently, since the Fubini--Study distance satisfies
\[
F(\psi,\psi_0)=\cos^2\rho(\psi,\psi_0),
\]
the probability of returning to a small Fubini--Study ball about the initial
ray decreases as a high power of the radius.

For example, when \(N=32\),
\[
\mathbb P(F>0.9)=10^{-31},
\qquad
\mathbb P(F>0.99)=10^{-62}.
\]
In the infinite-dimensional limit this probability vanishes for every fixed
\(\varepsilon<1\). Thus, even before detector-defined equivalence classes are
introduced, the loss of the realized \({\bf (RM)}\) Hamiltonian history produces
an effective irreversibility: a statistically similar continuation is not a
reversal of the realized path.

\begin{figure}[h]
\centering
\includegraphics[width=0.78\textwidth]{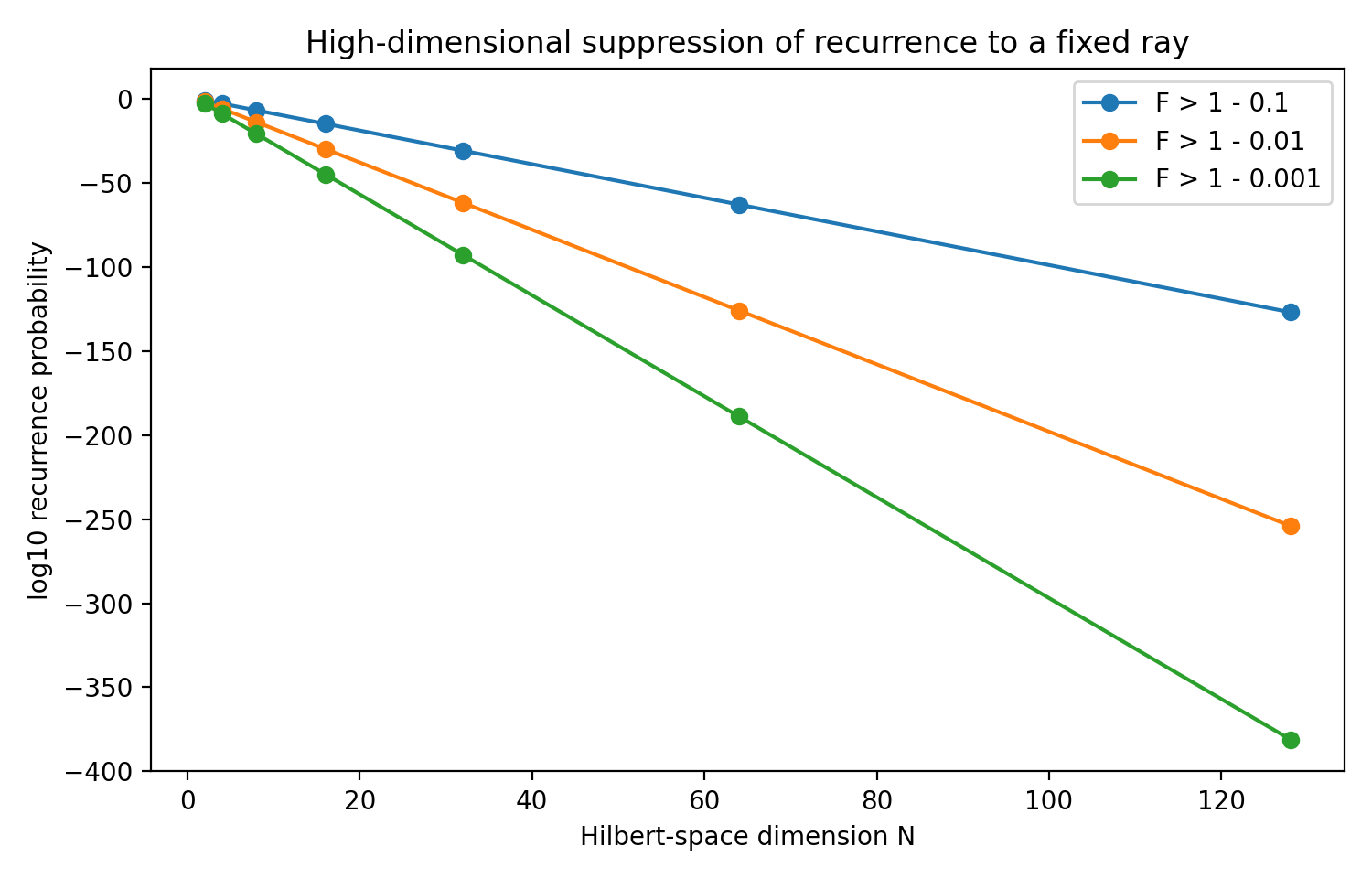}
\caption{High-dimensional suppression of recurrence to a fixed ray. The
probability that a random state in \(\mathbb C^N\) has fidelity
\(F>1-\varepsilon\) with a fixed initial state is \(\varepsilon^{N-1}\), and
therefore decreases rapidly with \(N\).}
\label{fig:rm-basic-irreversibility-dimension}
\end{figure}

Figure~\ref{fig:rm-basic-irreversibility-dimension} shows this suppression as
a function of the Hilbert-space dimension.

\begin{figure}[h]
\centering
\includegraphics[width=0.78\textwidth]{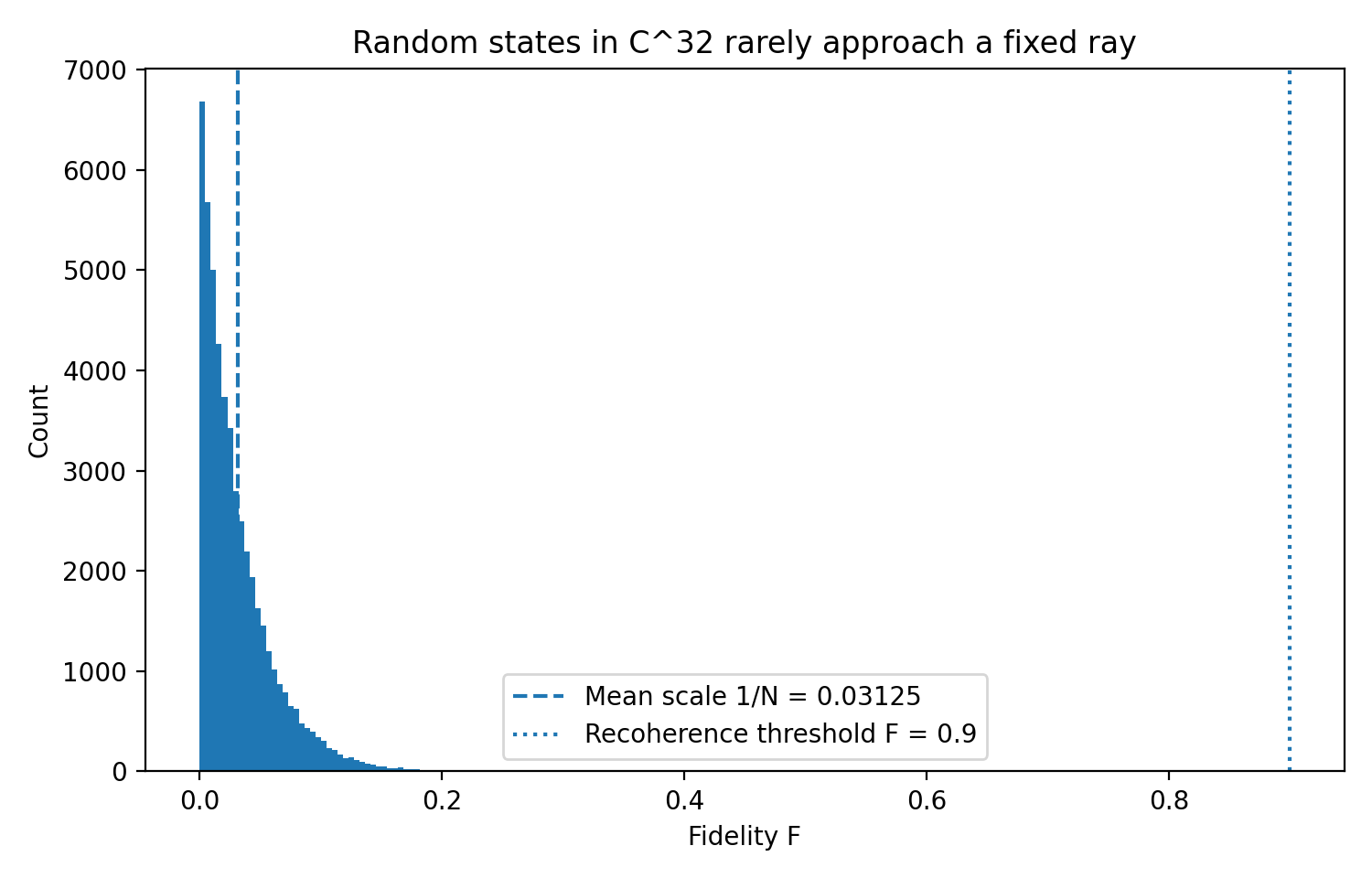}
\caption{Distribution of fidelities with a fixed initial state for random
states in \(\mathbb C^{32}\). The typical fidelity is of order \(1/N\), and
return to a high-fidelity neighborhood of the initial ray has negligible
probability.}
\label{fig:rm-basic-irreversibility-histogram}
\end{figure}

The same point is seen in a Monte Carlo simulation in \(\mathbb C^{32}\).
Figure~\ref{fig:rm-basic-irreversibility-histogram} shows the distribution of
fidelities with a fixed initial state for \(50000\) random states. No sampled
state had fidelity exceeding \(0.5\), \(0.75\), \(0.9\), or \(0.99\). The
expected number of states with \(F>0.9\) in this sample is only
\[
50000\cdot 10^{-31}=5\cdot 10^{-27}.
\]

Selected values are shown in Table~\ref{tab:rm-basic-irreversibility}.

\begin{table}[h]
\centering
\caption{Suppression of recurrence probabilities in high-dimensional projective
Hilbert space.}
\label{tab:rm-basic-irreversibility}
\begin{tabular}{ccc}
\toprule
\(N\) & \(\mathbb P(F>0.9)\) & \(\mathbb P(F>0.99)\) \\
\midrule
\(4\)  & \(10^{-3}\)  & \(10^{-6}\) \\
\(8\)  & \(10^{-7}\)  & \(10^{-14}\) \\
\(16\) & \(10^{-15}\) & \(10^{-30}\) \\
\(32\) & \(10^{-31}\) & \(10^{-62}\) \\
\(64\) & \(10^{-63}\) & \(10^{-126}\) \\
\bottomrule
\end{tabular}
\end{table}

This is the first, purely state-space source of irreversibility. It is not a
property of detector records. It follows from the fact that the \({\bf (RM)}\)
process takes place in a high-dimensional, and ultimately infinite-dimensional,
projective Hilbert space.

\subsection{Time reversal and the GUE Hamiltonian history}

The second source concerns the distinction between exact unitary inversion and
antiunitary time reversal. Let
\[
\psi_n=U_n\cdots U_1\psi_0,
\qquad
U_k=e^{-iH_k\Delta t},
\]
where the \(H_k\) are independent GUE Hamiltonians. If the full realized
sequence
\[
H_1,\ldots,H_n
\]
is retained, then the path can be reversed exactly by applying
\[
U_1^{-1}\cdots U_n^{-1}
\]
to \(\psi_n\). Since
\[
U_k^{-1}=e^{iH_k\Delta t}=e^{-i(-H_k)\Delta t},
\]
the inverse unitary is generated by the Hamiltonian \(-H_k\), which is again
distributed according to GUE.

However, this exact inverse is not the same as antiunitary time reversal. In a
basis in which spinless time reversal is represented by complex conjugation,
the time-reversed Hamiltonian is \(H^*\). A typical GUE Hamiltonian is not
invariant under this operation:
\[
H^*\neq H.
\]
Thus the effective GUE interaction is not constrained by antiunitary
time-reversal symmetry. Moreover, the Hamiltonian \(H^*\) does not generate the
inverse unitary path. The exact inverse path requires \(-H\) in the reverse
order, whereas antiunitary time reversal involves complex conjugation of the
Hamiltonian.

We test this distinction numerically. The Hilbert space has dimension
\[
N=32.
\]
Starting from an initial state \(\psi_0\), we generate a realized
\({\bf (RM)}\) path of length
\[
n=180
\]
using GUE Hamiltonians and unitary steps of strength
\[
\epsilon=0.22.
\]
From the final state \(\psi_n\), we compare three procedures.

First, we apply the exact inverse of the realized unitary path. This returns
the state to \(\psi_0\), up to numerical precision.

Second, we replace the inverse Hamiltonian history by the conjugated
Hamiltonian history \(H_k^*\), applied in reverse order. Since this is not the
inverse generator, the state does not return to \(\psi_0\).

Third, we discard the realized Hamiltonian history and attempt reversal using
fresh independently sampled GUE Hamiltonians. This represents the situation in
which the ensemble law is retained, but the realized path information is lost.

\begin{figure}[h]
\centering
\includegraphics[width=0.78\textwidth]{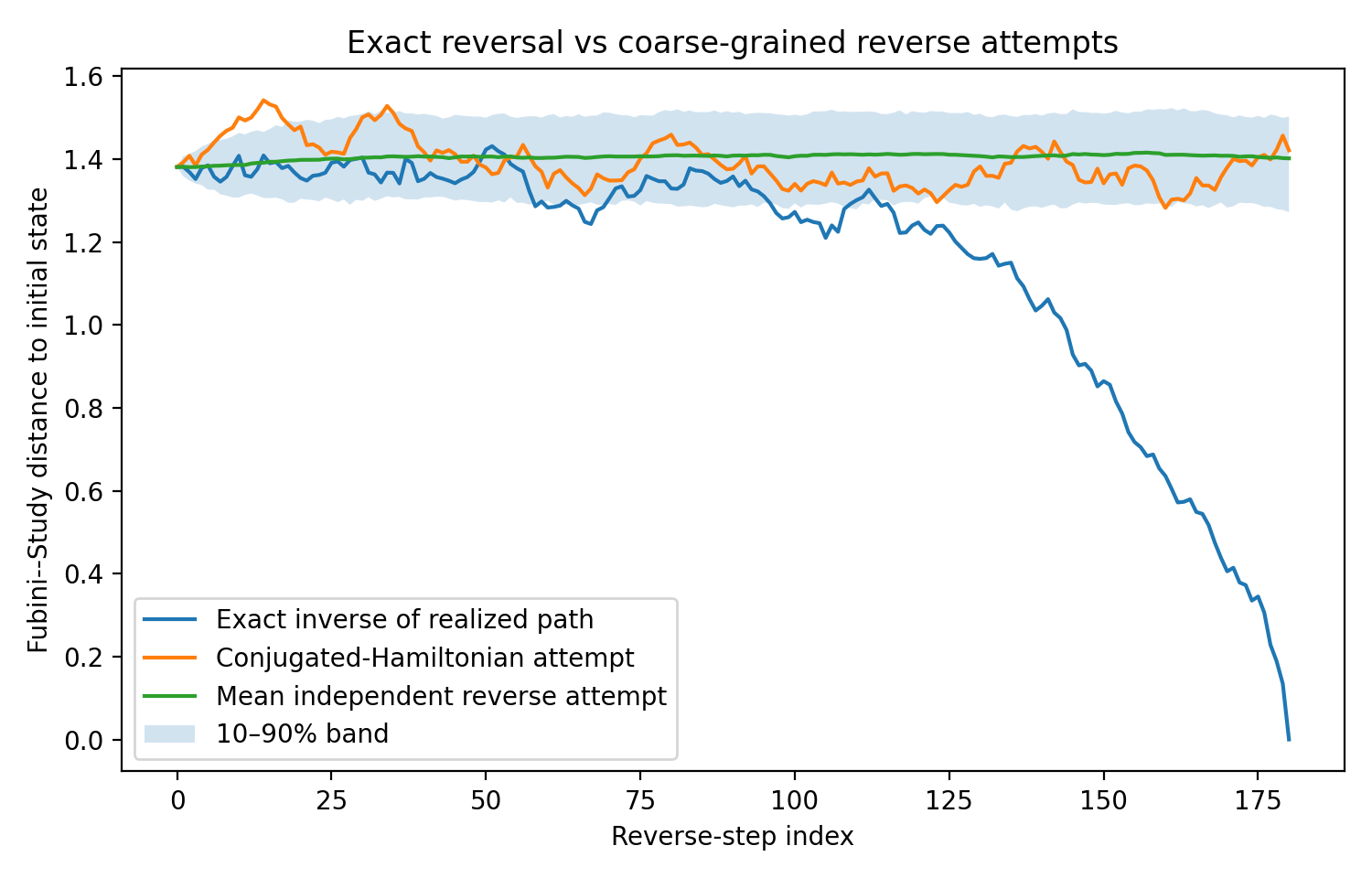}
\caption{Exact reversibility of a realized \({\bf (RM)}\) path and effective
irreversibility after loss of path information. The exact inverse of the
realized unitary path returns to the initial state. Replacing the inverse
history by the conjugated Hamiltonian history \(H_k^*\), or by independently
sampled GUE Hamiltonians, does not reconstruct the initial state.}
\label{fig:rm-irreversibility-distance}
\end{figure}

The comparison is shown in Figure~\ref{fig:rm-irreversibility-distance}. The
exact inverse path returns to zero Fubini--Study distance from \(\psi_0\). By
contrast, both the conjugated-Hamiltonian attempt and the independent reverse
attempts remain far from the initial state.

\begin{figure}[h]
\centering
\includegraphics[width=0.78\textwidth]{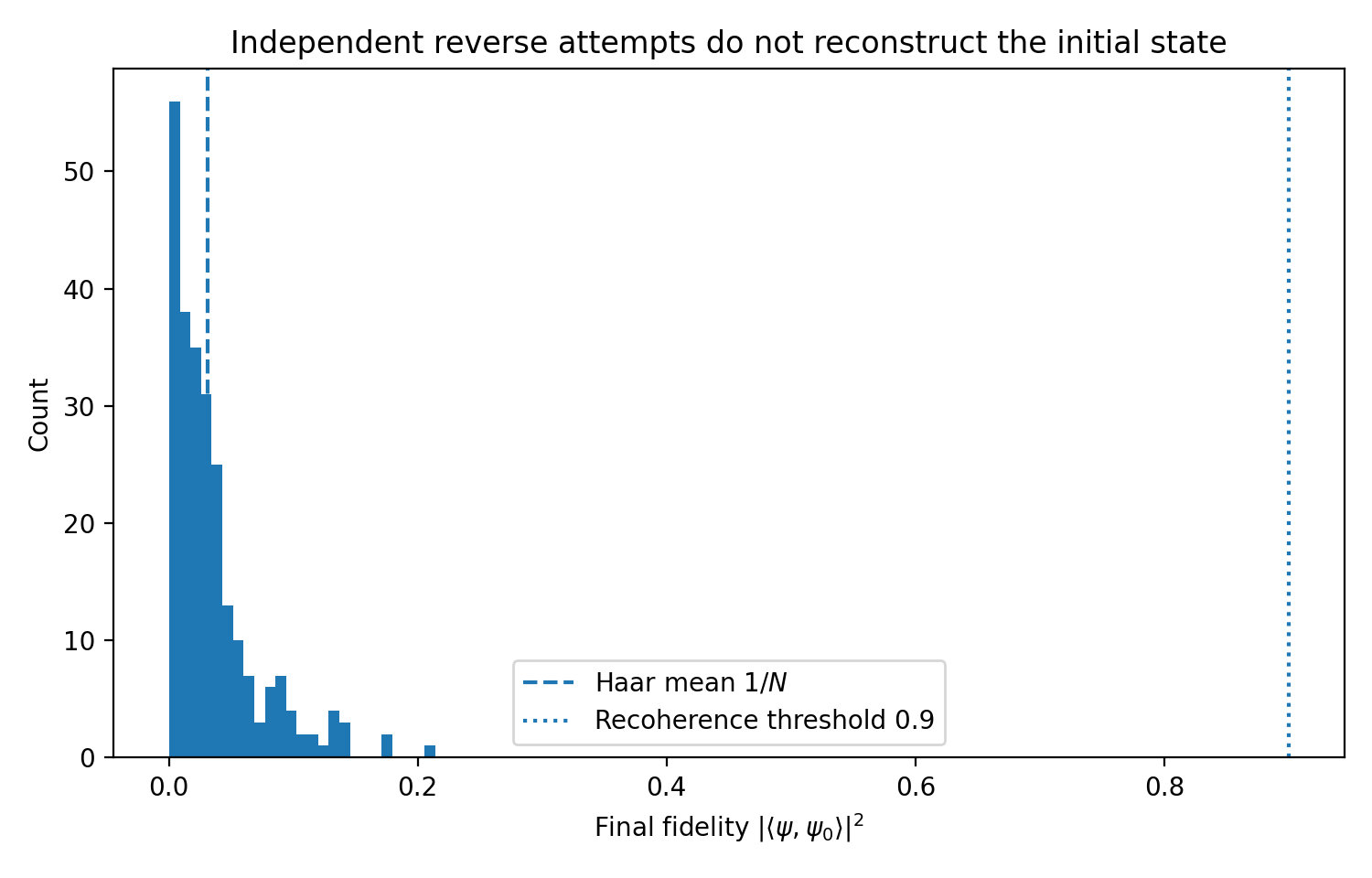}
\caption{Final fidelities with the initial state for independent reverse
attempts using freshly sampled GUE Hamiltonians. The distribution remains near
the random-state scale \(1/N\), and no recoherence event with fidelity greater
than \(0.9\) was observed.}
\label{fig:rm-irreversibility-fidelity}
\end{figure}

The final-state fidelities for the independent reverse attempts are shown in
Figure~\ref{fig:rm-irreversibility-fidelity}. The mean final fidelity was
\[
0.035447,
\]
close to the random-state scale
\[
\frac{1}{N}=0.03125.
\]
No independent reverse attempt returned to fidelity larger than
\[
0.9
\]
with the initial state.

The numerical results are summarized in Table~\ref{tab:rm-irreversibility}.

\begin{table}[h]
\centering
\caption{Summary of the \({\bf (RM)}\) irreversibility simulation.}
\label{tab:rm-irreversibility}
\begin{tabular}{lc}
\toprule
Quantity & Value \\
\midrule
Hilbert-space dimension \(N\) & \(32\) \\
Number of \({\bf (RM)}\) steps & \(180\) \\
Step strength \(\epsilon\) & \(0.22\) \\
Independent reverse attempts & \(250\) \\
Final Fubini--Study distance after exact inverse & \(0.000000\) \\
Final Fubini--Study distance after \(H^*\)-history attempt & \(1.420865\) \\
Mean final fidelity, independent reverse attempts & \(0.035447\) \\
Median final fidelity, independent reverse attempts & \(0.024513\) \\
Maximum final fidelity, independent reverse attempts & \(0.213899\) \\
Random-state scale \(1/N\) & \(0.03125\) \\
Recoherence threshold & \(0.9\) \\
Recoherence events at final time & \(0\) \\
Recoherence events at any time & \(0\) \\
\bottomrule
\end{tabular}
\end{table}

This is the second source of irreversibility. It is not a violation of
microscopic unitarity, but it shows that antiunitary time reversal and exact
unitary inversion are different operations for typical GUE histories. For a
realized GUE Hamiltonian, complex conjugation produces the time-reversed
Hamiltonian \(H^*\), whereas exact reversal of the unitary step requires
\(-H\). Thus the effective dynamics is not time-reversal symmetric in the sense
of being generated by the same Hamiltonian history viewed backward.

\subsection{Measurement-related irreversibility}

The third source of irreversibility is measurement-specific. A measurement
record does not retain the exact final ray and does not retain the exact
\({\bf (RM)}\) path that led to it. It retains only a finite-resolution
detector-defined equivalence class. Thus the many microscopic state-space paths
leading to the same record are represented only through their aggregate path
weight.

The particle-device simulations above provide the numerical realization of
this mechanism. In those simulations the relevant state space is a
tensor-product Hilbert space, and the detector record is represented by a
macroscopic equivalence class of the device. The projected \({\bf (RM)}\)
increments in the particle and device coordinates were first checked to be
Gaussian and approximately independent in the induced Fubini--Study metric. In
the simulation of the four coordinates
\[
(\tau_p,s_p;\tau_D,s_D),
\]
the normalized variances were close to one, and the largest off-diagonal
correlation was only
\[
0.00521.
\]
Thus the tensor-product \({\bf (RM)}\) motion has the expected local isotropic
form in the detector coordinates.

The measurement-related irreversibility appears when the device coordinate is
coarse-grained to a macroscopic record class. Let \(Q_D\) denote the projection
onto the recorded device class and \(Q_D^\perp=I-Q_D\). The aggregate weight
outside the recorded device class is
\[
W_{\rm out}^{(D)}
=
\|(I\otimes Q_D^\perp)\Psi\|^2.
\]
The device-stability simulation shows that this quantity becomes negligible in
the device limit. As the ratio \(\delta_D/R_D\) decreases, the aggregate weight
outside the recorded device class rapidly tends to zero. For example, in the
simulation the mean outside-class weight decreases from
\[
4.75\times 10^{-1}
\]
at \(\delta_D/R_D=0.70\) to
\[
8.61\times 10^{-4}
\]
at \(\delta_D/R_D=0.15\), and to approximately
\[
7.31\times 10^{-7}
\]
at \(\delta_D/R_D=0.05\).
This is shown in Figures~\ref{fig:sim15-device-retention}
and~\ref{fig:sim15-device-outside-weight}.

\begin{figure}[h]
\centering
\includegraphics[width=0.78\textwidth]{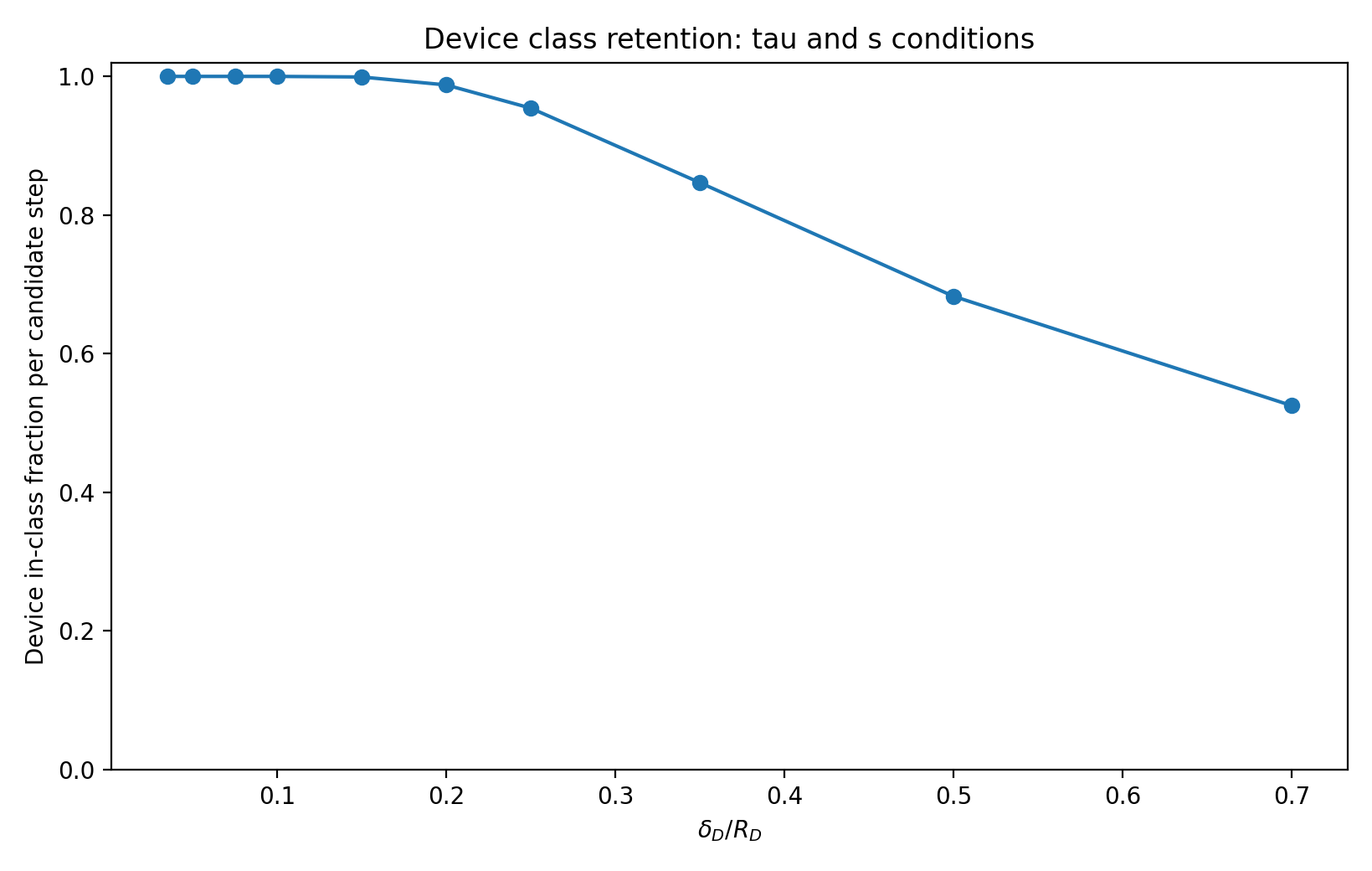}
\caption{Retention of the recorded device equivalence class in the
particle-device simulation. As the device becomes macroscopic relative to its
resolution scale, the state remains in the recorded device class with
overwhelming aggregate path weight.}
\label{fig:sim15-device-retention}
\end{figure}

\begin{figure}[h]
\centering
\includegraphics[width=0.78\textwidth]{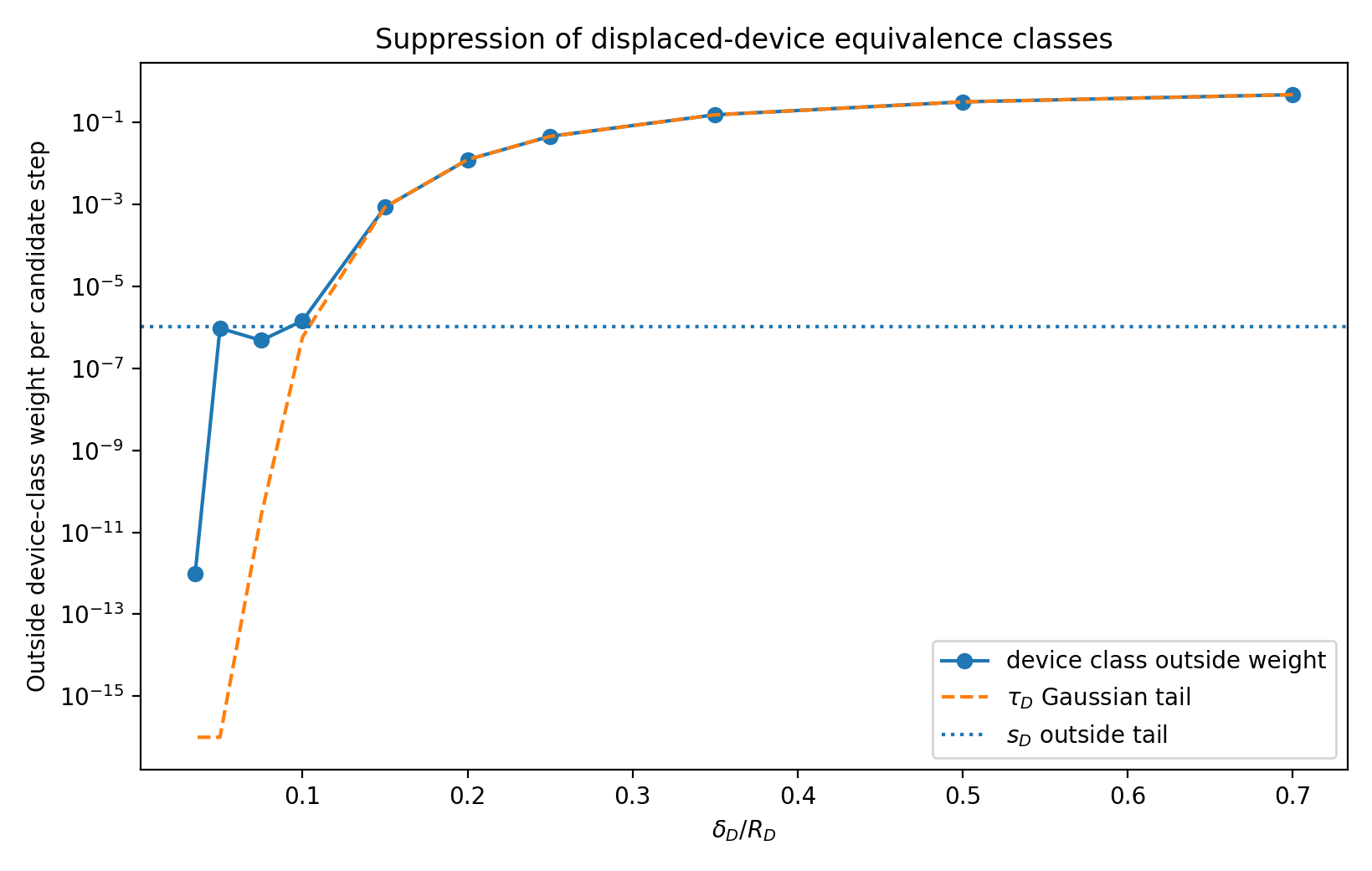}
\caption{Aggregate weight outside the recorded device class as a function of
\(\delta_D/R_D\). In the device limit, the weight of paths leading to displaced
device classes becomes negligible.}
\label{fig:sim15-device-outside-weight}
\end{figure}

At the same time, the particle outcome statistics remain governed by the
calibrated aggregate path weights. In the stable-device simulation, the
observed frequencies agree with the initial particle weights. For example, the
frequencies for the upper outcome were
\[
0.10046,\quad 0.25296,\quad 0.34896,\quad 0.50106,\quad
0.64528,\quad 0.79930,\quad 0.90148
\]
for initial weights
\[
0.10,\quad 0.25,\quad 0.35,\quad 0.50,\quad
0.65,\quad 0.80,\quad 0.90,
\]
respectively. Thus the particle can undergo state reduction with the expected
Born weights while the device remains in a stable macroscopic equivalence
class.

\begin{figure}[h]
\centering
\includegraphics[width=0.78\textwidth]{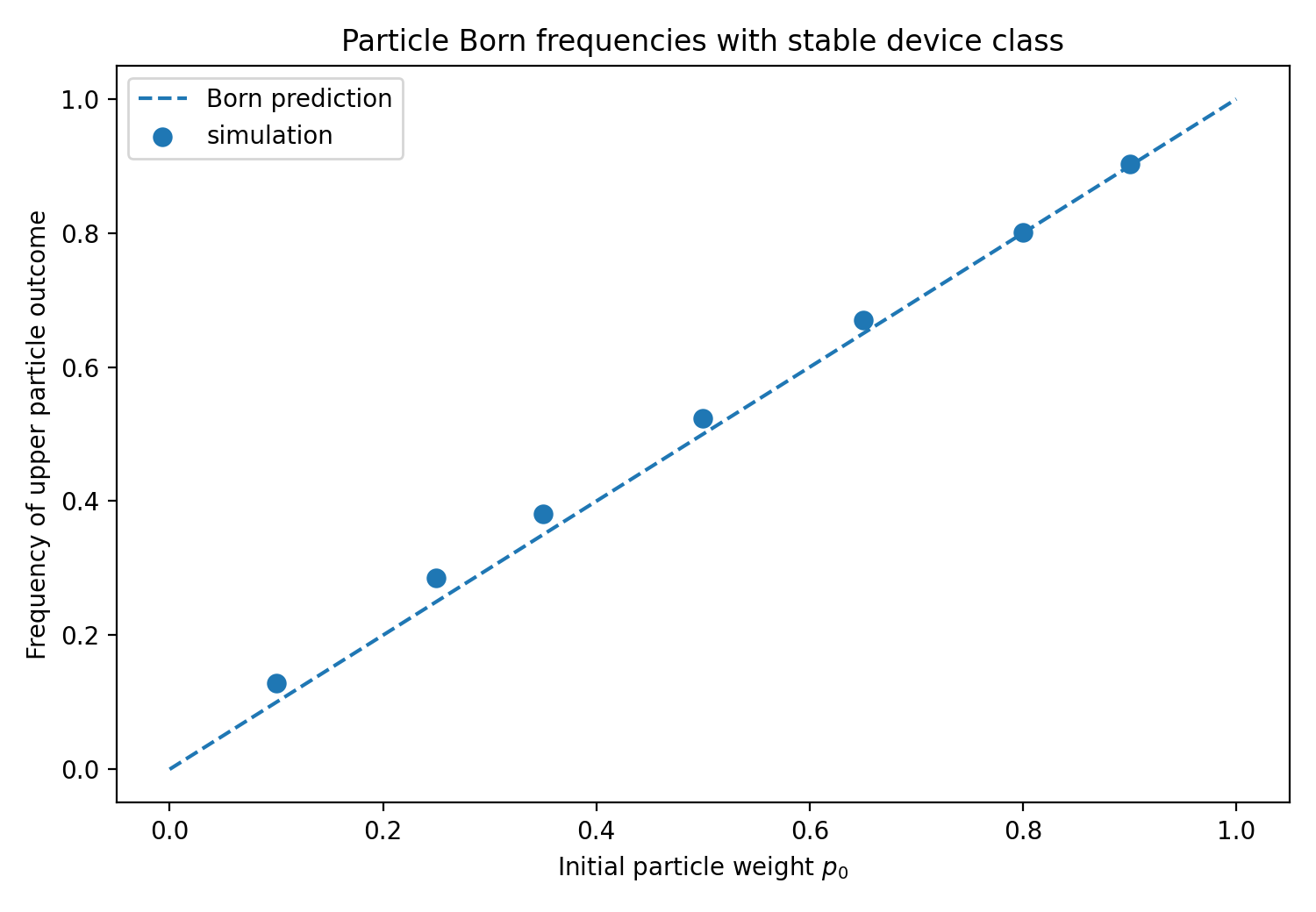}
\caption{Particle outcome frequencies in the presence of a stable device
equivalence class. The observed frequencies agree with the initial particle
weights, while the device remains in the recorded macroscopic class.}
\label{fig:sim15-particle-born-stable-device}
\end{figure}

These simulations show the measurement-related part of the irreversibility.
Once a detector-defined equivalence class has been recorded, the exact
state-space path is no longer part of the record. The record retains only the
finite-resolution device class, and the aggregate weight of paths leading to
macroscopically displaced device classes becomes negligible in the device
limit. Reversing the measurement would therefore require reconstructing both
the exact \({\bf (RM)}\) Hamiltonian history and the microscopic state within
the recorded equivalence class. This information has been discarded by the
record description.

\subsection{Interpretation}

The preceding simulations identify three distinct sources of effective
irreversibility in the {\bf (RM)} framework.

First, the dynamics takes place in projective Hilbert space rather than in
ordinary classical configuration space. Once the realized {\bf (RM)} path
is not retained, recurrence to a prescribed ray is suppressed by the
high-dimensional geometry of state space. In finite dimension this suppression
is already extremely strong, and in the infinite-dimensional limit the
probability of returning to a fixed Fubini--Study neighborhood of the initial
ray vanishes.

Second, typical GUE Hamiltonians are not constrained by antiunitary
time-reversal symmetry. Exact unitary inversion of a realized step is generated
by \(-H\), whereas antiunitary time reversal involves the conjugated
Hamiltonian \(H^*\). These are different operations for a typical GUE
Hamiltonian. Thus the {\bf (RM)} interaction is not simply a
time-symmetric Hamiltonian history viewed backward.

Third, measurement records introduce an additional coarse-graining. A realized
{\bf (RM)} path is reversible if the full Hamiltonian history is retained.
However, a measurement record does not retain that history, nor does it retain
the exact final ray. It retains only a finite-resolution detector-defined
equivalence class. Thus the many microscopic {\bf (RM)} paths leading to
the same record are represented only through their aggregate path weight.

The third source is the one most closely comparable to the irreversibility
usually invoked in decoherence theory. In decoherence theory, the combined
system--environment state evolves unitarily and remains reversible in
principle. Effective irreversibility appears because phase information is
dispersed into many environmental degrees of freedom and then becomes
inaccessible when the environment is traced out or ignored. Recoherence would
require reconstructing an enormous number of system--environment correlations.

The similarity is that both descriptions preserve microscopic unitarity while
making reversal effectively impossible after inaccessible information has been
discarded. The difference is in what is discarded and what is retained. In
decoherence, the inaccessible information is carried by environmental
correlations, and the effective description is usually given by a reduced
density matrix. In the {\bf (RM)} framework, the inaccessible information is
the realized state-space path and Hamiltonian history, while the retained
object is a finite-resolution detector-defined equivalence class.

Thus the {\bf (RM)} measurement record is not merely a diagonal element of a
reduced density matrix. It is a stable finite-resolution class in projective
state space, reached by an aggregate of {\bf (RM)} paths with a calibrated
path weight. The effective arrow of time in the {\bf (RM)} description of
measurement therefore combines high-dimensional state-space recurrence
suppression, lack of antiunitary time-reversal symmetry for typical GUE
histories, and coarse-graining to detector-defined equivalence classes.

\section{Conclusion}
\label{sec:final-conclusion}

The simulations provide numerical support for the main links in the
\({\bf (RM)}\) framework. GUE Hamiltonians generate homogeneous and isotropic
infinitesimal motion in complex projective state space, whereas GOE
Hamiltonians do not. The tangential projection of this motion to localized
classical submanifolds produces Gaussian increments and Brownian scaling in the
classical coordinate. In the \((\tau,s)\)-coordinates, the same GUE-induced
motion is Gaussian with respect to the induced Fubini--Study metric and gives
the stochastic motion associated with localization and detector-defined
equivalence classes.

The microscopic simulations show how detector-defined outcomes arise from the
same state-space diffusion. Reduction-coordinate walks reproduce Born-rule
frequencies, while finite-resolution screen records reproduce the coherent and
which-slit double-slit patterns. The simulations also show how a recorded
endpoint class becomes effectively stable under continuing state-space motion,
giving the equivalence-class version of a Zeno effect.

The macroscopic simulations show the complementary classical regime. When the
tangential diffusion between successive localizations is small compared with
the detector resolution, and when environmental recording returns the state
frequently to the localized sector, the recorded positions remain concentrated
around a Newtonian trajectory. Thus macroscopic classical motion appears as a
conditioned stochastic process on the localized sector of projective Hilbert
space.

The tensor-product simulations extend the picture to particle-device and
two-particle systems. The projected GUE increments in the particle and device
coordinates are orthogonal and Gaussian in the induced Fubini--Study metric.
In the device limit, the particle state can move toward its detector-defined
outcome class while the device state remains in the same macroscopic
equivalence class with overwhelming probability. Components corresponding to
displaced device classes have vanishingly small weight, giving a stable
measurement record at the level of finite-resolution equivalence classes. The
irreversibility of such records is effective: exact unitary paths are reversible
if the full Hamiltonian history is retained, but recurrence to a prescribed ray
becomes negligible in high-dimensional state space once that history is lost.

Taken together, the simulations illustrate how Brownian measurement errors,
Born-rule outcome frequencies, Zeno stability of records, irreversibility of
finite-resolution measurement records, and macroscopic Newtonian motion arise
as different coarse-grained manifestations of random-matrix-induced unitary
diffusion in projective Hilbert space.

\end{document}